\documentclass[aps,prb,twocolumn,superscriptaddress,floatfix,showpacs,10pt,letterpaper]{revtex4-1}

\usepackage{graphicx}
\usepackage{graphics} 

\allowdisplaybreaks[1]
\pdfoutput=1

\newcommand{\rw}{\text{w}}
\newcommand{\RZ}{\R/\Z}
\newcommand{\LSPT}{\text{LSPT}}
\newcommand{\PSPT}{\text{PSPT}}
\newcommand{\HSPT}{\text{HSPT}}
\newcommand{\iTOL}{\text{iTO}_L}
\newcommand{\iTO}{\text{iTO}}

\usepackage{amsthm}
\usepackage{euscript}
\usepackage{epstopdf} 

\newtheoremstyle{wenthm}
  {3pt}
  {3pt}
  {\slshape}
  {}
  {\bfseries}
  {:}
  {.5em}
  {}

\theoremstyle{wenthm}

\theoremstyle{definition}

\begin{document}

\setlength{\extrarowheight}{1mm}

\begin{titlepage}

\title{
Construction of bosonic symmetry-protected-trivial states\\  
and their topological invariants
via $G\times SO(\infty)$ non-linear $\sigma$-models }

\author{Xiao-Gang Wen} 
\affiliation{Department of Physics, Massachusetts Institute of Technology, Cambridge, Massachusetts 02139, USA}
\affiliation{Perimeter Institute for Theoretical Physics, Waterloo, Ontario, N2L 2Y5 Canada}

\begin{abstract} 
It has been shown that the bosonic symmetry-protected-trivial (SPT)
phases with pure gauge anomalous boundary can all be realized via non-linear
$\sigma$-models (NL$\sigma$Ms) of the symmetry group $G$ with various
topological terms.  Those SPT phases (called the pure SPT phases) can be
classified by group cohomology ${\cal H}^d(G,\mathbb{R}/\mathbb{Z})$.  But
there are also SPT phases 
with mixed gauge-gravity anomalous boundary (which will be called the mixed SPT
phases).  Some of the mixed SPT states were also referred as the
beyond-group-cohomology SPT states. In this paper, we show that those
beyond-group-cohomology SPT states are actually within another type of group
cohomology classification.  More precisely,  we show that both the pure and the
mixed SPT phases can be realized by $G\times SO(\infty)$ NL$\sigma$Ms with
various topological terms.  Through the group cohomology ${\cal H}^d[G\times
SO(\infty),\mathbb{R}/\mathbb{Z}]$, we find that the set of our constructed
SPT phases  in $d$-dimensional space-time are described by $ 
 E^d(G)\rtimes
\oplus_{k=1}^{d-1} {\cal H}^k(G,\text{iTO}_L^{d-k})\oplus {\cal
H}^d(G,\mathbb{R}/\mathbb{Z}) $ where $G$ may contain time-reversal.  Here
$\text{iTO}_L^d$ is the set of the topologically-ordered phases in
$d$-dimensional space-time that have no topological excitations, and one has
$\text{iTO}_L^1=\text{iTO}_L^2=\text{iTO}_L^4=\text{iTO}_L^6=0$,
$\text{iTO}_L^3=\mathbb{Z}$, $\text{iTO}_L^5=\mathbb{Z}_2$,
$\text{iTO}_L^7=2\mathbb{Z}$.  
For $G=U(1)\rtimes Z_2^T$ (charge conservation and time-reversal symmetry) ,
we find that the mixed SPT phases beyond ${\cal H}^d[U(1)\rtimes
Z_2^T,\mathbb{R}/\mathbb{Z}]$ are described by $\mathbb{Z}_2$ in 3+1D,
$\mathbb{Z}$ in 4+1D, $3\mathbb{Z}_2$ in 5+1D, and $4\mathbb{Z}_2$ in 6+1D.
Our construction also gives us the topological invariants that fully
characterize the corresponding SPT and iTO phases.  Through several examples,
we show how can the universal physical properties of SPT phases be obtained
from those topological invariants.

\end{abstract}


\maketitle

\end{titlepage}

{\small \setcounter{tocdepth}{1} \tableofcontents }

\section{Introduction and results}
\label{intro}

\subsection{Gapped quantum liquid without  topological excitations}

In 2009, in a study of the Haldane phase\cite{H8364} of spin-1 chain using
space-time tensor network,\cite{GW0931} it was found that, from the
entanglement point of view, the Haldane state is really a trivial product
state. So the non-trivialness of Haldane phase must be contained in the way how
symmetry and short-range entanglement\cite{CGW1038} get intertwined.  This led
to the notion of \emph{symmetry protected trivial (SPT) order} (also known as
\emph{symmetry protected topological order}).  Shortly after, the concept of
SPT order allowed us to classify\cite{CGW1107,CGW1128,SPC1139} all 1+1D
gapped phases for interacting bosons/spins and
fermions.\cite{PBT1039,FK1009,TPB1102,FK1103,PBT1225} This result is quickly
generalized to higher dimensions where a large class of SPT phases is
constructed using group cohomology theory.\cite{CLW1141,CGL1314,CGL1204} 

Such a higher-dimension construction is based on $G$ non-linear $\si$-model
(NL$\si$M)\cite{CGL1314,CGL1204,LW1305}
\begin{align}
\label{Lg}
\cL= \frac{1}{\la} (\prt g)^2 + \ii L^d_\text{top}(g^{-1}\prt g), \ \
g(x)\in G,
\end{align}
with topological term $L^d_\text{top}$ in $\la\to \infty$ limit.  Since the
topological term $L^d_\text{top}$ is classified by the elements in group
cohomology class $\cH^d(G,\RZ)$,\cite{CGL1314,CGL1204,LW1305} this allows us to
show that such kind of SPT states are classified by $\cH^d(G,\RZ)$. (See
Appendix \ref{gcoh} for an introduction of group cohomology.) Later, it was
realized that there also exist time-reversal protected SPT states that are
beyond the $\cH^d(G,\RZ)$ description.\cite{VS1306,WS1334,BCF1372} 

We like to point out that there are many other ways to construct SPT states,
which include Chern-Simons theories,\cite{LV1219,SL1301} NL$\si$Ms of symmetric
space,\cite{VS1306,X1321,OCX1325,XS1372,BRX1315,YX1420}  projective
construction,\cite{YW1328,MW1469,LMY1476} domain wall decoration,\cite{CLV1407}
string-net,\cite{BCF1372} layered construction,\cite{WS1334} higher gauge
theories,\cite{YG1494,XY1486,BX1571} \etc.

SPT states are \emph{gapped quantum liquids},\cite{CGW1038,ZW1490}
characterized by having no \emph{topological excitations},\cite{LW1384,KW1458}
and having no \emph{topological order}.\cite{Wtop,WNtop,Wrig,KW9327} $E_8$
bosonic quantum Hall state\cite{LV1219,PMN1372} described by the $E_8$
$K$-matrix\cite{BW9045,R9002,FZ9117,FK9169,WZ9290,FS9333,W9505} is also a
gapped quantum liquid with no topological excitations, but it has a non-trivial
topological order. We will refer such kind of topologically ordered states as
invertible topologically ordered (iTO) states\cite{KW1458,F1478} (see Table
\ref{invTop}).  Bosonic SPT and iTO states are simplest kind
of gapped quantum liquids.  In this paper, we will try to develop a systematic
theory for those phases.  The main result is \eqn{SPTHH} which generalizes the
$\cH^d(G,\RZ)$ description of the SPT phases, so that the new description also
include the time-reversal protected SPT phases beyond the $\cH^d(G,\RZ)$
description.  This result is derived in Section \ref{pmSPT}. Applying
\eqn{SPTHH} to simple symmetry groups,  we obtain Table \ref{SPT} for the SPT
phases produced by NL$\si$Ms.

\begin{table*}[t]
\caption{
The L-type bosonic iTO phases realized by the $SO(\infty)$ NL$\si$Ms in
$d$-dimensional space-time form an Abelian group $\si\iTOL^d$.  (The meaning
of ``L-type'' is defined in Section \ref{RPtopinv}, and one can ignore such a
qualifier in the first reading.) More general L-type bosonic iTO phases
realized by the NL$\si$Ms , Chern-Simons theories \etc form a bigger Abelian
group $\iTOL^d$.  The generating topological invariants $W^d_\text{top}(\Ga)$
are also listed.
} \label{invTop}
 \centering
 \begin{tabular}{ |c|cc|cc| }
 \hline
 dim. & $\si\iTOL^d$ & $W^d_\text{top}$ &   $\iTOL^d$ & $W^d_\text{top}$\\
\hline
$d=0+1$ & 0 & & 0 &\\
$d=1+1$ & 0 & & 0 &\\
$d=2+1$ & $\Z$ & $\om_3$ & $\Z$ & $\frac13 \om_3$ \\
$d=3+1$ & 0 & & 0 & \\
$d=4+1$ & $\Z_2$ & $\frac12  \rw_2\rw_3$ &  $\Z_2$ & $\frac12  \rw_2\rw_3$\\
$d=5+1$ & 0 & & 0 &\\
$d=6+1$ & $2\Z$ & $\om^{p_1^2}_7,\ \om^{p_2}_7$ & $2\Z$  &
$\frac{\om_7^{p_1^2}-2\om_7^{p_2}}{5}$,
$\frac{-2\om_7^{p_1^2}+5\om_7^{p_2}}{9} $
 \\
\hline
 \end{tabular}
\end{table*}

\begin{table*}[t]
 \caption{The L-type bosonic SPT phases realized by the $G\times SO(\infty)$
NL$\si$Ms, which are described by $E^d(G)\rtimes \oplus_{k=1}^{d-1}
H^k(BG,\iTOL^{d-k})\oplus \cH^d(G,\RZ) $.  The results in black are the pure
SPT phases described by $\cH^d(G,\RZ)$ first discovered in \Ref{CGL1314}.  The
pure SPT states have boundaries that carry only pure ``gauge'' anomaly.  The
results in blue are the mixed SPT phases described by $\oplus_{k=1}^{d-1}
H^k(BG,\iTOL^{d-k})$.  The results in red are the extra mixed SPT phases
described by $E^d(G)$.  The mixed SPT states have boundaries that carry mixed
gauge-gravity anomaly.
} 
\label{SPT}
 \centering
 \begin{tabular}{ |c|c|c|c|c|c|c|c| }
 \hline
symmetry & 0+1D & 1+1D & 2+1D &3+1D & 4+1D & 5+1D & 6+1D  \\
\hline
$Z_n$ & $\Z_n$ & 0 & $\Z_n$ & 0 & ${\Z_n}\oplus\blue{\Z_n}$ & $\blue{\Z_{\<n,2\>}}$ & ${\Z_n}\oplus\blue{\Z_n\oplus \Z_{\<n,2\>}}$  \\
$Z_2^T$ & 0 & $\Z_2$ & 0 & ${\Z_2}\oplus\blue{\Z_2}$ & 0 & ${\Z_2}\oplus \blue{2\Z_2}$ & $\blue{\Z_2}$ \\
$U(1)$ & $\Z$ & 0 & $\Z$ & 0 & ${\Z}\oplus \blue{\Z}$ & 0 & ${\Z}\oplus \blue{\Z\oplus \Z_2}$  \\
$U(1)\rtimes Z_2=O_2$ & $\Z_2$ & $\Z_2$ & $\Z\oplus \Z_2$ & ${\Z_2}$ & ${2\Z_2}\oplus \blue{\Z_2}$ & ${2\Z_2}\oplus \blue{2\Z_2}$ & ${\Z\oplus 2\Z_2}\oplus \blue{\Z\oplus 3\Z_2}$ \\
$U(1)\times Z_2^T$ &0 & $2\Z_2$ & 0 & ${3\Z_2}\oplus \blue{\Z_2}$ & 0 & ${4\Z_2}\oplus \blue{3\Z_2}$ & $\blue{2\Z_2}\oplus \red{\Z_2}$ \\
$U(1)\rtimes Z_2^T$ &$\Z$ &$\Z_2$ & $\Z_2$ & ${2\Z_2}\oplus \blue{\Z_2}$ & ${\Z\oplus \Z_2}\oplus \blue{\Z}$ & ${2\Z_2}\oplus \blue{2\Z_2}$ & ${2\Z_2}\oplus \blue{3\Z_2}\oplus \red{\Z_2}$ \\
\hline
 \end{tabular}
\end{table*}

\subsection{Probing SPT phases and topological invariants}

The above is about the construction of SPT states.  But how to probe and
measure different SPT orders in the ground state of a generic system?  The SPT
states have no topological order. Thus, their fixed-point partition function
$Z_\text{fixed}(M^d)$ on a closed space-time manifold $M^d$ is trivial
$Z_\text{fixed}(M^d)=1$,\cite{KW1458}  and cannot be used to probe different
SPT orders.  However, if we add the $G$-symmetry twists\cite{W1447,HW1339} to
the space-time by gauging the on-site symmetry $G$,\cite{LG1220,RZ1232,SCR1325} we may get a
non-trivial fixed-point partition function $Z_\text{fixed}(M^d,A)\in U(1)$
\emph{which is a pure $U(1)$ phase}\cite{KW1458} that depends on $A$. Here $A$
is the background non-dynamical gauge field that describes the symmetry twist.
The fixed-point partition function $Z_\text{fixed}(M^d,A)$ is robust against
any smooth change of the local Lagrangian $\del \cL$ that preserve the
symmetry, and is a topological invariant. Such type of topological invariants
should completely describe the SPT states that have no topological order.  In
this paper, we will express such universal fixed-point partition function in
terms of topological invariant $W^d_\text{top}$ (which is a $d$-form, or more precisely, a $d$-cocycle):
\begin{align}
	Z_\text{fixed}(M^d,A) =\ee^{\ii \int_{M^d} 2\pi W^d_\text{top}(A,\Ga)}
\end{align}
where $\Ga$ is the connection on $M^d$.  We will use $W^d_\text{top}$ to
characterize the SPT phases.

Even without the symmetry, the fixed-point partition function
$Z_\text{fixed}(M^d)$ can still be a pure $U(1)$ phase that depend on the
topologies of space-time. In this case, the fixed-point partition function
describes an iTO state.\cite{KW1458}  Thus, we can also use $W^d_\text{top}$ to
characterize the iTO phases.  We believe that the function, $\ee^{\ii
\int_{M^d} 2\pi W^d_\text{top}(A,\Ga)}$, that maps various closed space-time
manifolds $M^d$ with  various $G$-symmetry twist $A$ to the $U(1)$ value,
completely characterizes the iTO phases and the SPT phases.\cite{W1447} So in
this paper, we will often use $W^d_\text{top}$ to label/describe iTO and SPT
phases.

We like to point out that the topological invariant $W^d_\text{top}(A)$ is
given by a cocycle $\om_d$ in $\cH^d(G\times SO(\infty),\RZ)$.  Eq.
(\ref{Wdomd}) tells us how to calculate $\ee^{\ii \int_{{M^d}} 2\pi
W^d_\text{top}(A)}$, from $\om_d$, the space-time manifold $M^d$, and the
symmetry-twist $A$.  So $\ee^{\ii \int_{{M^d}} 2\pi W^d_\text{top}(A)}$ is well
defined.

\begin{table}[t]
\caption{The L-type $U(1)$ SPT phases.  
} \label{U1SPT}
 \centering
 \begin{tabular}{ |c|c|c| }
 \hline
 $d=$ & $\LSPT_{U(1)}^d$ & generators $W^d_\text{top}$ \\[1mm]
\hline
$0+1$ & $\Z$ & $a$ \\[1mm]
\hline
$1+1$ & 0 &   \\[1mm]
\hline
$2+1$ & $\Z$ &  $ac_1$ \\[1mm]
\hline
$3+1$ & 0 &    \\[1mm]
\hline
$4+1$ & $\Z\oplus \blue{\Z}$ &  $ac_1^2$, $\blue{\frac13 ap_1}$ \\[1mm]
\hline
$5+1$ & 0 &    \\[1mm]
\hline
$6+1$ & $\Z \oplus \blue{\Z\oplus \Z_2}$ &  $ac_1^3$, 
$\blue{ \frac13 ac_1p_1}$, $\blue{\frac12 \rw_2\rw_3c_1}$  \\[1mm]
\hline
 \end{tabular}
\end{table}

\subsection{Simple SPT phases and their physical properties}

In Tables \ref{U1SPT}, \ref{ZnSPT},  \ref{UTspinSPT}, \ref{BTISPT},
\ref{Z2TSPT}, \ref{O2SPT}, we list the generators  $W^d_\text{top}(A,\Ga)$ of those
topological invariants for simple SPT phases.  The $U(1)$-symmetry twist on the
space-time $M^d$ is described by a vector potential one-form $A$ and the $Z_n$
-symmetry twist is described by a vector potential one-form $A_{Z_n}$ with
vanishing curl that satisfies $\oint A_{Z_n}=0$ mod $2\pi/n$.  However, in the
tables, we use the normalized one form $a\equiv A/2\pi$ and $a_1\equiv n
A_{Z_n}/2\pi$.  Also in the table, $c_1=\dd a$ is the first Chern-Class,
$\rw_i$ is the Stiefel-Whitney classes and $p_1$ the first Pontryagin classes
for the tangent bundle of $M^d$.  The results in black are for the pure SPT
phases (which are defined as the SPT phases described by $\cH^d(G,\RZ)$).  The
results in blue are for the mixed SPT phases described by $\oplus_{k=1}^{d-1}
H^k(BG,\iTOL^{d-k})$.  The results in red are for the extra mixed SPT phases
described by $E^d(G)$ (see \eqn{SPTHH}).

Those topological invariants fully characterize the
corresponding topological phases. All the universal physical
properties\cite{LG1220,VS1306,MKF1331,W1447,YW1427,BRX1424,WSW1456} of the
topological phases can be derived from those topological invariants. This is
the approach used in \Ref{W1447}. In the following, we will discuss some of the
simple cases as examples.  We find that the topological invariants allow us to
``see'' and obtain many universal physical properties easily.

\subsubsection{$U(1)$ SPT states in Table \ref{U1SPT}}

The 0+1D $U(1)$ SPT phases are classified by $k\in \Z$ with a gauge topological
invariant
\begin{align}
	W^1_\text{top}(A)= k \frac{A}{2\pi}.
\end{align}
It describes a $U(1)$ symmetric ground state with charge $k$.  The $\Z$ class
of 2+1D $U(1)$ SPT phases are generated by $ W^3_\text{top}=ac_1$, or
\begin{align}
	W^3_\text{top}(A)= \frac{A\dd A}{(2\pi)^2},
\end{align}
where $A\dd A$ is the wedge product of one-form $A$ and two-form $\dd A$: $A\dd
A =A \wedge \dd A$.  Those SPT states have even-integer Hall conductances
$\si_{xy}= \frac{\text{even}}{2\pi}$.\cite{LV1219,CW1235,LW1305,SL1301}

The above are the pure $U(1)$ SPT states whose boundary has only pure $U(1)$
anomalies.  The $\Z$ class of 4+1D $U(1)$ SPT phases introduced in
\Ref{WGW1489} are mixed SPT states. The generating state is described by
(see Appendix \ref{giTO3})
\begin{align}
\label{W5U1}
W^5_\text{top}(A,\Ga)= \frac13 \frac{A p_1}{2\pi} = - \frac13 \bt(A/2\pi) \om_3 =  
-\frac13 \frac{\dd A}{2\pi} \om_3
\end{align}
where $\om_3$ is a gravitational Chern-Simons three-form: $\dd \om_3=p_1$.
Also $\bt$ is the natural map $\bt: \cH^d(G,\RZ) \to \cH^{d+1}(G,\Z)$ that maps
$a \in \cH^1(U(1),\RZ)$ to $\bt(a)=c_1\in \cH^2(U(1),\Z)$.  One of the physical
properties of such a state is its dimension reduction: we put the state on
space-time $M^5=M^2\times M^3$ and put $2\pi$ $U(1)$ flux through $M^2$. In
large $M^3$ limit, the effective theory on $M^3$ is described by effective
Lagrangian $W^3_\text{top}(\Ga)= -\frac13 \om_3$, which is a $E_8$
quantum Hall state with chiral central charge $c=8$.  If $M^3$ has a boundary,
the boundary will carry the gapless chiral edge state of $E_8$ quantum
Hall state.  Note that the boundary of $M^3$ can be viewed as the core of a
$U(1)$ monopole (which forms a loop in four spatial dimensions).  So the core
of a $U(1)$ monopole will carry the gapless chiral edge state of $E_8$
quantum Hall state.

Since the monopole-loop in 4D space can be viewed as a boundary of $U(1)$
vortex sheet in 4D space, the above physical probe also leads to a mechanism of
the $U(1)$ SPT states: we start with a $U(1)$ symmetry breaking state. We then
proliferate the $U(1)$ vortex sheets to restore the $U(1)$ to obtain a trivial
$U(1)$ symmetric state.  However, if we bind the $E_8$ state to the vortex
sheets, proliferate the new $U(1)$ vortex sheets will produce a non-trivial
$U(1)$ SPT state. In general, a probe of SPT state will often leads to a
mechanism of the SPT state.

If the mixed $U(1)$ SPT state is realized by a continuum field theory, then we
can have another topological invariant: we can put the state on a spatial
manifold of topology $\C P^2$ or $T^4=(S^1)^4$. Since $ \int_{\C P^2}\frac 13
p_1 -\int_{T^4}\frac 13 p_1 =1$, we find that the ground state on $\C P^2$ and
on $T^4$ will carry different  $U(1)$ charges (differ by one unit).  We like to
stress that the above result is a field theory result, which requires the
lattice model to have a long correlation length much bigger than the lattice
constant.

\begin{table}[t]
\caption{The L-type $Z_2$ SPT phases.  
} \label{ZnSPT}
 \centering
 \begin{tabular}{ |c|c|c| }
 \hline
 $d=$ & $\LSPT_{Z_2}^d$ & generators $W^d_\text{top}$ \\[1mm]
\hline
$0+1$ & $\Z_2$ & $\frac 12 a_1$ \\[1mm]
\hline
$1+1$ & 0 &   \\[1mm]
\hline
$2+1$ & $\Z_2$ &  $\frac12 a_1^3$ \\[1mm]
\hline
$3+1$ & 0 &    \\[1mm]
\hline
$4+1$ & $\Z_2\oplus \blue{\Z_2}$ &  
        $\frac12 a_1^5$, $\blue{\frac12 a_1p_1}$\\[1mm]
\hline
$5+1$ & $\blue{\Z_2}$ &  $\blue{\frac12 a_1\rw_2\rw_3}$  \\[1mm]
\hline
$6+1$ & $\Z_2 \oplus \blue{2 \Z_2}$ &  
$\frac12 a_1^7$, $\blue{ \frac12 a_1^3p_1}$, $\blue{\frac12 a_1^2 \rw_2\rw_3}$ \\[1mm]
\hline
 \end{tabular}
\end{table}

\subsubsection{$Z_2$ SPT states in Table \ref{ZnSPT}}
\label{secZ2SPT}

The 2+1D $Z_2$ SPT state described by
\begin{align}
	W^3_\text{top}(A_{Z_2}) = \frac12 a_1^3 
\end{align}
is the first discovered SPT state beyond 1+1D.\cite{CLW1141} Here
$a_1=A_{Z_2}/2\pi$ is the $Z_2$-connection that describes the $Z_2$-symmetry
twist on space-time.  However, $a_1^3$ is not the wedge product of three
one-forms: $a_1^3\neq a_1\wedge a_1 \wedge a_1$. $a_1^3$ is the cup product
$a_1^3\equiv a_1\cup a_1 \cup a_1$, after we view $a_1$ as a 1-cocycle in
$H^1(M^3,\Z_2)$.  The cup-product of cocycles generalizes the wedge product of
differential forms.  

But how to compute the action amplitude $\ee^{\ii \int_{M^3}
2\pi W^3_\text{top}(A_{Z_2})} =\ee^{\ii \pi \int_{M^3} a_1^3 }$ that involves
cup-products?  One can use the defining relation \eqn{Wdomd} to compute
$\ee^{\ii \int_{M^3} 2\pi W^3_\text{top}(A_{Z_2})}$.
First, we note that
the cocycle $a_1 \in \cH^1(Z_2,\Z_2)$ is given by
\begin{align}
 a_1(1)=1, \ \ \
 a_1(-1)=-1. 
\end{align}
where $\{1,-1\}$ form the group $Z_2$. The cocylce for the cup-product, $a_1^3$,
is simply given by 
\begin{align}
 a_1^3(g_0,g_1,g_2)= a_1(g_0) a_1(g_1) a_1(g_2),
\end{align}
which is a cocycle in $\cH^3(Z_2,\Z_2)$.  Then $\om_3(g_0,g_1,g_2)=\frac12
a_1^3(g_0,g_1,g_2)$ is a cocycle in $\cH^3(Z_2,\RZ)$, that describes our $Z_2$
SPT state.  This allows us to use \eqn{Wdomd} to compute $\ee^{\ii \int_{M^3}
2\pi W^3_\text{top}(A_{Z_2})}$.

However, there are simpler ways to compute $\ee^{\ii \int_{M^3}
2\pi W^3_\text{top}(A_{Z_2})}$.  According to the Poincar\'e duality, an $i$-cocycle
$x_i$ in a $d$-dimensional manifold $M^d$ is dual to a $(d-i)$-cycle (\ie a
$(d-i)$-dimensional closed sub-manifold), $X^{d-i}$, in $M^d$.  In our case,
$a_1$ is dual to a 2D closed surface $N^2$ in $M^3$, and the 2D closed surface
is the surface across which we perform the $Z_2$ symmetry twist.  We will
denote the Poincar\'e dual of $x_i$ as $[x_i]^*=X^{d-i}$.  Under the Poincar\'e
duality, the cup-product has a geometric meaning: Let $X^{d-i}$ be the dual of
$x_i$ and $Y^{d-j}$ be the dual of $y_j$.  Then the cup-product of $x_i\cup
y_i$ is a $(i+j)$-cocycle $z_{i+j}$, whose dual is a $(d-i-j)$-cycle
$Z^{d-i-j}$. We find that $Z^{d-i-j}$ is simply the intersection of $X^{d-i}$
and $Y^{d-j}$: $Z^{d-i-j}=X^{d-i}\cap Y^{d-j}$.  In other words
\begin{align}
 x_i\cup y_i=z_{i+j} \ \ \longleftrightarrow \  \ [x_i]^*\cap [y_i]^*=[z_{i+j}]^* .
\end{align}
So to calculate $\int_{M^3} a_1^3$, we need to choose three different 2D
surfaces $N^2_1$, $N^2_2$, $N^2_3$ that describe the equivalent $Z_2$
symmetry twists $a_1$.  Then
\begin{align}
\label{a13N123}
 \int_{M^3} a_1^3 = \text{number of the points in }
N^2_1 \cap N^2_2 \cap N^2_3  \text{ mod } 2.
\end{align}

\begin{figure}[tb]
\begin{center}
\includegraphics[scale=0.6]{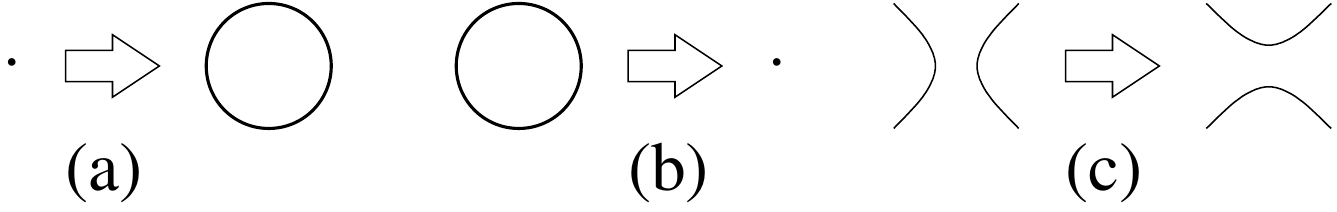} \end{center}
\caption{ 
(a) a loop creation.
(b) a loop annihilation.
(c) a line reconnection.
}
\label{loops}
\end{figure}

There is another way to calculate $\int_{M^3} a_1^3$. Let $N^2$ be a 2D surface
in space-time $M^3$ that describe the $Z_2$ symmetry twist $a_1$.  We choose
the space-time $M^3$ to have a form $M^2\times S^1$ where $S^1$ is the time
direction. At each time slice, the surface of symmetry twist, $N^2$, becomes
loops in the space $M^2$.  Then (see Fig. \ref{loops})
\begin{align}
\label{a13loop}
 \int_{M^3} a_1^3 = &\text{ number of loop creation/annihilation } +
\nonumber\\
&\text{ number of line reconnection} \ \ \ \text{ mod }2 ,
\end{align}
as we go around the time loop $S^1$.  (Such a result leads to the picture in
\Ref{LG1220}.)

\begin{figure}[tb]
\begin{center}
\includegraphics[scale=0.6]{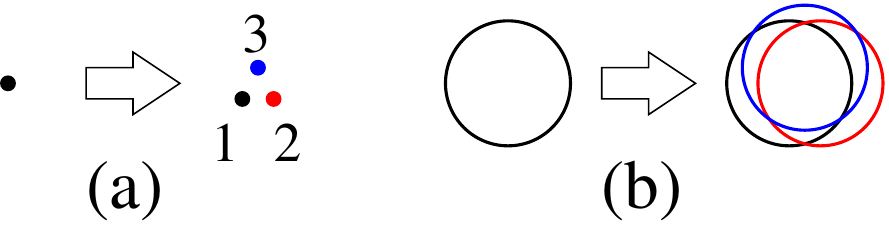} \end{center}
\caption{ 
(Color online)
(a) a point is split into three points.
(b) a surface $N^2$ is split into three surfaces  $N^2_1$, $N^2_2$, $N^2_3$.
}
\label{split}
\end{figure}

\begin{figure}[tb]
\begin{center}
\includegraphics[scale=0.55]{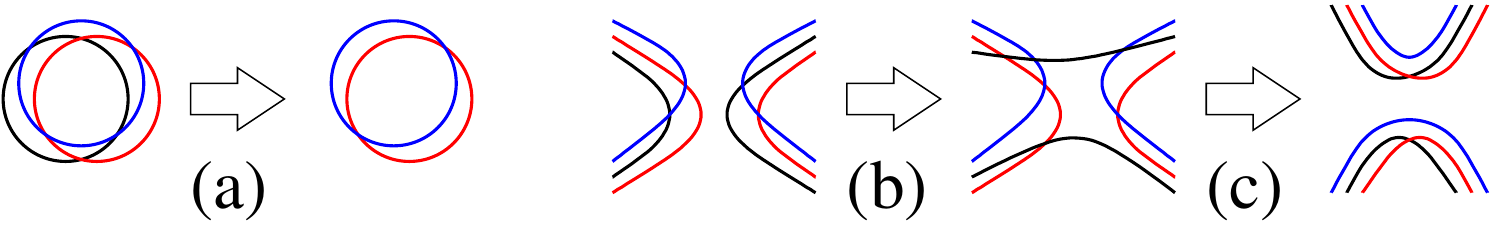} \end{center}
\caption{ 
(Color online)
\textbf{Loop annihilation}: (a)  as we shrink the black circle to a point, the black
line sweeps across the intersection of red and blue line once.  This means that
$N^2_1$, $N^2_2$, $N^2_3$ intersect once in the loop annihilation/creation
process.  \textbf{Line reconnection}: as we deform the black lines in process (b), the
black lines sweep across the intersection of red and blue lines once.  But in
process (c), no line sweeps across the intersection of the other
two lines.  This means that $N^2_1$, $N^2_2$, $N^2_3$ intersect once in the
line reconnection process.
}
\label{deform}
\end{figure}

To show the relation between \eqn{a13N123} and \eqn{a13loop}, we split each
point on $N^2$ into three points 1, 2, 3 (see Fig. \ref{split}), which split
$N^2$ into three nearby 2D surfaces $N_1^2$, $N_2^2$, and $N_3^2$.  Then from
Fig.  \ref{deform}, we can see the relation between \eqn{a13N123} and
\eqn{a13loop}.

\begin{figure}[tb]
\begin{center}
\includegraphics[scale=0.6]{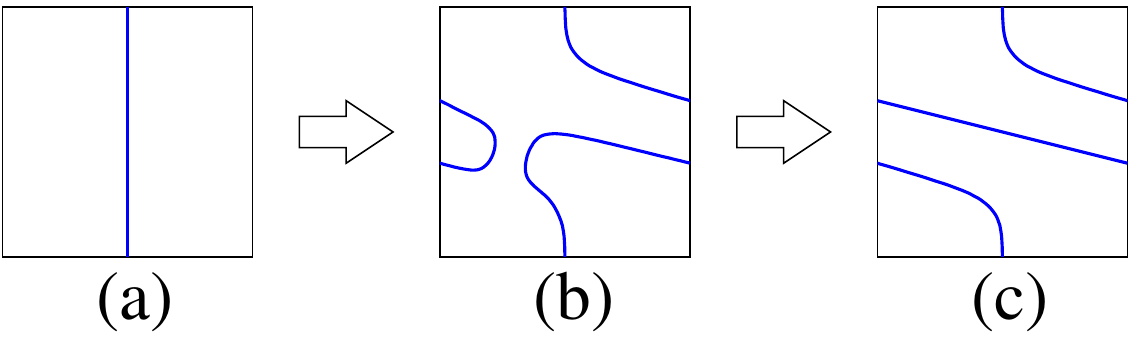} \end{center}
\caption{ 
(Color online)
(a) a $Z_2$ symmetry twist on a torus.
(c) the $Z_2$ symmetry twist obtained from (a) by double Dehn twist.
(a$\to$b$\to$c) contains a line reconnection.
}
\label{Z2torus}
\end{figure}

Eqn. (\ref{a13loop}) is consistent with the result in \Ref{HW1339} where we
considered a space-time $T^2\times I$, where $I=[0,1]$ is an 1D line segment for
time $t\in [0,1]=I$.  Then we added a $Z_2$ symmetry twist on a torus $T^2$ at
$t=0$ (see Fig. \ref{Z2torus}a).  Next, we evolved such a $Z_2$-twist at $t=0$
to the one described by Fig. \ref{Z2torus}c at $t=1$, via the process Fig.
\ref{Z2torus}a $\to$ Fig. \ref{Z2torus}b $\to$ Fig.  \ref{Z2torus}c.  Last, we
clued the tori at $t=0$ and at $t=1$ together to form a closed space-time,
\emph{after we do a  double Dehn twist on one of the tori}.  \Ref{HW1339}
showed that the value of the topological invariant on such a space-time with
such a  $Z_2$-twist is non-trivial: $\int_{M^3} a_1^3 =1$ mod 2, through an
explicit calculation.  In this paper, we see that the non-trivial value comes
from the fact that there is  one line-reconnection in the process Fig.
\ref{Z2torus}a $\to$ Fig. \ref{Z2torus}b $\to$ Fig.  \ref{Z2torus}c. 

Using the result \eqn{a13loop}, we can show that the end of the $Z_2$-symmetry
twist line (which is called the monodromy defect\cite{W1447}) must carry a
fractional spin $\frac14 $ mod 1 and a semion fractional
statistics.\cite{LG1220}

Let us use $\big |\bmm \includegraphics[scale=0.33]{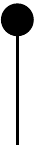}\emm\big
\>_\text{def}$ to represent the many-body wave function with a monodromy
defect.  We first consider the spin of such a defect to see if the spin is
fractionalized or not.\cite{FFN0683,Wang10}  Under a $360^\circ$ rotation, the
monodromy defect (the end of $Z_2$-twist line)  is changed to $\big
|\bmm \includegraphics[scale=0.33]{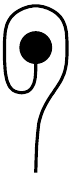}\emm \big \>_\text{def} $.  Since $\big
|\bmm \includegraphics[scale=0.33]{def1}\emm \big \>_\text{def} $ and $\big
|\bmm \includegraphics[scale=0.33]{def3}\emm \big \>_\text{def} $ are alway
different even after we deform and reconnect the $Z_2$-twist lines,
$\big |\bmm \includegraphics[scale=0.33]{def1}\emm \big \>_\text{def} $ is not
an eigenstate of $360^\circ$ rotation and does not carry a definite spin.

To construct the  eigenstates of $360^\circ$ rotation, let us make another
$360^\circ$ rotation to $\big |\bmm \includegraphics[scale=0.33]{def3}\emm \big
\>_\text{def}$.  To do that, we first use the line reconnection move in Fig.
\ref{loops}c, to change $\big |\bmm \includegraphics[scale=0.33]{def3}\emm \big
\>_\text{def} \to - \big |\bmm \includegraphics[scale=0.33]{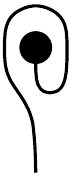}\emm \big
\>_\text{def} $.  A $360^\circ$ rotation on $\big |\bmm
\includegraphics[scale=0.33]{def2}\emm \big \>_\text{def} $ gives us $\big
|\bmm \includegraphics[scale=0.33]{def1}\emm \big \>_\text{def} $.

We see that a $360^\circ$ rotation changes $(\big |\bmm
\includegraphics[scale=0.33]{def1}\emm \big \>_\text{def}, \big |\bmm
\includegraphics[scale=0.33]{def3}\emm \big \>_\text{def} )$ to $( \big |\bmm
\includegraphics[scale=0.33]{def3}\emm \big \>_\text{def}, -\big |\bmm
\includegraphics[scale=0.33]{def1}\emm \big \>_\text{def} )$.  We find that
$\big |\bmm \includegraphics[scale=0.33]{def1}\emm \big \>_\text{def} + \ii
\big |\bmm \includegraphics[scale=0.33]{def3}\emm \big \>_\text{def} $ is the
eigenstate of the $360^\circ$ rotation with eigenvalue $-\ii$, and $\big
|\bmm \includegraphics[scale=0.33]{def1}\emm \big \>_\text{def} - \ii \big
|\bmm \includegraphics[scale=0.33]{def3}\emm \big \>_\text{def} $ is the other
eigenstate of the $360^\circ$ rotation with eigenvalue $\ii$.  So the
defect $\big |\bmm \includegraphics[scale=0.33]{def1}\emm \big \>_\text{def}
+ \ii \big |\bmm \includegraphics[scale=0.33]{def3}\emm \big \>_\text{def} $
has a spin $-1/4$, and the defect $\big |\bmm
\includegraphics[scale=0.33]{def1}\emm \big \>_\text{def} - \ii \big |\bmm
\includegraphics[scale=0.33]{def3}\emm \big \>_\text{def} $ has a spin $1/4$.

\begin{figure}[tb]
\centerline{
\includegraphics[height=0.7in]{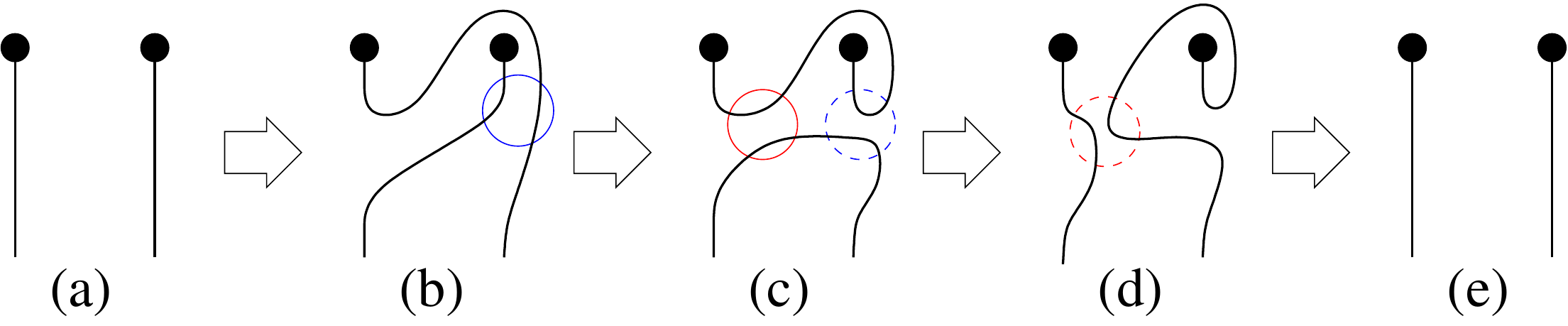}
}
\caption{
(Color online)
Deformation of the $Z_2$-twist lines and two reconnection moves, plus an
exchange of two defects and a $360^\circ$ rotation of one of the defects, change the configuration (a) back to itself.  Note that
from (a) to (b) we exchange the two defects, and from (d) to (e) we rotate of
one of the defect by $360^\circ$.  The combination of those moves do not
generate any phase, since the number of the reconnection move is even.
}
\label{exch}
\end{figure}

If one believes in the spin-statistics theorem, one may guess that the defects
$\big |\bmm \includegraphics[scale=0.33]{def1}\emm \big \>_\text{def} + \ii
\big |\bmm \includegraphics[scale=0.33]{def3}\emm \big \>_\text{def} $ and
$\big |\bmm \includegraphics[scale=0.33]{def1}\emm \big \>_\text{def} - \ii
\big |\bmm \includegraphics[scale=0.33]{def3}\emm \big \>_\text{def} $ are
semions.  This guess is indeed correct.  Form Fig. \ref{exch}, we see that we
can use deformation of $Z_2$-twist lines and two reconnection moves to generate
an exchange of the two defect and a $360^\circ$ rotation of one of the defect.
Such operations allow us to show that Fig.  \ref{exch}a and  Fig. \ref{exch}e
have the same amplitude, which means that an exchange of two defects followed
by a $360^\circ$ rotation of one of the defect do not generate any phase.  This
is nothing but the spin-statistics theorem.  

The above understanding of geometric meaning of the topological invariant $\frac12
a_1^3$ in terms of $Z_2$-twist domain wall also leads to a mechanism of the
$Z_2$ SPT state.  Consider a quantum Ising model on 2D triangle lattice
\begin{align} 
H=-J\sum_{\<ij\>} \si^z_i\si^z_j - g\sum_i \si^x_i, 
\end{align}
where $\si^{x,y,z}$ are the Pauli matrices and $\<ij\>$ are nearest neighbors.
Such a model can be described by the path integral of the domain walls between
$\si^z=1$ and $\si^z=-1$ in space-time.  However, all domain walls in
space-time have an amplitude of $+1$.

In order to have the non-trivial $Z_2$ SPT state, we need to modify the domain
wall  amplitudes in the path integral to allow them to have values $\pm 1$.
The $\pm 1$ is assigned based on the following rules: as time evolves, a
domain-wall-loop creation/annihilation will contribute to a $-1$ to the
domain-wall amplitude.  A domain-wall-line reconnection will also contribute to
a $-1$ to the domain-wall amplitude.  Those additional $-1$'s can be
implemented through local Hamiltonian.  We simply need to modify the $ -\sum_i
\si^x_i$ term which create the fluctuations of the domain-walls:
\begin{align}
 H=-J\sum_{\<ij\>} \si^z_i\si^z_j - g\sum_i \si^x_i
\Big(-\ee^{\ii \frac{\pi}{4} 
\sum_{\mu=1}^6 
(1-\si^z_{i,\mu} \si^z_{i,\mu+1})
}\Big), 
\end{align}
where $\sum_{\mu=1}^6 
(1-\si^z_{i,\mu} \si^z_{i,\mu+1})$ is the sum over all six spins
neighboring the site-$i$. (In fact, we can set $J=0$).
The factor $-\ee^{\ii \frac{\pi}{4} 
\sum_{\mu=1}^6 
(1-\si^z_{i,\mu} \si^z_{i,\mu+1})
}$ contributes to a $-1$ when the spin flip generated by $\si^x$
creates/annihilates a small loop of domain walls or causes a reconnection
of the  domain walls. The factor $-\ee^{\ii \frac{\pi}{4} 
\sum_{\mu=1}^6 
(1-\si^z_{i,\mu} \si^z_{i,\mu+1})
}$ contributes to a $+1$ when the spin flip 
only deform the shape of the  domain walls.
This is the Hamiltonian obtained in \Ref{LG1220}.

\begin{figure}[tb]
\centerline{
\includegraphics[scale=0.6]{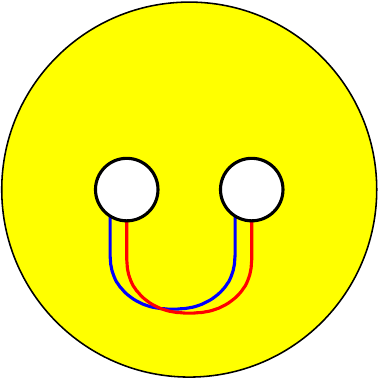}
}
\caption{
(Color online)
Two identical $Z_2$ monodromy defects on $S^2$.  The boundary across which we do
the $Z_2$ twist is split into the red and blue curves. Note that the
splitting is identical at the two monodromy defects.
The red and blue lines crosses once, indicating that
$\int_{S^2} a_1^2=1$.
}
\label{S2def}
\end{figure}

Now let us switch to the 4+1D $Z_2$ SPT described by
(see Appendix \ref{giTO3})
\begin{align}
	W^5_\text{top}(A)= \frac12 a_1p_1 
=\bt(a_1) \om_3=a_1^2\om_3,
\end{align}
which is a new mixed SPT phase first discovered in this paper.  Here $\bt$ is
the natural map $\bt: \cH^d(G,\Z_2) \to \cH^{d+1}(G,\Z)$ that maps $a_1
\in \cH^1(Z_2,\Z_2)$ to $\bt(a_1)=Sq^1(a_1)=a_1^2\in \cH^2(Z_2,\Z)$ (see
Appendix \ref{calgen} and also \eqn{W5U1}).  
We note that $\int_M \frac 23 p_1 =0 $ mod 2. Hence
we can rewrite $p_1 = \frac 13 p_1 +\frac 23 p_1 =\frac 13 p_1$
if we concern about mod 2 numbers.
The above topological invariant can be rewritten as
\begin{align}
	W^5_\text{top}(A)= \frac12 a_1\frac 13p_1 
=a_1^2\frac 13\om_3.
\end{align}

One of the physical properties of such a $Z_2$ SPT state is its dimension
reduction:
we put the state on space-time $M^5=S^2\times M^3$ and 
choose the $Z_2$-twist $a_1$ to create two \emph{identical} monodromy defects
on $S^2$  (see Fig. \ref{S2def}). (The physics of two \emph{identical}
monodromy defects was discussed in detail in \Ref{W1447} and here we follow a
similar approach. Also we may embed $Z_2$ into $U(1)$ and view the $Z_2$
monodromy defect as the $U(1)$ $\pi$-flux.) For such a design of $S^2$ and
$a_1$, we have $\int_{S^2} a_1^2=1$ mod 2 (see Fig. \ref{S2def}).  We then take
the large $M^3$ limit, and examine the induced the effective theory on $M^3$.
The induced effective Lagrangian must have a form $\cL = 2\pi k \frac 13\om_3$
with $k=1$ mod 2, which describes a topologically ordered state with chiral
central charge $8k$.  If $M^3$ has a boundary, the boundary will carry the
gapless chiral edge state of chiral central charge $8k$.  

We like to remark that adding two $Z_2$ monodromy defects to $S^2$ is not a
small perturbation.  Inducing a $E_8$ bosonic quantum Hall state on $M^3$
by a large perturbation on $S^2$ does not imply the parent state on $S^2\times
M^3$ to be non-trivial.  Even when the  parent state is trivial, an large
perturbation on $S^2$ can still induce  a $E_8$ state on $M^3$.  However,
what we have shown is that \emph{two identical} $Z_2$ monodromy defects on
$S^2$ induce an \emph{odd} numbers of $E_8$ states on $M^3$. This can
happen only when the parent state on $S^2\times M^3$ is non-trivial.

We may choose another  dimension reduction by  putting the state on space-time
$M^5=S^1\times M^4$ and adding a $Z_2$-twist by threading a $Z_2$-flux line
through the $S^1$.  We then take the large $M^4$ limit. The effective theory on
$M^4$ will be described by effective Lagrangian $\cL_\text{eff}= \pi \frac 13 p_1$.
When $M^4$ has a boundary, $\prt M^4\neq \emptyset$, the system on the
$M^3=\prt M^4$ must has chiral central charge $c=4$ mod 8.  In order words,
if the 4-dimensional space has a 3-dimensional boundary $S^1\times M^2$ and if
we thread a $Z_2$-flux line through the $S^1$, then the state on $M^2$ will
have a gravitational response described by a gravitational Chern-Simons
effective Lagrangian $\cL_{eff}=k \pi \frac 13\om_3$, with $k=1$ mod 2.  Such a
state on $M^2$ is either gapless or have a non-trivial topological order,
regardless if the symmetry is broken on the boundary or not.  

Let us assume that the $Z_2$ SPT state has a gapped symmetry breaking boundary.
The above result implies that if we have a symmetry breaking domain wall on
$S^1$, then the induced boundary state on $M^2$ must be topologically ordered
with a chiral central charge $c=4$ mod 8. (The mod 8 comes from the possibility
that the modified the local Hamiltonian at the domain wall may add several copy
of $E_8$ bosonic quantum Hall states.) We see that a $Z_2$ symmetry breaking
domain wall on the boundary carries a 2+1D topologically ordered state with a
chiral central charge $c=4$ mod 8.

\begin{table}[t]
 \caption{The L-type $Z_2^T$ SPT phases.
} \label{Z2TSPT}
 \centering
 \begin{tabular}{ |c|c|c| }
 \hline
 $d=$ & $\LSPT_{Z_2^T}^d$ & generators $W^d_\text{top}$ \\[1mm]
\hline
$0+1$ & 0 &  \\[1mm]
 \hline
$1+1$ & $\Z_2$ & $\frac12 \rw_1^2$ \\[1mm]
 \hline
$2+1$ & 0 &   \\[1mm]
 \hline
$3+1$ & $\Z_2\oplus \blue{\Z_2}$ & $\frac12 \rw_1^4$, $\blue{\frac12 p_1}$  \\[1mm]
 \hline
$4+1$ & 0 &  \\[1mm]
 \hline
$5+1$ & $\Z_2\oplus \blue{2\Z_2}$ &  $\frac12 \rw_1^6$,  $\blue{\frac12 \rw_1^2 p_1}$ $\blue{\frac12 \rw_1 \rw_2\rw_3}$\\[1mm]
 \hline
$6+1$ & $\blue{\Z_2}$ &  $\blue{\frac12 \rw_1^2 \rw_2\rw_3}$  \\[1mm]
\hline
 \end{tabular}
\end{table}

\begin{table}[t]
\caption{The L-type $U(1)\times Z_2^T$ SPT phases.  
} \label{UTspinSPT}
 \centering
 \begin{tabular}{ |c|c|c| }
 \hline
 $d=$ & $\LSPT_{U(1)\times Z_2^T}^d$ & generators $W^d_\text{top}$ \\[1mm]
\hline
$0+1$ & $0$ & $\frac 12 \rw_1$ \\[1mm]
\hline
$1+1$ & $2\Z_2$ & $\frac 12 \rw_1^2$, $\frac 12 c_1$ \\[1mm]
\hline
$2+1$ & $0$ &  \\[1mm]
\hline
$3+1$ & $3\Z_2\oplus \blue{\Z_2}$ &  
$\frac12 c_1^2$, $\frac12 \rw_1^2c_1$, $\frac12 \rw_1^4$,
$\blue{\frac12 p_1}$  \\[1mm]
\hline
$4+1$ & $ 0$ &  
         \\[1mm]
\hline
$5+1$ & $4\Z_2 $ & 
$\frac12 c_1^3$, $\frac12 \rw_1^2c_1^2$, $\frac12 \rw_1^4c_1$, $\frac12 \rw_1^6$
     \\
 & $\blue{3\Z_2}$ & 
     $\blue{\frac12 c_1p_1}$, $\blue{\frac12 \rw_1^2p_1}$, $\blue{\frac12 \rw_1\rw_2\rw_3}$  \\[1mm]
\hline
$6+1$  & $\blue{2 \Z_2}\oplus \red{\Z_2}$ &  
$\blue{ \frac12 c_1\rw_2\rw_3}$, $\blue{\frac12 \rw_1^2 \rw_2\rw_3}$, $\red{\frac12 \rw_1 c_1p_1}$
 \\[1mm]
\hline
 \end{tabular}
\end{table}

\begin{table}[t]
\caption{The L-type $U(1)\rtimes Z_2^T$ SPT phases.  } \label{BTISPT}
 \centering
 \begin{tabular}{ |c|c|c| }
 \hline
 $d=$ & $\LSPT_{U(1)\rtimes Z_2^T}^d$ & generators $W^d_\text{top}$ \\[1mm]
\hline
$0+1$ & $\Z$ & $a$ \\[1mm]
\hline
$1+1$ & $\Z_2$ & $\frac 12 \rw_1^2$ \\[1mm]
\hline
$2+1$ & $\Z_2$ & $\frac12 \rw_1 c_1$ \\[1mm]
\hline
$3+1$ & $2\Z_2\oplus \blue{\Z_2}$ &  
  $\frac12 c_1^2$, $\frac12 \rw_1^4$, $\blue{\frac12 p_1}$  \\[1mm]
\hline
$4+1$ & $ \Z\oplus\Z_2\oplus \blue{\Z}$ & $ac_1^2$, 
$ \frac12 \rw_1^3c_1$, 
   $\blue{\frac 13 ap_1}$       \\[1mm]
\hline
$5+1$ & $2\Z_2\oplus \blue{2\Z_2} $ & 
$\frac12 \rw_1^2c_1^2$, $\frac12 \rw_1^6$,
     $\blue{\frac12 \rw_1^2p_1}$, $\blue{\frac12 \rw_1\rw_2\rw_3}$  \\[1mm]
\hline
$6+1$  & $2\Z_2\oplus  \red{\Z_2}$ &  
$\frac12 \rw_1 c_1^3$, $\frac12 \rw_1^5 c_1$,
$\red{\frac12 \rw_1 c_1p_1}$
 \\
  & $\blue{3 \Z_2}$ & $\blue{ \frac12 \rw_1c_1 p_1}$, 
$\blue{ \frac12 c_1\rw_2\rw_3}$, $\blue{\frac12 \rw_1^2 \rw_2\rw_3}$
 \\[1mm]
\hline
 \end{tabular}
\end{table}

\subsubsection{
$Z_2^T$, 
$U(1)\times Z_2^T$, and 
$U(1)\rtimes Z_2^T$ 
SPT states in Tables
\ref{Z2TSPT},
\ref{UTspinSPT}, and
\ref{BTISPT}
}

The tables \ref{Z2TSPT}, \ref{UTspinSPT}, and \ref{BTISPT} list the so called
realizable topological invariants, which can be produced via our NL$\si$M
construction.  The potential topological invariants (which may or may not be
realizable) for those symmetries have been calculated in \Ref{K1459v2} using
cobordism approach and in \Ref{F1478} using spectrum approach.  For the
topological invariants that generate the $\Z_2$ classes, our realizable topological
invariants agree with the potential topological invariants obtained in \Ref{K1459v2}. For
the topological invariants that generate the $\Z$ classes, our realizable
topological invariants only form a subset of the potential topological invariant obtained
in \Ref{K1459v2} and in \Ref{F1478}.

\begin{figure}[tb]
\centerline{
\includegraphics[scale=0.8]{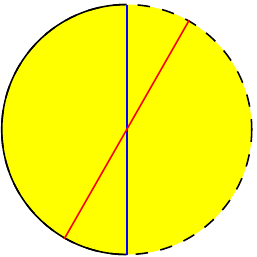}
}
\caption{
(Color online)
The shaded disk represents a 2-dimensional manifold $\R P^2$, where the
opposite points on the boundary are identified ($\hat {\v r} \sim -\hat {\v
r}$).  The blue and red lines are two non-contractible loops in $\R P^2$.
Consider a $Z_2$-twist $a_1$ described by $[a_1]^* =$ a contractible loop.
Then the blue and red lines represent the same $Z_2$ twist $a_1$.  For such a
$Z_2$-twist, we find that $\int_{\R P^2} a_1^2 =1$ since the  blue and red lines
cross once.  The above $Z_2$-twist is also the orientation reversing twist.  So
$a_1=\rw_1$ and we have $\int_{\R P^2} \rw_1^2 =1$.
}
\label{RP2}
\end{figure}

In 1+1D, all those time-reversal protected SPT phases contain one described by
\begin{align}
	W^2_\text{top}(A,\Ga) = \frac12 \rw_1^2 .
\end{align}
Here, we would like to remark that time-reversal symmetry and space-time mirror
reflection symmetry should be regarded as the same symmetry.\cite{K1467,K1459v2}
If a system has no time reversal symmetry, then we can only use orientable
space-time to probe it.  Putting a system with  no time reversal symmetry on a
non-orientable space-time is like adding a boundary to the system.  If a system
has a time reversal symmetry, then we can use non-orientable space-time to
probe it, and in this case, the $Z_2^T$-twist is described by $a_1=\rw_1$.
Since $\rw_1 \neq 0$ only on non-orientable manifolds, the $Z_2^T$-twist is
non-trivial only on non-orientable manifolds.  So we should use a
non-orientable space-time to probe the above time-reversal protected SPT phase.
In fact, the above topological invariant can be detected on $\R P^2$: $\int_{\R P^2}
\rw_1^2=1$ mod 2 (see Fig.  \ref{RP2}).

\begin{figure}[tb]
\centerline{
\includegraphics[scale=0.8]{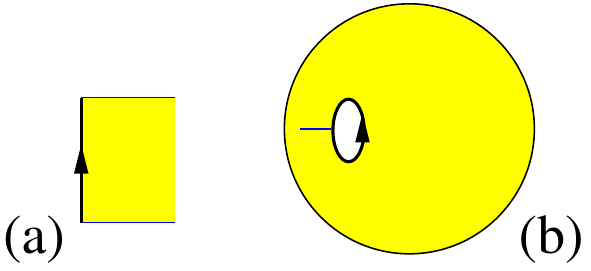}
}
\caption{
(Color online)
(a) The path integral of a single boundary point of 1D space 
over the time loop $S^1$.  The
shaded area represents the 1+1D space-time.  The two ends of the thick
line are identified to form a loop $S^1$.
The two blue lines are also identified.
(b) The loop $S^1$ is extended to a sphere with a hole.
The identified  blue line is also shown.
}
\label{Shole}
\end{figure}

In the following we will explain how the above topological invariant ensure the
degenerate ground states at the boundaries of 1D space.  We first consider the
partition of a single boundary point over a time loop $S^1$  (see Fig.
\ref{Shole}a).  Such a partition function on $S^1$ is defined by first
extending $S^1$ into a sphere with a hole $S^2_\text{hole}$ (see Fig.
\ref{Shole}b), and then we use the 1+1D partition function defined on
$S^2_\text{hole}$ (from the path integral of $\ee^{\ii \int_{S^2_\text{hole}}
W^2_\text{top}(A,\Ga)}$) to define the partition function on $S^1$.  We find
that such a partition function on $S^1$ is trivial $Z=1$.

\begin{figure}[tb]
\centerline{
\includegraphics[scale=1.0]{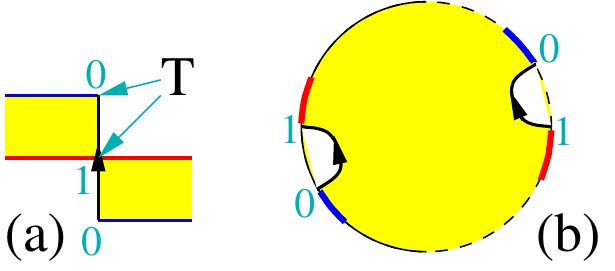}
}
\caption{
(Color online)
(a) The path integral of a single boundary point over the time loop $S^1$ with
two time-reversal transformations at point $0$ and $1$.  The shaded area
represents the 1+1D space-time.  The two ends of the thick line are identified
to form a loop $S^1$.  The two blue lines are also identified after a
horizontal reflection.  The two red lines on the two sides of the thick line
are identified as well after a horizontal reflection.  (b) The shaded disk
represents a 2-dimensional manifold $\R P^2$, where the opposite points on the
boundary are identified ($\hat {\v r} \sim -\hat {\v r}$).  The loop $S^1$ in
(a) is extended to the $\R P^2$ with a hole in (b).  The two red lines and the
two blue lines in (a) are also shown in (b).
}
\label{RP2hole}
\end{figure}

Now, we like to consider the partition of a single boundary point over a time
loop $S^1$, but now with two time-reversal transformations inserted  (see Fig.
\ref{RP2hole}a), where the time-reversal is implemented as mirror reflection in
the transverse direction.  Next, we extend Fig.  \ref{RP2hole}a  into a $\R
P^2$  with a hole $\R P^2_\text{hole}$ (see Fig.  \ref{RP2hole}b).  Since
(after taking the small hole limit) $\int_{\R P^2} W^2_\text{top}(A,\Ga)=\frac12
\int_{\R P^2} \rw_1^2 =\frac12 $ mod 1, we find that the partition function on
$S^1$ with two time-reversal transformations is non-trivial $Z=-1$. This
implies that $T^2=-1$ when acting on the states on a single boundary point.
The states on a single boundary must form Kramers doublets, and degenerate.

From the Tables \ref{Z2TSPT}, \ref{UTspinSPT}, and \ref{BTISPT}, we also see
that most generators of 3+1D time-reversal SPT states are pure SPT states
described by $\cH^4(G,\RZ)$. All mixed time-reversal SPT states are generated
by a single generator
\begin{align}
	W^4_\text{top}(A,\Ga) = \frac12 p_1 ,
\end{align}
which is a mixed $Z_2^T$ SPT state.\cite{VS1306}  In other words, all mixed
time-reversal SPT states can be obtained from the  pure SPT states by stacking
with one copy of the above mixed $Z_2^T$ SPT state.

\subsubsection{$U(1)\rtimes Z_2=O_2$ SPT states in Table \ref{O2SPT}}

\begin{table}[t]
\caption{The L-type $O_2$ SPT phases.  
} \label{O2SPT}
 \centering
 \begin{tabular}{ |c|c|c| }
 \hline
 $d=$ & $\LSPT_{O_2}^d$ & generators $W^d_\text{top}$ \\[1mm]
\hline
$0+1$ & $\Z_2$ & $\frac 12 a_1$ \\[1mm]
\hline
$1+1$ & $\Z_2$ &  $\frac 12 c_1$ \\[1mm]
\hline
$2+1$ & $\Z\oplus \Z_2$ & $a c_1$, $\frac12 a_1^3$ \\[1mm]
\hline
$3+1$ & $\Z_2$ &   $\frac12 a_1^2c_1$  \\[1mm]
\hline
$4+1$ & $2\Z_2\oplus \blue{\Z_2}$ &  
        $\frac12 a_1^5$, $\frac12 a_1c_1^2$, $\blue{\frac12 a_1p_1}$\\[1mm]
\hline
$5+1$ & $2\Z_2 \oplus \blue{2\Z_2}$ & $\frac12 c_1^3$, $\frac12 a_1^4c_1$,  
     $\blue{\frac12 c_1p_1}$,  $\blue{\frac12 a_1\rw_2\rw_3}$  \\[1mm]
\hline
$6+1$ & $\Z\oplus 2\Z_2  $ &  
$ac_1^3$, $\frac12 a_1^7$, $\frac12 a_1^3c_1^2$ \\
 & $\blue{\Z\oplus 3 \Z_2}$ &  
$\blue{ \frac 13 c_1^2\om_3}$, $\blue{\frac12 a_1^3p_1}$, 
$\blue{\frac12 a_1^2 \rw_2\rw_3}$,
$\blue{\frac12 c_1 \rw_2\rw_3}$,
 \\[1mm]
\hline
 \end{tabular}
\end{table}

The 1+1D $O_2$ SPT state is characterized by the following topological invariant
\begin{align}
	W^2_\text{top}(A)= \frac12 \frac{\dd A}{2\pi} .  
\end{align}
Let us explain how such a topological invariant ensure the degenerate ground
states at the boundaries of 1D space.  Let us consider a 1+1D space-time
$S^2_\text{hole}$ which is $S^2$ with a small hole (see Fig. \ref{Shole}b).
The partition function for $S^2_\text{hole}$ can be viewed as the effective
theory for the boundary $S^1=\prt S^2_\text{hole}$, which is the  partition
function for a single boundary point of 1D space over the time loop $S^1$ (see
Fig.  \ref{Shole}a).  Since the partition function on $S^2_\text{hole}$ changes
sign as we add $2\pi$ $U(1)$ flux to $S^2_\text{hole}$, this means that a
$2\pi$ $U(1)$ rotation acting on the states on a single boundary point will
change the sign of the states. So the states on a single boundary point must
form a projective representation of $O_2$ where the $2\pi$ $U(1)$ rotation is
represented by $-1$. Such a projective representation is alway even
dimensional, and the states on a single boundary point must have an even
degeneracy.

From the 2+1D topological invariants, we see that the 2+1D $O_2$ SPT is actually the
2+1D $U(1)$ SPT state (by ignoring $Z_2$) and the  2+1D $Z_2$ SPT state (by
ignoring $U(1)$).

In 3+1D, we have a pure 3+1D $O_2$ SPT state described by
\begin{align}
\label{O2W4}
W^4_\text{top}(A) = \frac12 a_1^2 c_1
,  
\end{align}
which is a SPT state for quantum spin systems.  

To construct a  physical probe for the above $U(1)\rtimes Z_2$ SPT state, we
first note that the topological invariant \eq{O2W4} is invariant under time reversal
(mod $2\pi$). So the corresponding $U(1)\rtimes Z_2$ SPT state is compatible
with time reversal symmetry.  If we assume the $U(1)\rtimes Z_2$ SPT state also
have the time reversal symmetry, then we can design the following probe for the
$U(1)\rtimes Z_2$ SPT state.  We choose the 3+1D space-time to be $S^2\times
M^2$, and put $2\pi$ $U(1)$ flux through $S^2$, where $S^2$ is actually a
lattice.  But such $2\pi$ flux is in a form a two \emph{identical} thin
$\pi$-flux and each $\pi$-flux going through a single unite cell in $S^2$.
Such a configuration has $\int_{S^2} c_1 =1$ mod 2, and at the same time, does
not break the $U(1)\rtimes Z_2$ symmetry.  

In the large $M^2$ limit, the dimension-reduced theory on $M^2$ is described by
a topological invariant $W^2_\text{top}= \frac12 a_1^2$.  However, due to an
identity $a_1^2=\rw_1a_1$ in 2-dimensional space, $\int_{M^2} a_1^2 =
\int_{M^2} \rw_1a_1 = 0$ mod 2, if $M^2$ is orientable (since $\rw_1=0$ iff the
manifold is orientable).  The topological invariant $W^2_\text{top}= \frac12
a_1^2$ can be detected only on non-orientable $M^2$.  This is where we need the
time reversal symmetry: in the presence of time reversal symmetry, we can use
non-orientable $M^2$ to probe the topological invariant.

Let $Z_2^t$ be the
symmetry group generated by the combined $Z_2$ transformation and time-reversal
$Z_2^T$ transformation.  Let $a_1^t$ be the $Z_2^t$-twist.  Then we have
$a_1^t=a_1=\rw_1$.  Thus the topological invariant can be rewritten as
$W^2_\text{top}= \frac12 \rw_1^2$, which describes a 1+1D SPT state
protected by time reversal symmetry $Z_2^t$.

We like to remark that threading two thin $\pi$-flux lines through $S^2$ is not
a small perturbation.  Inducing a $Z_2^t$ SPT state on $M^2$ by a large
perturbation on $S^2$ does not imply the parent state on $S^2\times M^2$ to be
non-trivial.  Even when the  parent state is trivial, an large perturbation on
$S^2$ can still induce  a $Z_2^t$ SPT state on $M^2$.  However, what we have
shown is that threading \emph{two identical} thin $\pi$-flux lines through
$S^2$ induces \emph{one} $Z_2^t$ SPT state on $M^2$. This can happen only when
the  parent state on $S^2\times M^2$ is non-trivial.


\subsection{Realizable and potential topological invariants}
\label{RPtopinv}

After discussing the physical consequences of various topological invariants,
let us turn to study the topological invariants themselves.  It turns out that
the topological invariants for iTO states satisfy many self consistent
conditions.  Solving those conditions allow us to obtain self consistent
topological invariants, which will be called \emph{potential} gauge-gravity
topological invariants.  \Ref{K1467,K1459v2,KTT1429,F1478,WGW1489} studied the
topological invariants from this angle and only the potential gauge-gravity
topological invariants are studied.  For example, when there is no symmetry,
the following type of potential gauge-gravity topological invariants were
found:
\\
(1) The 2+1D potential gravitational topological invariants are described by
$\Z$,\cite{KF9732,HLP1242,PMN1372,KW1458} which are generated by
\begin{align}
\label{cb3}
	W^3_\text{top}(\Ga)= \frac{1}{3}\om_3(\Ga)
\end{align}
where $\om_3(\Ga)$ is the
gravitational Chern-Simons term that is defined via $\dd \om_3=p_1$, with $p_i$
the $i^\text{th}$ Pontryagin class.
In \Ref{F1478}, it was suggested that
the 2+1D potential gravitational topological invariants
are generated by
\begin{align}
	W^3_\text{top}(\Ga)= \frac{1}{6}\om_3(\Ga).
\end{align}
\\
(2) The 4+1D
potential gravitational topological invariants are described by
$\Z_2$,\cite{K1467,K1459v2,KW1458} 
which are generated by
\begin{align}
\label{cb5}
	W^5_\text{top}(\Ga)= \frac12 \rw_2\rw_3
\end{align}
where $\rw_i$ is the $i^\text{th}$ Stiefel-Whitney class.
\\
(3)
The 6+1D potential gravitational topological invariants are described by
$2\Z$,\cite{KW1458}
which are generated by
\begin{align}
\label{cb7}
	W^7_\text{top}(\Ga) &=  \frac{\om^{p_1^2}_7-2\om^{p_2}_7}{5}
\nonumber\\
W^7_\text{top}(\Ga) &= \frac{-2\om^{p_1^2}_7+5\om^{p_2}_7}{9} 
\end{align}
where the gravitational Chern-Simons terms are defined by $\dd \om^{p_1^2}_7
=p_1^2$ and $\dd \om^{p_2}_7 =p_2$.

The potential topological invariants in \eqn{cb3}, \eqn{cb5}, and \eqn{cb7} have a close relation
to the orientated $d$-dimensional cobordism group
$\Om^{SO}_d$,\cite{K1467,K1459v2,KTT1429} which are Abelian groups generated by
the Stiefel-Whitney classes $\rw_i$ and the Pontryagin classes $p_i$.  For
example, $\Om^{SO}_4=\Z$ is generated by the Pontryagin class $\frac13 p_1$ and
$\Om^{SO}_8=2\Z$ by $\frac{\om^{p_1^2}_7-2\om^{p_2}_7}{5}$ and
$\frac{-2\om^{p_1^2}_7+5\om^{p_2}_7}{9}$.  Also $\Om^{SO}_5=\Z_2$ is generated
by Stiefel-Whitney class $\rw_2\rw_3$.  In this case, the set of potential
gravitational topological invariants in $d$-dimensional space-time (denoted as
$\text{P}\iTOL^d$) are exactly those Stiefel-Whitney classes and the Pontryagin
classes that describe the  cobordism group $\Om^{SO}_d$:
\begin{align}
 \text{Tor}(\text{P}\iTOL^d) &=  \text{Tor}(\Om^{SO}_d),
\nonumber\\
 \text{Free}(\text{P}\iTOL^d) &=  \text{Free}(\Om^{SO}_{d+1}).
\end{align}
Note that $\text{P}\iTOL^d$ and $\Om^{SO}_d$ are discrete Abelian groups.  Tor
and Free are the torsion part and the free part of the discrete Abelian groups.

However, we do not know if those potential gauge-gravity topological invariant
can all be realized or produced by local bosonic systems.  In this paper, we
will study this issue.  However, to address this issue, we need to first
clarify the meaning of ``realizable by local bosonic systems''.

We note that there are two types of local bosonic systems: L-type and
H-type.\cite{KW1458} L-type local bosonic systems are systems described by
local bosonic Lagrangians.  L-type systems have well defined partition
functions for space-time that can be any manifolds.  H-type local bosonic
systems are systems described by local bosonic Hamiltonians.  H-type systems
have well defined partition functions only for any space-time that are mapping
tori. (A mapping torus is a fiber bundle over $S^1$.) A L-type system always
correspond to a H-type system. However, a H-type system may not correspond to a
L-type system.  For example, SPT phases described by group cohomology and the
NL$\si$Ms are L-type topological phases (and they are also H-type topological
phases). The $E_8$ bosonic quantum Hall state is defined as a H-type
topological phase.  However, it is not clear if it is a L-type topological
phase or not.  In the following, we will argue that any quantum Hall state is
also a L-type topological phase (\ie realizable by space-time path integral,
that is well defined on any space-time manifold).


In this paper, we will only consider L-type bosonic quantum systems.  We will
study which potential gauge-gravity topological invariants are realizable by
L-type local bosonic systems.  We will use $SO(\infty)\times G$ NL$\si$Ms
\eq{Lg} to try to realize those potential gauge-gravity topological invariants.
After adding the $G$-symmetry twist and choose a curved space-time $M^d$, the
``gauged'' $SO(\infty)\times G$ NL$\si$Ms \eq{Lg}
becomes\cite{HW1267,HW1227,W1313}
\begin{align}
\label{LAGa}
\cL &= \frac{1}{\la} [(\prt +iA+i\Ga)  g)^2 + \ii L^d_\text{top}(g^{-1}(\prt +iA+i\Ga)g), 
\nonumber\\
& g(x) \in G\times SO,\ \ \ SO\equiv SO(\infty),
\end{align}
where the space-time connection $\Ga$ couples to $SO(\infty)$ and the ``gauge''
connection $A$ couples to $G$.  The induced gauge-gravity topological term
$L^d_\text{top}(g^{-1}(\prt +iA+i\Ga)g)$ is classified by group cohomology
$\cH^d[G\times SO,\RZ]$.  After we integrate out the matter fields $g$, the
above gauged NL$\si$M will produce a partition function that give rise to a
\emph{realizable gauge-gravity topological invariant} $W^d_\text{top}(A,\Ga)$ via
\begin{align}
	Z(M^d,A)=\ee^{\ii \int_{M^d} 2\pi W^d_\text{top}(A,\Ga)} .
\end{align}
(See \Ref{W8570} for a study of gauged topological terms described by
Free$[\cH^d(G,\RZ)]$ for continuous groups.)

The set of  potential gauge-gravity topological terms contain the set of
realizable gauge-gravity topological terms.  More precisely, the two sets are
related by a map
\begin{align}
	\{ L^d_\text{top}(g^{-1}(\prt +iA+i\Ga)g) \} \to \{ W^d_\text{top}(A,\Ga) \}.
\end{align}
However, the map may not be one-to-one and may not be surjective.  

For example, when there is no symmetry, we find that the following type of
realizable gauge-gravity topological invariants were generated by the above NL$\si$M
(see Table \ref{invTop}):\\ 
(1) Those 2+1D realizable gravitational topological invariants are described by
$\Z$, which are generated by
\begin{align}
	W^3_\text{top}(\Ga)= \om_3(\Ga).
\end{align}
The corresponding generating topological state has a chiral central charge
$c=24$ at the edge.  So the stacking of three $E_8$ bosonic quantum Hall states
can be realized by a well defined L-type local bosonic system.  It is not clear
if a single $E_8$ bosonic quantum Hall state can be realized by a L-type local
bosonic system or not.  However, we know that a single $E_8$ bosonic quantum
Hall state can be realized by a H-type local bosonic system.
\\
(2) Those 4+1D
realizable gravitational topological invariants are described by
$\Z_2$,
which are generated by
\begin{align}
	W^5_\text{top}(\Ga)= \frac12 \rw_2\rw_3.
\end{align}
(Note that $\cH^5(SO,\RZ)$ is also $\Z_2$ in this case.) In fact, we will show
that all the potential gauge-gravity topological invariants that generate a finite
group are realizable by the  $SO(\infty)$ NL$\si$Ms,
which are L-type local bosonic systems.
\\
(3) $\cH^6(SO,\RZ)=2\Z_2$, and there are four different types of $SO(\infty)$
NL$\si$Ms (with four different topological terms).  However, the four different
topological terms in the NL$\si$Ms all reduce to the same trivial gravitational
topological invariant $W^6_\text{top}(\Ga)$ after we integrate out the matter field
$g$, suggesting that all the four  NL$\si$Ms give rise to the same topological
order.
\\
(4)
Those 6+1D realizable gravitational topological invariants are described by $2\Z$,
which are generated by
\begin{align}
	W^7_\text{top}(\Ga) &=  \om^{p_1^2}_7,
&
	W^7_\text{top}(\Ga) &=  \om^{p_2}_7.
\end{align}
We see that only part of the potential gravitational topological invariants are
realizable by the $SO(\infty)$ NL$\si$Ms.  

However, it is possible that  $SO(\infty)$ NL$\si$Ms do not realize all
possible L-type iTO's.  In the following, we will argue that the 2+1D $E_8$
bosonic quantum Hall state is a L-type iTO.  $SO(\infty)$ NL$\si$M cannot
realize the $E_8$ state since it has a central charge $c=8$ and a topological
invariant $\frac 13 \om_3$.

In fact, we will argue that any quantum Hall state is a L-type topologically
ordered state.  Certainly, by definition, any quantum Hall state, being
realizable by some interacting Hamiltonians, is a H-type topologically ordered
state.  The issue is if we can have a path-integral description that can be
defined on any closed space-time manifold.  At first, it seems that such a
path-integral description does not exist and a quantum Hall state cannot be a
L-type topological order.  This is because quantum Hall state is defined with
respect to a non-zero background magnetic field -- an closed two-form field
($B=\dd A$) in 2+1D space-time. This seems imply that  a path-integral
description of quantum Hall state exist only on space-time that admits a
every-where non-zero closed two-form field.

However, as stressed in \Ref{WW0808,WW0809}, a quantum Hall state of filling
fraction $\nu=p/q$ always contains an $n$-cluster structure.  Also, the closed
two-form field $B=\dd A$ in 2+1D space-time may contain ``magnetic monopoles''.
If those ``magnetic monopoles'' are quantized as multiples of $n q$, they will
correspond to changing magnetic field by $n q$ flux quanta each time.  Changing
magnetic field by $n q$ flux quanta and changing particle number by $p$
$n$-clusters is like adding a product state to a gapped quantum liquid discussed
in \Ref{ZW1490}, which represents a ``smooth'' change of the quantum Hall state.
Since every-where non-zero closed two-form field $B=\dd A$ with ``magnetic
monopoles'' can be defined on any 2+1D space-time, we can have a path-integral
description of any quantum Hall state, such that the path-integral is well
defined on any space-time manifold.  We conclude that quantum Hall states, such
as the $E_8$ state,  are L-type topologically ordered states.  Therefore, the
gravitational topological invariant 
\begin{align}
	W^3_\text{top}(\Ga)= \frac 13 \om_3
\end{align}
is realizable by a 2+1D L-type iTO, \ie a $E_8$ state (see Table \ref{invTop}).


\subsection{A construction of L-type realizable pure and mixed SPT phases}

Now, let us include symmetry and discuss SPT phases  (\ie L-type topological
phases with short range entanglement).  We like to point out that some SPT
states are characterized by boundary effective theory with anomalous
symmetry,\cite{W1313,KT1430,KT1417} which is commonly referred as gauge anomaly
(or 't Hooft anomaly).  Those SPT states are classified by group cohomology
$\cH^d(G,\RZ)$ of the symmetry group $G$.  We also know that the boundaries of
topologically ordered states\cite{Wtop,WNtop,Wrig,KW9327} realize and
(almost\footnote{For example, the pure 2+1D gravitational anomalies described
by \emph{unquantized} thermal Hall conductivity are not classified by
topologically ordered states.}) classify all pure gravitational
anomalies.\cite{KW1458} So one may wonder, the boundary of what kind of order
realize mixed gauge-gravity anomalies?  The answer is SPT order. This is
because the mixed gauge-gravity anomalies are present only if we have the
symmetry.  Such SPT order is also beyond the $\cH^d(G,\RZ)$ description, since
the mixed gauge-gravity anomalies are beyond the pure gauge anomalies.  We will
refer this new class of SPT states as \emph{mixed SPT states} and refer the SPT
states with only the pure  gauge anomalies as \emph{pure SPT states}.  We would
like to mention that the gauge anomalies and mixed gauge-gravity anomalies have
played a key role in the classification of free-electron topological
insulators/superconductors.\cite{RSF0957,RML1204}

The main result of this paper is a classification of both pure and mixed SPT
states realized by the NL$\si$Ms: 
\begin{align}
\label{siSPTHH}
&\ \ \ \
\si\LSPT_G^d
\\
&=\frac{\oplus_{k=1}^{d-1} H^k[BG,\cH^{d-k}(SO,\RZ)]
\oplus \cH^d(G,\RZ)
}{\La^d(G)}
,
\nonumber\\
&=E^d(G)\rtimes \Big[\oplus_{k=1}^{d-1} H^k(BG,\si\iTOL^{d-k})
\oplus \cH^d(G,\RZ)
\Big]
,
\nonumber 
\end{align}
where $\si\LSPT_G^d$ is the Abelian group formed by the L-type $G$ SPT phases
in $d$-dimensional space-time produced by the NL$\si$Ms, and $\si\iTOL^d$ is
the Abelian group formed by the L-type iTO phases in $d$-dimensional space-time
\emph{produced by the NL$\si$Ms}.  Also $\La^d(G)$ is a subgroup of
$\oplus_{k=1}^{d-1} H^k[BG,\cH^{d-k}(SO,\RZ)] \oplus \cH^d(G,\RZ)$.

Replacing  $\si\iTOL^d$ by  $\iTOL^d$ -- the Abelian group formed by the L-type
iTO phases in $d$-dimensional space-time, we obtain more general SPT states
described by $\LSPT_G^d$: \frm{
\begin{align}
\label{SPTHH}
&\ \ \ \
\LSPT_G^d
\\
&=E^d(G)\rtimes \Big[\oplus_{k=1}^{d-1} H^k(BG,\iTOL^{d-k})
\oplus \cH^d(G,\RZ)
\Big]
,
\nonumber 
\end{align}
}
If $G$ contains time-reversal transformation, it
will have a non-trivial action $\RZ \to -\RZ$ and $\iTOL^{d-k} \to
-\iTOL^{d-k}$.  Also, $BG$ is the classifying space of $G$ and $H^k(BG,\M)$ is
the topological cohomology class on $BG$. 

Note that stacking two topological phases $\cC_1$ and $\cC_2$ together will
produce another topological phase $\cC_3$. We denote such a stacking operation
as $\cC_1 \boxplus \cC_2=\cC_3$.  Under $\boxplus$, the topological phases form
a commutative monoid.\cite{KW1458}  In general, a  topological phase $\cC$ may
not have an inverse, \ie we can not find another topological phase $\cC'$ such
that $\cC\boxplus \cC'=0$ is a trivial product state.  This is why  topological
phases form a commutative monoid, instead of an Abelian group.  However, a
subset of topological phases can have inverse and form an Abelian group.  Those
topological phases are called invertible.\cite{KW1458,F1478} One can show that
a topological phase is  invertible iff it has no topological
excitations.\cite{KW1458,F1478} Therefore, all SPT phase are invertible.  Some
topological orders are also invertible, which are called invertible topological
orders (iTO).  SPT phases and iTO phases form Abelian groups under the stacking
$\boxplus$ operation.  So for SPT states and iTO states, we can replace
$\boxplus$ by $+$:
\begin{align}
 \cC_1\boxplus \cC_2=\cC_1+\cC_2.
\end{align}
So $\LSPT_G^d$ and $\iTOL^d$ can be viewed as modules over the ring $\Z$, and
they can appear as the coefficients in group cohomology.

The result \eq{siSPTHH} can be understood in two ways. It means that the SPT
states constructed from NL$\si$Ms are all described by $\oplus_{k=1}^{d-1}
H^k[BG,\cH^{d-k}(SO,\RZ)]\oplus \cH^d(G,\RZ)$, but in a many-to-one fashion;
\ie $\oplus_{k=1}^{d-1} H^k[BG,\cH^{d-k}(SO,\RZ)]\oplus \cH^d(G,\RZ)$ contain
a subgroup $\La^d(G)$ that different elements in $\La^d(G)$ correspond to the
same SPT state.  It also means that the constructed  SPT states are described
by $\oplus_{k=1}^{d-1} H^k(BG,\si\iTOL^{d-k})\oplus \cH^d(G,\RZ)$, but in a
one-to-many fashion; \ie each element of $\oplus_{k=1}^{d-1}
H^k(BG,\si\iTOL^{d-k})\oplus \cH^d(G,\RZ)$ correspond to several SPT states that
form a group $E^d(G)$.  The group $\La^d(G)$ and $E^d(G)$ can be calculated but
we do not have a simple expression for them (see Section \ref{pmSPT}).

In \eqn{SPTHH}, $\LSPT_G^d$ includes both pure and mixed SPT states.  The group
cohomology class $\cH^d(G,\RZ)$ describes the pure SPT phases, and the group
cohomology class $E^d(G)\rtimes \oplus_{k=1}^{d-1} H^k(BG,\iTOL^{d-k})$
describes the mixed SPT phases.  We would like to mention that an expression of
the form \eqn{SPTHH} was first proposed in \Ref{W1447} in a study of
topological invariants of SPT states.  We see that our NL$\si$Ms construction
can produce mixed SPT phases with and without time reversal symmetry.  We have
used \eqn{SPTHH} to compute the SPT phases for some simple symmetry groups (see
Table
\ref{SPT}).  

The formal group cohomology methods employed for obtaining the result
\eq{SPTHH} directly shed light on the physics of these phases.  The SPT states
described by $\cH^1(G,\iTOL^{d-1})$ in \eqn{SPTHH} can be constructed using the
decorated domain walls proposed in \Ref{CLV1407}.  Other  SPT states described
$H^k(BG,\iTOL^{d-k})$ can be obtained by a generalization of the
decorated-domain-wall construction,\cite{LGW1418,WGW1489,GWW1568} which will be
called the nested construction.\cite{GW14} The formal methods also lead to
physical/numerical probes for these
phases.\cite{LG1220,VS1306,MKF1331,W1447,YW1427,BRX1424,WSW1456} In addition,
these methods are easy to generalize to fermionic systems\cite{GW1248,GW14},
and provide answers for the physically important situation of continuous
symmetries (like charge conservation).

We also studied the potential SPT phases (\ie might not realizable) for a
non-on-site symmetry -- the mirror-reflection symmetry $Z_2^M$. The Abelian
group formed by those SPT phases is denoted as $\PSPT_{Z_2^M}^d$. Following
\Ref{K1467,K1459v2,KTT1429}, we find that $\PSPT_{Z_2^M}^d$ is given by a
quotient of the unoriented cobordism groups $\Om^{O}_d$ 
\frm{
\begin{align}
\PSPT^d_{Z_2^M}=\Om^{O}_d/\bar\Om^{SO}_d, 
\end{align}
}
where $\bar\Om^{SO}_d$ is the orientation invariant subgroup of $\Om^{SO}_d$
(\ie the manifold $M^d$ and its orientation reversal $-M^d$ belong to the same
oriented cobordism class).  It is interesting to see 
\begin{align}
\label{Z2TZ2M}
\PSPT^d_{Z_2^M}=\LSPT^d_{Z_2^T}
\end{align}
(see Table \ref{Z2TSPT}).

We want to remark that, in this paper, the time reversal transformation is
defined as the complex conjugation transformation (see Section \ref{gterm}),
without the $t\to -t$ transformation.  The mirror reflection correspond to the
$t\to -t$ transformation.  The time-reversal symmetry used in
\Ref{K1467,HMC1402,K1459v2,KTT1429} is actually the mirror-reflection symmetry
$Z_2^M$ in this paper.  The two ways to implement time-reversal symmetry should
lead to the same result as demonstrated by \eqn{Z2TZ2M}, despite the involved
mathematics, the cobordism approach and NL$\si$M approach, are very different.


\subsection{Discrete gauge anomalies, discrete
mixed gauge-gravity anomalies, and invertible discrete gravitational anomalies}

First, let us explain the meaning of \emph{discrete anomalies}.  All the
commonly known anomalies are discrete in the sense that different anomalies
form a discrete set. However, there are continuous gauge/gravitational
anomalies labeled by one or more continuous parameters.\cite{W1313,KW1458} In
this section, we only consider discrete anomalies.

Since the boundaries of SPT states realize all pure gauge anomalies, as a
result,  group cohomology  $\cH^d(G,\RZ)$ systematically describe all the
perturbative and global gauge anomalies.\cite{W1313,KT1430} For topological
orders, we found that they can be systematically described by tensor category
theory\cite{FNS0428,LWstrnet,CGW1038,GWW1017,KK1251,GWW1332,KW1458} and tensor
network,\cite{VC0466,GLS0918,BAV0919}   and those theories also  systematically
describe all the perturbative and global gravitational anomalies.\cite{KW1458}

Or more precisely, the discrete pure bosonic gauge anomalies in $d$-dimensional
space-time are described by $\cH^{d+1}(G,\RZ)$.  The discrete invertible pure
bosonic gravitational anomalies in $d$-dimensional space-time are described
$\iTO_L^{d+1}\simeq$Free$(\Om^{d+2}_{SO})\oplus$Tor$(\Om^{d+1}_{SO})$.  The
discrete mixed bosonic gauge-gravity anomalies  are described by $E^d(G)\rtimes
\oplus_{k=1}^{d} H^k(BG,\iTOL^{d-k+1})$.  

In Table \ref{invTop},  we list the generators  of the topological invariants
$W^d_\text{top}(\Ga)$. Those topological invariants describe various bosonic
invertible gravitational anomalies in one lower dimension.  For example,
$W^3_\text{top}(\Ga)=\frac 13 \om_3$ describes the well known perturbative
gravitational anomaly in 1+1D chiral boson theories.  The topological invariant
$W^4_\text{top}(\Ga)=\frac12 \rw_2\rw_3$ implies a new type of bosonic global
gravitational anomaly in 4+1D bosonic theories.  In Tables \ref{U1SPT},
\ref{ZnSPT},  \ref{UTspinSPT}, \ref{BTISPT}, \ref{Z2TSPT}, \ref{O2SPT}, we list
the generators  of the topological invariants $W^d_\text{top}(A,\Ga)/2\pi$ for
some simple groups.  Those topological invariants describe various bosonic
anomalies for those groups at one low dimensions. For example,
$W^3_\text{top}(\Ga)=\frac{1}{(2\pi)^2}A\dd A$ describes the well known
perturbative $U(1)$ gauge anomaly in 1+1D chiral boson theories.  The
topological invariant $W^4_\text{top}(A_{O_2})=\frac12 a_1^2 c_1$ implies a new
type of bosonic global $O_2$ gauge anomaly in 2+1D bosonic theories.  In fact,
all the non-$\Z$-type topological invariants in the Tables give rise to  new
type of bosonic global gauge/gravity/mixed anomalies in one lower dimension.

Note that the invertible anomalies are the usual anomalies people talked about.
They can be canceled by other anomalies. The anomalies, defined by the absence
of well defined realization in the same dimension, can be non-invertible (\ie
cannot be canceled by any other anomalies).\cite{KW1458} All pure gauge and
mixed gauge-gravity anomalies are invertible, but most gravitational anomalies
are not invertible.\cite{KW1458}

\subsection{The relations between the H-type and the L-type topological phases}

We have introduced the concept of potential SPT phases $\PSPT_G^d$ (which may
or may not be  realizable), H-type SPT phases $\HSPT_G^d$ (which are realizable
by H-type local quantum systems), and L-type SPT phases $\LSPT_G^d$ (which are
realizable by L-type local quantum systems).  Those SPT phases are related
\begin{align}
 \LSPT_G^d & \subset \PSPT_G^d,
\nonumber\\
 \HSPT_G^d & \subset \PSPT_G^d,
\nonumber\\
 \LSPT_G^d & \to \HSPT_G^d .
\end{align}
where $\subset$ represents subgroup and $\to$ is a group homomorphism.
Similarly, we also introduced the concept of potential iTO phases $\iTO_P^d$
(which may or may not be  realizable), H-type iTO phases $\iTO_H^d$ (which are
realizable by H-type local quantum systems), and L-type iTO phases $\iTO_L^d$
(which are realizable by L-type local quantum systems).  Those iTO phases are
related
\begin{align}
 \iTO_L^d & \subset \iTO_P^d,
\nonumber\\
 \iTO_H^d & \subset \iTO_P^d,
\nonumber\\
 \iTO_L^d & \to \iTO_H^d .
\end{align}
In condensed matter physics, we are interested in $\iTO_H^d$ and $\HSPT_G^d$.
(A study on the H-type topological phases can be found in \Ref{KW1458,F1478}.)
But in this paper, we will mainly discuss $\iTO_L^d$ and $\LSPT_G^d$. The SPT
states constructed in \Ref{CLW1141,CGL1314,CGL1204} belong to $\LSPT_G^d$ (and
they also belong to $\HSPT_G^d$). The SPT states constructed in
\Ref{VS1306,WS1334,BCF1372,XY1486} belong to $\HSPT_G^d$.  In
\Ref{K1467,K1459v2,KTT1429,F1478,WGW1489} only the potential SPT states
$\PSPT_G^d$ are studied.

\subsection{The organization of this paper}

In Section II, we review the NL$\si$M construction of the pure SPT states.  In
Section III, we generalize the NL$\si$M construction to cover the mixed SPT
states and iTO states.  In Section IV, a construction L-type iTO orders is
discussed.  Using such a construction Section V, we proposed a
construction the pure and the mixed SPT states of the L-type.  In Section
VI, we discussed the L-type SPT states protected by the mirror reflection
symmetry.

\section{Group cohomology and the L-type pure SPT states}

\label{lattG}

A L-type pure SPT state in $d$-dimensional space-time ${M^d}$ can be realized
by a NL$\si$M with the symmetry group $G$ as the target space
\begin{align} 
Z=\int D[g] \ee^{- \int_{M^d} \Big[\frac{1}{\la} [\prt
g(x)]^2 + \ii L^d_\text{top}(g^{-1}\prt g) \Big]}, \ \ \ g(x)\in G 
\end{align} 
in large $ \la$ limit.  Here we treat the space-time as a (random)
lattice which can be viewed as a $d$-dimensional complex.  The space-time
complex has vertices, edges, triangles, tetrahedrons etc.  The field $g(x)$ live
on the vertices and $ \partial g(x)$ live on the edges.  So $\int_{M^d} $ is in
fact a sum over the vertices, edges, and other simplices of the lattice.  $
\partial $ is the lattice difference between vertices connected by edges.  The
above action $S$ actually defines a lattice theory.\cite{CGL1314,CGL1204} 

Under renormalization group transformations, $\la$ flows to infinity.  So the
fixed point action contains only the topological term.  In this
section, we will describe such a fixed-point theory on a space-time
lattice.\cite{LG1220,MR9162,HW1227}. The space-time lattice is a triangulation
of the space-time.  So we will start by describing such a triangulation.

\subsection{Discretize space-time}
\label{disltgauge}

\begin{figure}[t]
\begin{center}
\includegraphics[scale=0.6]{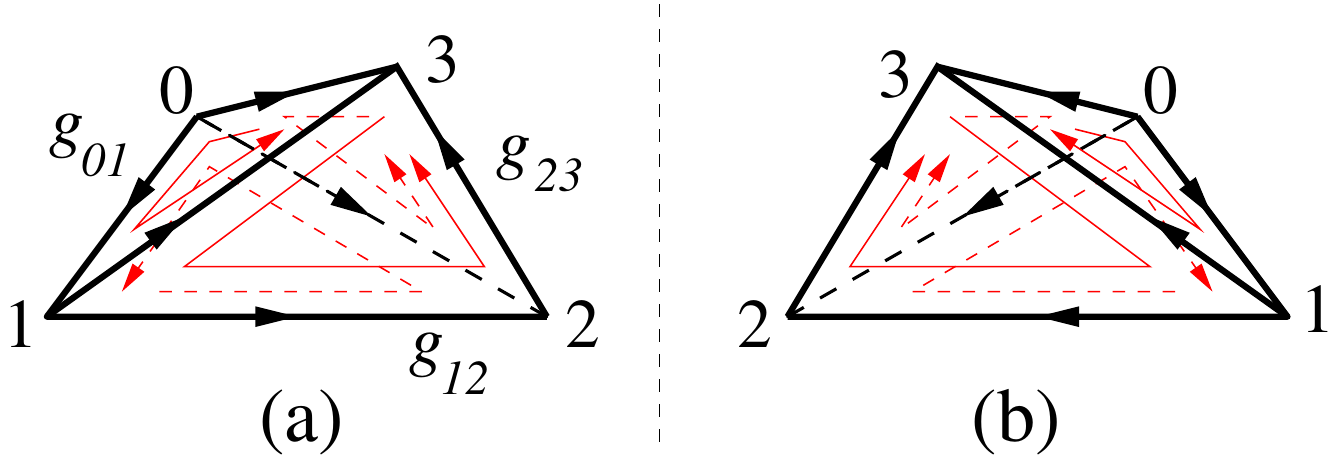} \end{center}
\caption{ (Color online) Two branched simplices with opposite orientations.
(a) A branched simplex with positive orientation and (b) a branched simplex
with negative orientation.  }
\label{mir}
\end{figure}

Let $M^d_\text{tri}$ be a triangulation of the $d$-dimensional space-time.  We
will call the triangulation $M^d_\text{tri}$ as a space-time complex, and a
cell in the complex as a simplex.  In order to define a generic lattice theory
on the space-time complex $M^d_\text{tri}$, it is important to give the
vertices of each simplex a local order.  A nice local scheme to order  the
vertices is given by a branching structure.\cite{C0527,CGL1314,CGL1204} A
branching structure is a choice of orientation of each edge in the
$d$-dimensional complex so that there is no oriented loop on any triangle (see
Fig. \ref{mir}).

The branching structure induces a \emph{local order} of the vertices on each
simplex.  The first vertex of a simplex is the vertex with no incoming edges,
and the second vertex is the vertex with only one incoming edge, \etc.  So the
simplex in  Fig. \ref{mir}a has the following vertex ordering: $0,1,2,3$.

The branching structure also gives the simplex (and its sub simplices) an
orientation denoted by $s_{ij \cdots k}=1,*$.  Fig. \ref{mir} illustrates two
$3$-simplices with opposite orientations $s_{0123}=1$ and $s_{0123}=*$.  The
red arrows indicate the orientations of the $2$-simplices which are the
subsimplices of the $3$-simplices.  The black arrows on the edges indicate the
orientations of the $1$-simplices.

\subsection{$G$ NL$\si$M on a space-time lattice}
\label{gterm}

In our lattice NL$\si$M, the degrees of freedom live on the vertices of the
space-time complex, which are described by $g_i \in G$ where $i$ labels the
vertices.

The action amplitude $\ee^{-S_\text{cell}}$ for a $d$-cell $(ij \cdots k)$ is a
complex function of $g_i$: $A_{ij \cdots k}(\{g_i\})$.  The total action
amplitude $\ee^{-S}$ for a configuration (or a path) is given by
\begin{align}
\label{lattS}
\ee^{-S}=
\prod_{(ij \cdots k)} [A_{ij \cdots k}(\{g_i\})]^{s_{ij \cdots k}}
\end{align}
where $\prod_{(ij \cdots k)}$ is the product over all the $d$-cells $(ij \cdots
k)$.  Note that the contribution from a $d$-cell $(ij \cdots k)$ is $A_{ij
\cdots k}(\{g_i\})$ or $A^*_{ij \cdots k}(\{g_i\})$ depending on the
orientation $s_{ij \cdots k}$ of the cell.  Our lattice $G$ NL$\si$M is defined
by following imaginary-time path integral (or partition function)
\begin{align}
 Z_\text{gauge}=\sum_{ \{g_i\} }
\prod_{(ij \cdots k)} [A_{ij \cdots k}(\{g_i\})]^{s_{ij \cdots k}}
\end{align}
where the  action amplitude $A_{ij \cdots k}(\{g_i\})$ is invariant or
covariant under the $G$-symmetry transformation $g_i \to g'_i =gg_i$, $g \in
G$:
\begin{align}
\label{symmV}
A_{ij \cdots k}(\{gg_i\})= A_{ij \cdots k}^{S(g)}(\{g_i\}) 
\end{align}

Note that here we allow $G$ to contain time-reversal symmetry. In H-type theory
(\ie in Hamiltonian quantum theory) the time-reversal transformation is
implemented by complex conjugation without reversing the time $t\to -t$ (there
is no time to reverse in Hamiltonian quantum theory).  Generalizing that to
L-type theory, we will also implement  time-reversal transformation by complex
conjugation without reversing the time $t\to -t$.  This is the implementation
used in \Ref{CGL1314,CGL1204}.  $S(g)$ in \eqn{symmV} describes the effect of
complex conjugation.  $S(g)=1$ if $g$ contains no time-reversal transformation
and $S(g)=*$ if $g$ contains a time-reversal transformation. 

The fixed-point theory contains only the pure topological term.  Such a pure
topological term can be constructed from a group cocycle $\nu_d \in
\cH^d(G,\R/\Z)$.  Note that a group cocycle $\nu_d( g_0, g_1, \cdots ,g_d),\
g_i\in G$ is a map from $G^{d+1}$ to $\RZ$ (see Appendix \ref{gcoh}). We can
express the action amplitude $A_{ij \cdots k}(\{g_i\})$ that correspond to a
pure topological term as\cite{CGL1314,CGL1204}
\begin{align}
& A_{01 \cdots d}(\{g_i\})
=\ee^{2\pi \ii \nu_d( g_0, g_1, \cdots , g_d)}.
\end{align}

Due to the symmetry condition \eq{scond}, the  action amplitude $A_{ij \cdots
k}(\{g_i\})$ is invariant/covariant under the $G$-symmetry transformation.  Due
to the cocycle condition \eq{cccond}, the total action amplitude  on a closed
space-time $M^d$ is always equal to 1:
\begin{align}
\label{Zeq1}
\ee^{\ii \int_{M^d} L^d_\text{top}(g^{-1}\prt g)}=\prod_{(ij \cdots k)}
[A_{ij \cdots k}(\{h_{ij}\},\{g_i\})]^{s_{ij \cdots k}}=1.
\end{align}
Also two  cocycles different by a coboundary (see \eqn{cb}) can be smoothly
deformed into each other without affecting the condition \eq{Zeq1}.  In other
words, the connected components of the fixed-point theories that satisfy the
condition \eq{Zeq1} are described by $\cH^d(G,\R/\Z)$.  This way, we show that
the fixed-points of the $G$ NL$\si$Ms are classified by the
elements of $\cH^d(G,\R/\Z)$.

We like to remark that for continuous group, the cocycle $\nu_d( g_0, g_1,
\cdots ,g_d)$ do not need to be continuous function of $g_i$. It can be a
measurable function.

\subsection{Adding the $G$-symmetry twist}
\label{Gtwist}

The above bosonic system may be in different SPT phases for different choices
of the topological term (\ie for different choices of group cocycles $\nu_d \in
\cH^d(G,\R/\Z)$).  But how can we be sure that the system is indeed in
different SPT phases?  One way to address such a question is to find measurable
topological invariants, and show that different cocycles give rise to different
values for the topological invariants.

In this section, we will assume that the symmetry group does not contain
time-reversal. In this case, the universal topological invariants for SPT state
can be constructed systematically by twisting (or ``gauging'') the on-site
symmetry \cite{LG1220,HW1267,HWW1295,HW1227} and study the gauged bosonic model
\begin{align}
 Z(A)=\int D[g] \ee^{\int_{{M^d}} \Big[\frac{1}{ \la} 
 [(\prt -\ii A)g]^2 + \ii L^d_\text{top}(g^{-1}(\prt-\ii A) g)\Big]}.
\end{align}
Note that the gauge field $A$ just represents space-time dependent coupling
constants, which is not dynamical (\ie we do not integrate out the gauge field
$A$ in the path integral).  Since the SPT state is gapped for large $\la$, in
large space-time limit, the partition function has a form
\begin{align}
 Z(A)= \ee^{-\eps_0V_\text{space-time} }
 \ee^{\ii \int_{{M^d}} 2\pi W^d_\text{top}(A)}
\end{align}
where $\eps_0$ is the ground state energy density and $V_\text{space-time}$ is
the volume of the space-time manifold ${M^d}$.  The term $\int_{{M^d}}
2\pi W^d_\text{top}(A)$ represents the volume independent term in the partition
function and is conjectured to be universal (\ie independent of any small local
change of the Lagrangian that preserve the symmetry).\cite{KW1458} Such a term
is called \emph{the realizable gauge topological term (or topological invariant)}, which is referred as
the SPT invariant in \Ref{W1447,HW1339}. The SPT invariants are the topological
invariants that are believed to be able to characterize and distinguish any
SPT phases.

The  topological invariant is gauge invariant, \ie for any closed space-time
manifold ${M^d}$
\begin{align}
\label{gaugeinv}
&
\int_{{M^d}} W^d_\text{top}(A^g) - \int_{{M^d}} W^d_\text{top}(A) 
= 0 \text{ mod } 1,
\nonumber\\
&\ \ \  A^g = g^{-1}A g+\ii g^{-1}dg ,
\end{align}
where we have treated $A$ as the gauge field one form.  Also, as a topological
invariant,  $W^d_\text{top}(A)$ does not dependent on the metrics of the space-time.
For example  $W^d_\text{top}(A)$ can be a Chern-Simons term $\frac{2k}{4\pi}
\Tr A\dd A,\ \ k\in \Z$ in 2+1D or a $\th$-term $\frac{\th}{(2\pi)^2}  \dd A\dd
A$ in 3+1D.  The presence of non-trivial topological invariant
$Z_\text{fixed}(A)=\ee^{\ii \int_{{M^d}} 2\pi W^d_\text{top}(A)}$ indicates the
presence of non-trivial SPT phase.

In the above, we described the symmetry twist in the continuous field theory.
On lattice, the symmetry twist can be achieved by introducing $h_{ij}\in G$ for
each edge $ij$ in the space-time complex $M^d_\text{tri}$.  
The twisted theory (\ie the ``gauged'' theory) is described by the total action
amplitude $\ee^{-S}$ 
\begin{align}
\label{eS}
\ee^{-S}=
\prod_{(ij \cdots k)} [\t A_{ij \cdots k}(\{h_{ij}\},\{g_i\})]^{s_{ij \cdots k}}
\end{align}
The imaginary-time path integral (or partition function) is given by
\begin{align}
 Z(\{h_{ij}\})=\sum_{\{g_i\} }
\prod_{(ij \cdots k)} [\t A_{ij \cdots k}(\{h_{ij}\},\{g_i\})]^{s_{ij \cdots k}}.
\end{align}
We see that only $g_i$ are dynamical. $h_{ij}$ are non-dynamical background
probe fields.
The above action amplitude  $\prod_{(ij \cdots k)} [\t A_{ij \cdots
k}(\{h_{ij}\},\{g_i\})]^{s_{ij \cdots k}}$ on closed  space-time complex ($\prt
M^d=\emptyset $) should be invariant under the ``gauge'' transformation
\begin{align}
\label{gaugeT}
 h_{ij} \to g'_{ij}=h_i h_{ij} h_j^{-1}, g_i \to g'_i =h_ig_i\ \ \ h_i \in G.
\end{align}
and covariant under the global symmetry transformation
\begin{align}
\label{symmT}
 h_{ij} \to h'_{ij}=g h_{ij} g^{-1}, g_i \to g'_i =gg_i\ \ \ g \in G :
\end{align}
\begin{align}
 \t A_{ij \cdots k}(\{h_{ij}\},\{g_i\}) =
 \t A_{ij \cdots k}^{S(g)}(\{h'_{ij}\},\{g'_i\}) .
\end{align}

The gauged action amplitudes $\t A_{ij \cdots k}(\{h_{ij}\},\{g_i\})$ is
obtained from the ungauged  action amplitudes $A_{ij \cdots k}(\{g_i\})$ in the
following way (where we assume $G$ is discrete):
\begin{align}
& \t A_{01 \cdots d}(\{h_{ij}\},\{g_i\})=0,
\text{ if } h_{ij}h_{jk}\neq h_{ik} ,
\nonumber\\
& \t A_{01 \cdots d}(\{h_{ij}\},\{g_i\})=
 A_{01 \cdots d}(h_0 g_0, h_1 g_1, \cdots , h_d g_d),
\end{align}
where $h_i$ are given by
\begin{align}
 h_0 &=1, & h_1 &=h_0 h_{01},
&
 h_2 &=h_1 h_{12}, & h_3 &=h_2 h_{23},  \cdots .
\end{align}

At a fixed-point, the twisted action amplitude $A_{ij \cdots
k}(\{h_{ij}\},\{g_i\})$ is given by
\begin{align}
\label{Vnud}
& \t A_{01 \cdots d}(\{h_{ij}\},\{g_i\})=
\ee^{2\pi \ii \nu_d( h_0 g_0, h_1 g_1, \cdots , h_d g_d)},
\nonumber\\ &
=
\ee^{2\pi \ii 
\om_d(g_0^{-1}h_{01} g_1, \cdots, g_{d-1}^{-1}h_{d-1,d}g_d )},
\text{ if } h_{ij}h_{jk}= h_{ik},
\nonumber
\end{align}
where $\om_d$ is the inhomogeneous cocycle corresponding to $\nu_d$
\begin{align}
 \om_d(h_{01},h_{12},\cdots,h_{d-1,d})=\nu_d(h_0,h_1,\cdots,h_d) .
\end{align}
By rewriting the partition function as (see \eqn{Vnud})
\begin{align}
 Z(\{h_{ij} \})=\sum_{ \{g_i\} }
\prod_{(ij \cdots k)} [A_{ij \cdots k}(\{g_i^{-1}h_{ij}g_j\},\{1\})]^{s_{ij \cdots k}}
\end{align}
we find that the partition function is explicitly gauge invariant and
symmetric. 

The topological invariant $W^d_\text{top}(A)$ is given by the fixed-point partition
function for the twisted theory 
\begin{align}
	\ee^{\ii \int_{{M^d}} 2\pi W^d_\text{top}(A)}
= Z_\text{fixed}(\{h_{ij} \}) =Z_\text{fixed}(A) . 
\end{align}
The twisted fixed-point partition function $Z_\text{fixed}(\{h_{ij} \})$ or
$Z_\text{fixed}(A)$ is non-trivial and depend on the symmetry twist $h_{ij}$
(or gauge connection $A$).  We see that different realizable topological invariants
$W^d_\text{top}(A)$ are classified and given explicitly by the elements of
group cohomology $\cH^d(G,\R/\Z)$:
\begin{align}
\label{Wdomd}
\ee^{\ii \int_{{M^d}} 2\pi W^d_\text{top}(A)}
= \prod_{(ij \cdots k)} [\ee^{2\pi \ii \om_{d}(\{h_{ij}\})}]^{s_{ij \cdots k}}
\end{align}
where $\om_{d}(h_1,\cdots,h_d)$ is an inhomogeneous cocycle in
$\cH^d(G,\R/\Z)$, and $\{h_{ij}\}$ on the edges complex $M^d_\text{tri}$ define
the symmetry twist $A$ in space-time $M^d$.  Eq. (\ref{Wdomd}) tells us how to
calculate $\ee^{\ii \int_{{M^d}} 2\pi W^d_\text{top}(A)}$, given cocycle $\om_d$,
the space-time manifold $M^d_\text{tri}$, and the symmetry-twist $A=\{h_{ij}\}$.

We can also see this within the field theory.  The realizable gauge topological
invariant $W^d_\text{top}(A)$ and the NL$\si$M topological term $
L^d_\text{top}(g^{-1}(\prt-\ii A) g)$ are directly related:
\begin{align}
	L^d_\text{top}(g^{-1}(\prt-\ii A) g)= 2\pi W^d_\text{top}( A).
\end{align}
Since the  NL$\si$M topological terms $
L^d_\text{top}(g^{-1}(\prt-\ii A) g)$ are classified by the group cohomology
$\cH^d(G,\RZ)$ of the symmetry group  $G$.  The realizable gauge topological
invariants $W^d_\text{top}( A)$ are also classified by $\cH^d(G,\RZ)$.

The gauge topological term (or topological invariant) $W^d_\text{top}(A)$ can be defined for both continuous
and discrete symmetry groups $G$. In general, it is a generalization of the
Chern-Simons term.\cite{DW9093,HW1267,HWW1295} It describes the response of the
quantum ground state.  We hope that the ground states in different quantum
phases will produce different responses, which correspond to different classes
of gauge topological terms, that cannot be smoothly deformed into each other.
So we can use such a term to study and classify pure SPT phases.  

We would like to point out that there are two kinds of topological invariants.
The topological invariants correspond to Tor$[\cH^d(G,\RZ)]$
are called \emph{locally-null topological invariants}.
They have the following defining properties:\\
(1) $\int_{M^d} W^d_\text{top}(A)$ are well defined for any symmetry twists $A$.\\
(2) $\int_{M^d} W^d_\text{top}(A)$ does not depend on any small smooth change
of the symmetry twist:
\begin{align}
	\int_{M^d} W^d_\text{top}(A+\del A)=\int_{M^d} W^d_\text{top}(A)
\end{align}
The topological invariants correspond to Free$[\cH^d(G,\RZ)]$
are called \emph{Chern-Simons topological invariants}.
The Chern-Simons topological invariants is only well defined for some
symmetry twists $A$. In general, only the difference 
\begin{align}
	\int_{\t M^d} W^d_\text{top}(\t A)- \int_{M^d} W^d_\text{top}(A) 
\end{align}
is well defined, provided that there exist an $(d+1)$-dimensional manifold
$N^{d+1}$ such that $\prt N^{d+1} = \t M^d \cup (-M^d)$ and the gauge
connections $A$ on $M^d$ and $\t A$ on $\t M^d$ can be extended to $N^{d+1}$
(see Appendix \ref{agterm}).

Now two questions naturally arise:\\
(1) how to write down the most general topological invariants
$W^d_\text{top}(A)$ (\ie  the most general topological invariants) which are
self consistent?  We will call such topological invariants as \emph{potential
topological invariants}.\\
(2) can we show that every potential topological invariant can be induced by
some symmetric local bosonic model, after we gauge the on-site symmetry?\\
In Appendix \ref{agterm}, we will 
address these two questions.  We find that the potential gauge topological
invariants $W^d_\text{top}(A)$ are described by $H^{d+1}(BG,\Z)$,
which are all realizable since $\cH^d(G,\RZ)=H^{d+1}(BG,\Z)$.

\subsection{$G\times G'$ pure SPT states}
\label{GGpSPT}

In this section, we will study $G\times G'$ pure SPT states described by group
cohomology $\cH^d(G\times G',\RZ)$.  This result will be useful for later
discussions.  First, we can use the following version of K\"unneth
formula\cite{W1313,W1447}
\begin{align}
\label{kunnH}
& \cH^d(G\times G',\RZ)  \simeq 
\cH^d(G,\RZ)\oplus \cH^d( G',\RZ)\oplus
\nonumber \\
&\ 
\oplus_{k=1}^{d-1} H^k[BG, \cH^{d-k}(G',\RZ)] 
\end{align}
to compute $\cH^d(G\times G',\RZ)$.  In addition, the above  K\"unneth formula
can help us to construct topological invariants to probe the
$G\times G'$ SPT order.\cite{W1447}  

For example, a $G$ SPT order in $d$-dimensional space-time can be
probed by a map $W^d_\text{top}$, that maps a closed space-time
$M^d$ with a $G$-symmetry twist $A$ to a number in $\RZ$: 
\begin{align}
\label{WMA}
\int_{M^d} W^d_\text{top}(A) \in \RZ.
\end{align}
Such a map is nothing but the topological invariant that we discussed before.
At the same time, the topological invariant can also be viewed as a cocycle in
$H^d[BG, \RZ]$, since it is a map for the $G$-bundles (\ie the $G$-symmetry
twists) on $M^d$ to $\RZ$, and the  $G$-bundles on $M^d$ is classified by the
embedding of $M^d$ into the classifying space.  Different SPT states will lead
to different maps. We believe that the map $W^d_\text{top}$ fully
characterizes the $G$ SPT states described by $\cH^d[G, \RZ]$ (see Appendix
\ref{agterm}).\cite{W1447,HW1339}

Similarly, for the $G\times G'$ pure SPT states described by $ H^k[BG,
\cH^{d-k}(G',\RZ)]$, they can also be  probed by a map $W^d_{\cH^k}$, that maps
a closed space-time $M^k$ with a $G$-symmetry twist $A_G$ on $M^k$ to an
element in $H^{d-k}(G',\RZ)$. This is simply a dimension reduction: we consider
a pace-time of the form $M^k\times M^{d-k}$, add a $G$-symmetry twist $A_G$ on
$M^k$, and then take a large $M^{d-k}$ limit. The system can be viewed as a
$(d-k)$-dimensional $G'$ SPT state on $M^{d-k}$, which is described by an
element in $H^{d-k}(G',\RZ)$.  Such a dimension reduction can be formally
written as 
\begin{align} 
\int_{M^k} W^d_{\cH^k}(A_G) \in \cH^{d-k}(G',\RZ) .  
\end{align} 
which has the same structure as \eqn{WMA}.
The  map $W^d_{\cH^k}$ can be viewed as a cocycle in $H^k[BG,
\cH^{d-k}(G',\RZ)]$.  Such a map fully characterizes the $G\times G'$ pure SPT
states described by $H^k[BG, \cH^{d-k}(G',\RZ)]$.

The dimension reduction discussed above reveals the physical meaning of the
K\"unneth formula.  We will use such a physical picture to obtain the key
result of this paper.

\section{Constructing pure and mixed SPT states, as well as \lowercase{i}TO
states}

\subsection{SPT states, gauge anomalies and mixed gauge-gravity anomalies}

So far, we have reviewed the group cohomology approach
to pure SPT states.  It was pointed out in \Ref{W1313} that (a) the SPT orders
(described by $\cH^d(G,\RZ)$) and pure gauge anomalies in one lower dimension
are directly related and (b) the topological orders and gravitational anomalies
in one lower dimension are directly related. This suggests that the SPT orders
beyond $\cH^d(G,\RZ)$\cite{VS1306,WS1334,BCF1372,K1467,K1459v2,WGW1489,XY1486}
and mixed gauge-gravity anomalies are closely related.\cite{WGW1489} This line
of thinking gives us a deeper understanding of generic SPT states. In this
section, we are going to construct local bosonic models that systematically
realize iTO's, pure SPT orders (associated with pure gauge anomaly), and mixed
SPT orders (associated with mixed gauge-gravity anomaly).

\subsection{Realizable L-type SPT and iTO phases}
\label{Rggterm}

One of the key properties of SPT states is that they do not contain any
non-trivial topological excitations.\cite{CLW1141,CGL1314,CGL1204} In
\Ref{KW1458} it was conjectured that a gapped quantum liquid state has no
non-trivial topological excitations iff its fixed-point partition function is a
pure $U(1)$ phase. 

However, when we study the pure SPT orders described by $\cH^d(G,\RZ)$ using $G$ NL$\si$Ms, we only
add the symmetry twists,  which are associated with the $G$-bundles on the
space-time, to induce the non-trivial $U(1)$-phase-valued partition
function.  This is why we only get pure gauge anomalies in such an approach.
To get the gravitational anomalies and the mixed gauge-gravity anomalies, we
must include the space-time twist, described by the non-trivial tangent bundle
of the space-time as well.  The tangent bundle is a $SO_d\equiv SO(d)$ bundle.
Thus to include the gravitational anomalies and the mixed gauge-gravity
anomalies, as well as the pure gauge anomalies, we simply need to consider a
$SO_d\times G$ NL$\si$M with topological term $L^d_\text{top}(g^{-1}\prt g)$ where $g(x)\in
SO_d\times G$.  We can gauge the $G$ symmetry to probe the SPT states and the
pure gauge anomalies as before.  We can also choose non-flat space-time to
probe the SPT states (and the gravitational anomalies), that corresponds to
couple the $SO_d$ part of the NL$\si$M to the connection of the tangent bundle
of the space-time.  We will see that using $G\times SO_d$ NL$\si$Ms, we can
obtain a topological invariant $W^d_\text{top}(A,\Ga)$ that contains both the gauge
$G$-connection $A$ and the gravitational $SO_d$-connection $\Ga$.  Such kind of
bosonic NL$\si$M is capable of producing the pure SPT states that are
associated with pure gauge anomalies, as well as the mixed SPT states that are
associated with mixed gauge-gravity anomalies.  It can also produce iTO states,
if we choose a trivial symmetry group $G$.

Here we would like to remark that we can also use an $G\times SO_n$ NL$\si$M
with $n>d$ to produce the SPT states and iTO states.  The stability
consideration suggests the we should take $n=\infty$.  So we will use $G\times
SO$ NL$\si$M to study the new topological states, where $SO\equiv SO_\infty$.

Repeating the discussion in Section \ref{gterm}, we find that the realizable
gauge-gravity topological invariants $W^d_\text{top}(A,\Ga)$ in the $G\times SO$
NL$\si$M can be constructed from each element in the group cohomology class
$\cH^d(G\times SO,\RZ)$.  However, because of the restrictive relation between
the gravitational connection $\Ga$ and the topology of the space-time (see
Appendix \ref{Pggterm}), the correspondence is not one-to-one: different
elements in $\cH^d(G\times SO,\RZ)$ may produce the same realizable
gauge-gravity topological invariant $W^d_\text{top}(A,\Ga)$ after integrating out
the matter field $g$.  The reason is the following.  For two topological invariants
$W^d_\text{top}(A,\Ga)$ and $\t W^d_\text{top}(A,\Ga)$ obtained from two
cocycles $\nu_d$ and $\t\nu_d$ in $\cH^d(G\times SO,\RZ)$, it is possible that
\begin{align}
	\ee^{\ii \int_{M^d} 2\pi W^d_\text{top}(A,\Ga)} =
	\ee^{\ii \int_{M^d} 2\pi \t W^d_\text{top}(A,\Ga)}
\end{align}
on any closed space-time $M^d$.  In this case, we should view
$W^d_\text{top}(A,\Ga)$ and $\t W^d_\text{top}(A,\Ga)$ as the same topological
invariant.  (Note that the above two topological invariants can be distinguished if the
$SO$ connection $\Ga$ is not restricted to be the connection of the tangent
bundle of the space-time $M^d$.) Thus, $\cH^d(G\times SO,\RZ)$ contains a
subgroup $\La^d(G)$ such that the realizable gauge-gravity topological invariants
$W^d_\text{top}(A,\Ga)$ have an one-to-one correspondence with the elements in
$\cH^d(G\times SO,\RZ)/\La^d(G)$.  Those different NL$\si$Ms, that
produce different the realizable gauge-gravity topological invariants
$W^d_\text{top}(A,\Ga)$, realize different L-type topological phases with no
topological excitations.

In Appendix \ref{Pggterm}, we will discuss  potential gauge-gravity topological
invariants $W^d_\text{top}(A,\Ga)$.  We find that the locally-null potential
gauge-gravity topological invariants $W^d_\text{top}(A,\Ga)$ are described by a
subgroup of $\cH^d(G\times SO,\RZ)$, which are all realizable.  We also find
that Chern-Simons potential gauge-gravity topological invariants
$W^d_\text{top}(A,\Ga)$ are described by a subgroup of $H^{d+1}[B(SO\times
G),\Z(\frac1n)]$.  NL$\si$Ms can only realize those that are also in
$H^{d+1}[B(G\times SO), \Z]$.

\section{\lowercase{i}TO states}

Using Pontryagin class and Stiefel-Whitney class, one can show that different
L-type potential iTO phases (\ie may not be realizable)  are described by $\Z$
in 3-dimensions,\cite{PMN1372} $\Z_2$ in 4-dimensions,\cite{KW1458,K1459v2} and
$2\Z$ in 7-dimensions, where the dimensions $d$ are the space-time dimensions.
In this section, we will reexamine those results using the approaches discussed
above, and try to understand which L-type potential iTO can be realized by $SO$
NL$\si$Ms.  We will show that the above potential topologically ordered phases
described by Stiefel-Whitney class are always realizable, while only a subset
of those described by Pontryagin classes are realizable by $SO$ NL$\si$Ms. The
result is summarized in Table \ref{invTop}.

\subsection{Classification of $SO$ NL$\si$Ms}

Since we do not have any symmetry, the realizable gauge-gravity topological
invariants produced by the NL$\si$Ms are covered by $\cH^d(SO,\RZ)=H^{d+1}(BSO,\Z)$,
$d>1$.  In Appendix \ref{HBSOZ}, we calculated the ring $H^*(BSO,\Z)$.  In low
dimensions, we have
\begin{align}
\label{HBSOZ1}
 H^0(BSO,\Z) &=\Z,
\nonumber \\
 H^1(BSO,\Z) &=0,
\nonumber \\
 H^2(BSO,\Z) &=0,
\nonumber\\
 H^3(BSO,\Z) &=\Z_2, \text{ basis } \bt  (\rw_2),
\nonumber\\
 H^4(BSO,\Z) &=\Z, \text{ basis } p_1,
\nonumber\\
 H^5(BSO,\Z) &=\Z_2, \text{ basis } \bt  (\rw_4),
\\
 H^6(BSO,\Z) &=\Z_2, \text{ basis } \bt  (\rw_2) \bt  (\rw_2),
\nonumber\\
 H^7(BSO,\Z) &=2\Z_2, \text{ basis }\bt  (\rw_6) , \rw_2^2 \bt  (\rw_2) ,
\nonumber\\
 H^8(BSO,\Z) &=2\Z\oplus \Z_2, \text{ basis } p_1^2, p_2,  \bt  (\rw_2) \bt  (\rw_4).
\nonumber 
\end{align}
We note that due to the relation $\cH^d(SO,\RZ)=H^{d+1}(BSO,\Z)$, the
$d$-dimensional gauge-gravity topological invariants $W^d_\text{top}(A,\Ga)$ (with
values in $\RZ$) is promoted to $(d+1)$-dimensional topological invariants
$K^{d+1}(A,\Ga)$ (with values in $\Z$).  In the above, we also listed the
basis of those topological invariants, so that a generic topological invariant
$K^{d+1}(A,\Ga)$ is a superposition of those basis.  In the following, we
list $\cH^d(SO,\RZ)$ and the basis of their topological invariants
$W^d_\text{top}(A,\Ga)$: 
\begin{align}
\label{HBSOU1}
 \cH^0(SO,\RZ) &=0,
\nonumber \\
 \cH^1(SO,\RZ) &=0,
\nonumber\\
 \cH^2(SO,\RZ) &=\Z_2, \text{ basis } \frac12 \rw_2,
\nonumber\\
 \cH^3(SO,\RZ) &=\Z, \text{ basis } \om_3,
\nonumber\\
 \cH^4(SO,\RZ) &=\Z_2, \text{ basis } \frac12 \rw_4,
\\
 \cH^5(SO,\RZ) &=\Z_2, \text{ basis } \frac12 \rw_2(\rw_1\rw_2+\rw_3),
\nonumber\\
 \cH^6(SO,\RZ) &=2\Z_2, \text{ basis } \frac12 \rw_6 , \frac12 \rw_2^3 ,
\nonumber\\
 \cH^7(SO,\RZ) &=2\Z\oplus \Z_2, \text{ basis } \om_7^{p_1^2}, \om_7^{p_2},  
\frac12 (\rw_1\rw_2+\rw_3) \rw_4.
\nonumber 
\end{align}
The above basis give rise to the basis in \eqn{HBSOZ1} through the natural map
$\t\bt$: $\cH^d(G,\RZ)\to \cH^{d+1}(G,\Z)$ (see Appendix \ref{calgen}).

We see that $\cH^2(SO,\RZ)=\Z_2$, which implies that a realizable
gauge-gravity topological invariant exist in 1+1D, provided that we probe the $SO$
NL$\si$M by an arbitrary $SO$ bundle on an oriented 1+1D space-time manifold
$M^2$:
\begin{align}
	\int_{M^2}  W^2_\text{top}(\Ga_{SO})=
\int_{M^2} \frac m2  \rw_2^{SO}, \ \ \ \ \ \ \ m =0,1.
\end{align}
where $\Ga_{SO}$ is the connection of the $SO$ bundle on $M^2$ and
$\rw_i^{SO}$ are the Stiefel-Whitney classes for the $SO$ bundle. However, the
$SO$ bundle on $M^2$ is restricted: it must be the tangent bundle of
$M^2$. 
So we actually have
\begin{align}
	\int_{M^2}  W^2_\text{top}(\Ga)=
\int_{M^2} \frac m2 \rw_2, \ \ \ \ \ \ \ m =0,1.
\end{align}
where $\Ga$ is the connection of the tangent bundle on $M^2$ and $\rw_i$ are
the Stiefel-Whitney classes for the tangent bundle.  The  Stiefel-Whitney
classes for the tangent bundle have some special relations.  
In fact, we have\\
(1) a manifold is orientable iff $\rw_1=0$.\\
(2) a manifold admits a spin structure iff $\rw_2=0$.\\
Since all closed orientable 2-dimensional manifold is spin, thus both $\rw_1$
and $\rw_2$ vanish for tangent bundles of $M^2$. The realizable gauge-gravity
topological invariant cannot be probed by any oriented space-time $M^2$.  Thus, the
above realizable gauge-gravity topological invariants described by $\cH^2(SO,\RZ)$
collapse to zero in 1+1D.  There is no iTO in 1+1D (or in other words,
$\si\iTOL^2=H^2(SO,\RZ)/\La^2=0$).

\subsection{Relations between  Stiefel-Whitney classes}
\label{Rswc}

We see that to understand the realizable gauge-gravity topological invariants,
whether they collapse to zero or not,   it is important to understand all
relations that the Stiefel-Whitney classes must satisfy, when the
Stiefel-Whitney classes come from a tangent bundle.
To obtain such relations, let us first consider the Stiefel-Whitney classes for
an arbitrary $O$ vector bundle on a $d$-dimensional space.  

We note that the
total Stiefel-Whitney class $w=1+\rw_1+\rw_2+\cdots$ is related to the total Wu
class $u=1+u_1+u_2+\cdots$ through the total Steenrod square:
\begin{align}
 w=Sq(u),\ \ \ Sq=1+Sq^1+Sq^2+ \cdots .
\end{align}
Therefore, 
\begin{align}
\rw_n=\sum_{i=0}^n Sq^i u_{n-i}.
\end{align}
The Steenrod squares have the following properties:
\begin{align}
Sq^i(x_j) &=0, \  i>j, \ \ 
Sq^j(x_j) =x_jx_j,  \ \  Sq^0=1,
\end{align}
for any $x_j\in H^j(X^d,\Z_2)$.
Thus
\begin{align}
u_n=\rw_n+\sum_{i=1, 2i\leq n} Sq^i u_{n-i}.
\end{align}
This allows us to compute $u_n$ iteratively, using Wu formula
\begin{align}
\label{WuF}
Sq^i(\rw_j) &=0, \ \ i>j, \ \ \ \ \
Sq^i(\rw_i) =\rw_i\rw_i, 
\\
 Sq^i(\rw_j) &= \rw_i\rw_j+\sum_{k=1}^i 
\frac{(j-i-1+k)!}{(j-i-1)!k!}
\rw_{i-k} \rw_{j+k},\ \ i<j ,
\nonumber 
\end{align}
and the Steenrod relation 
\begin{align}
	Sq^n(xy)=\sum_{i=0}^n Sq^i(x)Sq^{n-i}(y).
\end{align}
We find
\begin{align}
u_0&=1, 
	\nonumber\\
u_1&=\rw_1, 
	\nonumber\\
u_2&=\rw_1^2+\rw_2, 
	\nonumber\\
u_3&=\rw_1\rw_2, 
	\nonumber\\
u_4&=\rw_1^4+\rw_2^2+\rw_1\rw_3+\rw_4, 
	\\
u_5&=\rw_1^3\rw_2+\rw_1\rw_2^2+\rw_1^2\rw_3+\rw_1\rw_4, 
	\nonumber\\
u_6&=\rw_1^2\rw_2^2+\rw_1^3\rw_3+\rw_1\rw_2\rw_3+\rw_3^2+\rw_1^2\rw_4+\rw_2\rw_4, 
	\nonumber\\
u_7&=\rw_1^2\rw_2\rw_3+\rw_1\rw_3^2+\rw_1\rw_2\rw_4, 
	\nonumber\\
u_8&=\rw_1^8+\rw_2^4+\rw_1^2\rw_3^2+\rw_1^2\rw_2\rw_4+\rw_1\rw_3\rw_4+\rw_4^2
	\nonumber\\
	&\ \ \ \ 
	+\rw_1^3\rw_5 +\rw_3\rw_5+\rw_1^2\rw_6+\rw_2\rw_6+\rw_1\rw_7+\rw_8. 
	\nonumber
\end{align}
We note that the Steenrod squares form an algebra:
\begin{align}
& Sq^aSq^b=\sum_{j=0}^{[a/2]} \frac{(b-j-1)!}{(a-2j)!(b-a+j-1)!}
Sq^{a+b-j} Sq^j, 
\nonumber\\
& 0<a<2b,
\end{align}
which leads to the relation $Sq^1Sq^1=0$ used in the last section.

If the $O$ vector bundle on $d$-dimensional space, $M^d$,
happen to be the tangent bundle of $M^d$, then
the Steenrod square and the Wu class satisfy
\begin{align}
\label{SqWu}
Sq^{d-j}(x_j)=u_{d-j} x_j,  \text{ for any } x_j \in H^j(X^d,\Z_2) .
\end{align}
(1) If we choose $x_j$ to be a combination of Stiefel-Whitney classes, the
above will generate many relations between Stiefel-Whitney classes.
\\
(2) Since $Sq^i(x_j)=0$ if $i>j$, therefore $u_ix_{d-i}=0$ for any $x_{d-i} \in
H^{d-i}(X^d,\Z_2)$ if $i>d-i$.  Thus, for $d$-dimensional manifold, the Wu
class $u_i=0$ if $2i>d$.  Also $Sq^n\cdots Sq^m(u_i)=0$ if $2i>d$.  This also
gives us relations among  Stiefel-Whitney classes.  
\\
(3) Last, there is another type of relation. In $4n$-dimension, the mod 2 reduction
of Pontryagin classes $p_{i_1}p_{i_2}\cdots$, $n=i_1+i_2+\cdots$, should be
regarded as zero. The reason is explained below the \eqn{w22}.  This lead to
the relations for $d$-dimensional manifold
\begin{align}
\label{w2i2pi}
 \rw_{2i_1}^2
 \rw_{2i_2}^2\cdots =0, \text{ if } 2i_1+2i_2+\cdots = d.
\end{align}
\frm{$\si\iTOL^d$ is given by $H^{d+1}(BSO,\Z)$ after quotient out all those relations.}

\subsection{iTO phases in low dimensions} \label{lowDiTO}

In 2-dimensional space-time $\cH^2(SO,\RZ)=H^3(BSO,\Z)=\Z_2$ which is generated
by $W^2_\text{top}=\frac12 \rw_2$.  So $\si\iTOL^2$ may be non-trivial.  The
relations $u_2=u_3=0$ give us
\begin{align}
	 \rw_1^2+\rw_2=0.
\end{align}
Since $M^2$ is oriented, $\rw_1=0$.
We see that $\rw_2=0$. $W^2_\text{top}$ vanishes, and there is no
realizable gauge-gravity topological invariant in 1+1D.  So $\si\iTOL^2=0$.

In 2+1D space-time, the corresponding $\cH^3(SO,\RZ)=\Z$ is generated by
$W^3_\text{top}=\om_3$.  There is no relation involving $\om_3$.  So
$\si\iTO_L^3=\Z$. The generating topological invariant $W^3_\text{top}(\Ga)=\om_3$
describes an iTO state with chiral central charge $c=24$.  

In 3+1D space-time, the corresponding $\cH^4(SO,\RZ)=\Z_2$ is generated by
the gauge-gravity topological invariant
$W^4_\text{top}=\frac12 \rw_4$.
The Wu classes $u_3=u_4=0$ can lead to
relations between the Stiefel-Whitney classes, which give us
\begin{align}
 \rw_1\rw_2=\rw_1^4+\rw_1\rw_3+\rw_2^2+\rw_4=0 .
\end{align}
Other relations can be obtained by applying the Steenrod squares to the above:
\begin{align}
 Sq^1(\rw_1\rw_3)=\rw_1\rw_3=0.
\end{align}
Additional relations can be obtained from \eqn{SqWu}
\begin{align}
Sq^1(\rw_3)=u_1\rw_3 &\to \rw_1\rw_3=\rw_1\rw_3
	\\
Sq^2(\rw_2)=u_2\rw_2 &\to \rw_2^2=\rw_1^2\rw_2+\rw_2^2.
	\nonumber 
\end{align}
We see that $\rw_4=\rw_2^2$, but nothing restricts $\rw_2^2$.  Naively, this
suggests that $\rw_2^2 \in H^4(M^4,\Z_2)$ is a realizable gauge-gravity
topological invariant in 3+1D: 
\begin{align} 
\label{w22}
W^4_\text{top}(\Ga)= \frac12 \rw_2^2 .
\end{align}
However, there is a relation between Pontryagin classes and Stiefel-Whitney
classes (see Appendix \ref{PandSW}):
\begin{align}
 \rw_{2i}^2 = p_i \text{ mod } 2.
\end{align}
on any closed oriented manifolds $M^{4i}$ of dimension $4i$.  Thus $\rw_2^2$ is
part of Pontryagin class $p_1$.  The topological invariant
$W^4_\text{top}(\Ga)= \frac12 \rw_2^2=\frac12 p_1$ is realizable, but also
smoothly connect to the trivial case via the Pontryagin class:
$W^4_\text{top}(\Ga)= \frac{\th}{2\pi} p_1$,  where $\th$ can go from $\pi$ to
$0$ smoothly.  There is no realizable gauge-gravity topological invariant in
3+1D that cannot connect to zero.  Thus $\si\iTOL^4=0$.  In general, such kind of
reasoning give rise to \eqn{w2i2pi}.

In 4+1D space-time, the corresponding $\cH^5(SO,\RZ)=\Z_2$ is generated by
the gauge-gravity topological invariant $W^4_\text{top}=\frac12
\rw_2(\rw_1\rw_2+\rw_3)$.
The Wu classes $u_3=u_4=u_5=0$ can lead to relations between the
Stiefel-Whitney classes, which give us
\begin{align}
 \rw_4+\rw_2^2=0 .
\end{align}
$\rw_4+\rw_2^2=0$ is just a generator of the relations.
Other relations can be obtained by applying the Steenrod squares:
\begin{align}
 Sq^1(\rw_4+\rw_2^2)=\rw_1 \rw_4+\rw_5=0.
\end{align}
Additional relations can be obtained from \eqn{SqWu}
\begin{align}
Sq^1(\rw_4)=u_1\rw_4 &\to \rw_1\rw_4+\rw_5=\rw_1\rw_4
	\\
Sq^2(\rw_3)=u_2\rw_3 &\to \rw_2\rw_3+\rw_1\rw_4+\rw_5=\rw_1^2\rw_3+\rw_2\rw_3.
	\nonumber 
\end{align}
We see that $\rw_5$ must vanishes, but nothing restricts $\rw_2\rw_3$.  
So we have an realizable gauge-gravity topological invariant in
4+1D:
\begin{align}
\label{w2w3}
W^5_\text{top}(\Ga)= \frac12 \rw_2\rw_3 .
\end{align}
Thus $\si\iTOL^5=\Z_2$.

In 5+1D space-time, the corresponding $\cH^6(SO,\RZ)=2\Z_2$ is generated by the
gauge-gravity topological invariant $W^4_\text{top}=\frac12 \rw_6,
\frac12 \rw_2^3 $.  The Wu classes $u_4=u_5=u_6=0$ give us
\begin{align}
 \rw_4+\rw_2^2=\rw_2\rw_4+\rw_3^2=0 .
\end{align}
Other relations can be obtained by applying the Steenrod squares:
\begin{align}
 Sq^1(\rw_4+\rw_2^2)&=\rw_1 \rw_4+\rw_5=0.
\nonumber\\
 Sq^2(\rw_4+\rw_2^2)&=\rw_1^2\rw_2^2+\rw_3^2+\rw_2\rw_4 +\rw_6=0.
\nonumber\\
 Sq^1Sq^1(\rw_4+\rw_2^2)&=0.
\end{align}
We see that $\rw_6$ must vanishes, and $\rw_2\rw_4=\rw_2^3=\rw_3^2$.
Additional relations can be obtained from \eqn{SqWu}
\begin{align}
Sq^1(\rw_5)=u_1\rw_5 &\to \rw_1\rw_5=\rw_1\rw_5
	\\
Sq^2(\rw_4)=u_2\rw_4 &\to \rw_2\rw_4+\rw_6=\rw_1^2\rw_4+\rw_2\rw_4,
	\nonumber \\
Sq^3(\rw_3)=u_3\rw_3 &\to \rw_3\rw_3=\rw_1\rw_2\rw_3.
	\nonumber 
\end{align}
We see that $\rw_2\rw_4=\rw_2^3=\rw_3^2=0$.
So $\si\iTOL^6=0$.

In 6+1D space-time, the corresponding $\cH^8(SO,\RZ)=2\Z\oplus \Z_2$ is
generated by the gauge-gravity topological invariant
$W^4_\text{top}=\om_7^{p_1^2}, \om_7^{p_2}, \frac12 (\rw_1\rw_2+\rw_3)\rw_4$.
The Wu classes $u_4=u_5=u_6=u_7=0$ give us
\begin{align}
 \rw_4+\rw_2^2=
 \rw_2\rw_4+\rw_3^2=0 .
\end{align}
Other relations can be obtained by applying the Steenrod squares
(setting $\rw_1=0$):
\begin{align}
 Sq^1(\rw_4+\rw_2^2)&=\rw_1 \rw_4+\rw_5=0,
\nonumber\\
 Sq^2(\rw_4+\rw_2^2)&=\rw_1^2\rw_2^2+\rw_3^2+\rw_2\rw_4 +\rw_6=0,
\nonumber\\
 Sq^1(\rw_2\rw_4+\rw_3^2)&=\rw_3 \rw_4+\rw_2 \rw_5=0.
\end{align}
Additional relations can be obtained from \eqn{SqWu}
(setting $\rw_1=0$):
\begin{align}
Sq^1(\rw_6)=u_1\rw_6 &\to \rw_7=0
	\\
Sq^1(\rw_2^3)=u_1\rw_2^3 &\to \rw_2^2\rw_3=0,
	\nonumber \\
Sq^2(\rw_2\rw_3)=u_3\rw_2\rw_3 &\to \rw_2^2\rw_3 +\rw_2 \rw_5.
	\nonumber 
\end{align}
We see that $\rw_2\rw_5=\rw_3\rw_4=\rw_2^2\rw_3=\rw_7=0$.
So $\si\iTOL^7=2\Z$.

\subsection{Relation to cobordism groups}

Two oriented smooth $n$-dimensional manifolds $M$ and $N$ are said to be
equivalent if $M\cup (-N)$ is a boundary of another manifold, where $-N$ is
the $N$ manifold with a reversed orientation.  With the multiplication given
by the disjoint union, the corresponding equivalence classes has a structure
of an Abelian group $\Om^{SO}_n$, which is called the cobordism group of
closed oriented smooth manifolds.  For low dimensions, we have\cite{OCob}\\
$\Om^{SO}_{0}=\Z$, generated by a point.\\ $\Om^{SO}_{1}=0$, since circles
bound disks.\\ $\Om^{SO}_{2}=0$, since all oriented surfaces bound
handlebodies.\\ $\Om^{SO}_{3}=0$.\\ $\Om^{SO}_{4}=\Z$, generated by $\C P^2$,
detected by $\frac13\int_M p_1$.\\ $\Om^{SO}_{5}=\Z_2$, generated by the Wu
manifold $SU(3)/SO(3)$,\\ {\white{.}} ~~~~~~ detected by the deRham invariant
or\\ {\white{.}} ~~~~~~ Stiefel-Whitney number $\int_M \rw_2 \rw_3$.\\
$\Om^{SO}_{6}=0$.\\ $\Om^{SO}_{7}=0$.\\ $\Om^{SO}_{8}=2\Z$ generated by $\C
P^4$ and $\C P^2\times \C P^2$.  \\

The potential gravitational topological invariants give us a map from closed
space-time $M^d$ to $U(1)$: $Z_\text{fixed}(M^d)=\ee^{\ii \int_{M^d}
2\pi W^d_\text{top}(\Ga)} \in U(1)$.  For locally-null  topological invariants, such a
map reduces to a map from $\Om^{SO}_d$ to $U(1)$.  In fact, $\ee^{\ii
\int_{M^d} 2\pi W^d_\text{top}(\Ga)}$ is an 1D representation of group
$\Om^{SO}_d$.  So the locally-null potential gravitational topological invariants
are described by 1D representations of the cobordism  group $\Om^{SO}_d$.
Since the locally-null potential  gravitational topological invariants are
discrete, so they are actually described by 1D representation of
Tor$(\Om^{SO}_d)$.  Since, for an Abelian group $G_A$, the set of its 1D
representations also form an Abelian group, which is $G_A$ itself.
Therefore, the discrete  locally-null potential gravitational topological
invariants in $d$-dimensional space-time are described by Tor$(\Om^{SO}_d)$.
Since all the locally-null potential gravitational topological invariants are
realizable, we find
\begin{align}
 \text{Tor}(\si\iTOL^d)=\text{Tor}(\Om^{SO}_d).
\end{align}
The Chern-Simons potential gravitational topological invariants in $d$-dimensional
space-time are described by Free$(\Om^{SO}_{d+1})$, since
Free$(\Om^{SO}_{d+1})$ is a subgroup of Free$H^{d+1}(BSO,\Z(\frac1n))$.  So the Chern-Simons
realizable gravitational topological invariants, described by
Free$\cH^{d}(SO,\RZ)=$Free$H^{d+1}(BSO,\Z)$, form a subgroup of $
\text{Free}(\Om^{SO}_d)$:
\begin{align}
 \text{Free}(\si\iTOL^d)\subset \text{Free}(\Om^{SO}_d).
\end{align}

\section{Pure and mixed SPT states}
\label{pmSPT}

\subsection{A generic result}

In this section, we are going to consider L-type SPT states protected by $G$
symmetry (which may contain time reversal symmetry) in $d$-dimensional
space-time. Those SPT states form an Abelian group $\LSPT_G^d$.  We only
consider  SPT states that are realized by $G\times SO$ NL$\si$Ms.  The
different $G\times SO$ NL$\si$Ms are characterized by their topological terms
which are classified by $\cH^d(G\times SO,\RZ)$.  Those topological terms
induced the realizable gauge-gravity topological invariants $W^d_\text{top}(A,\Ga)$
that are also ``classified'' by $\cH^d(G\times SO,\RZ)$.  Therefore, L-type SPT
states from NL$\si$Ms are ``classified'' by $\cH^d(G\times SO,\RZ)$, but in a
many-to-one fashion; \ie different elements in $\cH^d(G\times SO,\RZ)$ may
correspond to the same gauge-gravity topological invariant $W^d_\text{top}(A,\Ga)$
and the same SPT phase.

To understand this many-to-one correspondence, we note that the gauge-gravity
topological invariants $W^d_\text{top}(A,\Ga)$ should be fully detectable in the
following sense. The gauge-gravity topological invariants
$W^d_\text{top}(A,\Ga)$ can be regarded as map from a pair
$(M^d,A)$ to $\RZ$:
\begin{align}
	\int_{M^d} W^d_\text{top}(A,\Ga) = 
\frac{\th}{2\pi} \text{ mod } 1 .
\end{align}
where $M^d$ is a close space-time manifold with various topologies and $A$ is
a $G$ symmetry twist on $M^d$.  Two topological invariants are said to be
different if they produce different maps $(M^d,A)\to \RZ$ that cannot be
smoothly connected to each other.  However, there indeed exist gauge-gravity
topological invariants $ZW^d_\text{top}(A,\Ga)$ whose induced map $(M^d,A)\to \RZ$
can smoothly connected to 0 (see Appendix \ref{Pggterm}).  Then any two
topological invariants differ by $ZW^d_\text{top}(A,\Ga)$ should correspond to the
same SPT phase and should be identified.  We call $ZW^d_\text{top}(A,\Ga)=0$
a relation between topological invariants.  $ZW^d_\text{top}(A,\Ga)$ generate a
subgroup of $\cH^d(G\times SO,\RZ)$ which will be called $\La^d(G)$.  We see
that the distinct SPT phases, plus the iTO phases that are also produced by
the NL$\si$Ms, are classified by the quotient
\begin{align}
 \LSPT_G^d \oplus \si\iTOL^d = \cH^d(G\times SO,\RZ)/\La^d(G).
\end{align}
In the next subsection, we will discuss how to compute the subgroup $\La^d(G)$.

Using the K\"unneth formula \eq{kunnH}, we find that
\begin{align}
& \cH^d(G\times SO,\RZ)  \simeq 
\cH^d(G,\RZ)\oplus \cH^d( SO,\RZ)\oplus
\nonumber \\
&\ 
\oplus_{k=1}^{d-1} H^k[BG, \cH^{d-k}(SO,\RZ)] 
\end{align}
Clearly, the term $ \cH^d( SO,\RZ)$ describes iTO phases that do not require
any symmetry $G$.  So
\begin{align}
\oplus_{k=1}^{d-1} H^k[BG, \cH^{d-k}(SO,\RZ)] 
\oplus \cH^d(G,\RZ) 
\end{align}
should cover all the SPT states, \ie every cocycle
in $\oplus_{k=1}^{d-1} H^k[BG, \cH^{d-k}(SO,\RZ)] \oplus \cH^d(G,\RZ)$ is
realizable and describes a SPT state.  
In other words
\begin{align}
\LSPT_G^d =
\frac{\oplus_{k=1}^{d-1} H^k[BG, \cH^{d-k}(SO,\RZ)] 
\oplus \cH^d(G,\RZ) }{\La^d(G)}
\end{align}
For example the term $\cH^d(G,\RZ)$ describes pure SPT states.  Each element
in  $\cH^d(G,\RZ)$ correspond to distinct realizable SPT states (quotient is
not needed).  

Similarly, the term $H^k[BG, \cH^{d-k}(SO,\RZ)]$ describes mixed SPT states.
Every cocycle in $H^k[BG, \cH^{d-k}(SO,\RZ)]$ describes a mixed SPT state.  But
different  cocycles may correspond to the same SPT state.  This can be seen
from the dimension reduction discussed in Section \ref{GGpSPT}.  We put a
$G\times SO$ SPT state on $M^k\times M^{d-k}$ which is described by a cocycle
$\nu_d$ in $ \oplus_{k=1}^{d-1} H^k[BG,
\cH^{d-k}(SO,\RZ)]\oplus\cH^d(G,\RZ)\oplus \cH^d( SO,\RZ)$.  The cocycle $\nu_d$
can be viewed as a gauge-gravity topological invariant $W^d_\text{top}$ and vise
versa. 
Here we will consider a
mixed SPT states described by $W^{d,k}_\text{top}\in H^k[BG,
\cH^{d-k}(SO,\RZ)]$ in more detail.

Let us put a $G$-symmetry twist $A_G$ on $M^k$, but for the time being not any
$SO$-symmetry twist on $M^k$.  The decomposition $\oplus_{k=1}^{d-1} H^k[BG,
\cH^{d-k}(SO,\RZ)]$ implies that, in the large $M^{d-k}$ limit, we get an
$(d-k)$-dimensional topological state on $M^{d-k}$, described by a cocycle
$\nu_{d-k}$ in $\cH^{d-k}(SO,\RZ)$.  Formally, we can express the above
dimension reduction as
\begin{align}
\label{MkAG}
\int_{M^k,A_G}  W^{d,k}_\text{top} = \nu^{SO}_{d-k} \in \cH^{d-k}(SO,\RZ),
\end{align}
where $A_G$ represent the $G$-symmetry twist on $M^d$.  In particular, if we
choose $M^d$, $A_G$ and $W^{d,k}_\text{top}\in H^k[BG, \cH^{d-k}(SO,\RZ)]$
arbitrarily, we can produce any elements in $\cH^{d-k}(SO,\RZ)$.

However, due to the restrictive relation between the $SO$ connection and the
topology of $M^{d-k}$, different cocycles in $\cH^{d-k}(SO,\RZ)$ may correspond
to the same topological state.  So the distinct topological states are
described by a quotient $\cH^{d-k}(SO,\RZ)/\La^{d-k}$.  As we have discussed
before, the distinct topological states from $\cH^{d-k}(SO,\RZ)$ are nothing
but the $(d-k)$-dimensional iTO states that form $\si\iTOL^{d-k}$.  Therefore,
the distinct iTO states on  $M^{d-k}$ imply that the parent SPT states on
$M^d$ before the dimension reduction are distinct.  However, it is still
possible that different parent SPT states on $M^d$ lead to the same iTO state
on $M^{d-k}$.  So the SPT states are described by  $ H^k[BG, \si\iTOL^{d-k}]$
plus something extra.  This way, we conclude that the L-type realizable SPT
states are described by
\begin{align}
\label{LSPTHiTO}
&\ \ \ \
\si\LSPT_G^d 
\\
&=
\Big[E^d(G)\rtimes \oplus_{k=1}^{d-1} H^k(BG, \si\iTOL^{d-k})\Big] 
\oplus \cH^d(G,\RZ) ,
\nonumber 
\end{align}
which is one of the main results of this paper.  We like to point out that if
$G$ contains time-reversal transformation, it will have a non-trivial action
$\RZ \to -\RZ$ and $\si\iTOL^{d-k} \to -\si\iTOL^{d-k}$.  In the next subsection,
we will compute this extra group $E^d(G)$.

However, there is a mistake in the above derivation of \eqn{LSPTHiTO}.  Due to
the restrictive relation between the $SO$ connection and the topology of $M^k$,
we cannot set the $SO$-symmetry twist on $M^k$ to zero.  So the dimension
reduction is actually given by
\begin{align}
\label{MkAGa}
\int_{M^k,A_G,\Ga}  W^{d,k}_\text{top} &= \nu^\text{iTO}_{d-k} \in \si\iTOL^{d-k}, 
\end{align}
where $\Ga$ represent the $SO$-symmetry twist on $M^d$.  Due to the restrictive
relation between $(M^d,A_G)$ and $\Ga$, it is not clear that if we choose
$M^d$, $A_G$ and $W^{d,k}_\text{top}\in H^k[BG, \cH^{d-k}(SO,\RZ)]$
arbitrarily, we can still produce any elements in $\si\iTOL^{d-k}$.

In the following, we will show that we can indeed produce  any elements in
$\si\iTOL^{d-k}$.\\
(1) We note that the $SO$ tangent bundle of $M^k\times M^{d-k}$ splits
into an $SO''$ tangent bundle on $M^{d-k}$ and a $SO'$  tangent bundle  on
$M^k$.  So we can rewrite \eqn{MkAGa} as
\begin{align}
	\int_{M^k,A_G,\Ga'}  W^{d,k}_\text{top} &= \nu^{SO}_{d-k} \in \cH^{d-k}(SO,\RZ).
\end{align}
where $A_G,\Ga'$ is the $G\times SO'$ symmetry twist on $M^k$ and we put the
$SO''$ symmetry twist on $M^{d-k}$.  
This motivates us consider
a $G\times SO'\times SO''$ NL$\si$M and its topological terms.\\
(2) The natural group homomorphism $G\times SO'\times SO'' \to G\times SO$
via embedding $SO'\times SO''$ into  $SO$ leads to a ring homomorphism
$H^*[B(G\times SO),\Z]\to H^*[B(G\times SO'\times SO''),\Z]$.\\
(3) Due to the isomorphism $\cH^n(G,\RZ) \simeq H^{n+1}(BG,\Z)$, 
$W^{d,k}_\text{top}$ in $H^k[BG, \cH^{d-k}(SO,\RZ)]$
can be viewed as an element in $H^k[BG, H^{d-k+1}(BSO,\Z)]$. 
As a result, we can express
$W^{d,k}_\text{top}$ as a characteristic class in $H^k[BG, H^{d-k+1}(BSO,\Z)]$. For example, $W^{d,k}_\text{top}=
F^{G}_k F^{SO}_l F^{SO}_{d-k-l+1}$, where
$F^G_k$ is a characteristic class in $H^k(BG,\Z)$, and
$F^{SO}_n$ is a characteristic class in $H^n(BSO,\Z)$.
\\
(4) Using the above ring homomorphism,
we can map $W^{d,k}_\text{top}$ into an element
in $H^k[BG, H^{d-k+1}(B(SO'\times SO''),\Z)]$:
\begin{align}
&\ \ \ \
	W^{d,k}_\text{top}=F^{G}_k F^{SO}_l F^{SO}_{d-k-l+1} 
\nonumber\\
& \to
F^{G}_k (F^{SO'}_l+F^{SO''}_l) (F^{SO'}_{d-k-l+1}+F^{SO''}_{d-k-l+1})
\nonumber\\
&\ \ \ \ \ \ \in H^k[BG, H^{d-k+1}(B(SO'\times SO''),\Z)].
\end{align}
(5) 
Since the $SO'$-twist is only on $M^k$ and the $SO''$-twist is only on
$M^{d-k}$,
the above expression allows us to conclude that only the term
$F^{G}_k F^{SO''}_l F^{SO''}_{d-k-l+1}$ contribute to
$ \int_{M^k,A_G,\Ga'}  W^{d,k}_\text{top}$. Thus 
\begin{align}
\label{Gap0}
\int_{M^k,A_G,\Ga'}  W^{d,k}_\text{top} &= 
\int_{M^k,A_G,0}  W^{d,k}_\text{top} .
\end{align}
which reduces \eqn{MkAGa} to \eqn{MkAG} that leads to \eqn{LSPTHiTO}.
This completes our proof.

\subsection{A calculation of $\La^d(G)$ and $E^d(G)$}

The subgroup $\La^d(G)$ is generated by a set of relations in $
\cH^d(SO,\RZ)\oplus \cH^d(G,\RZ) \oplus_{k=1}^{d-1} H^k[BG,
\cH^{d-k}(SO,\RZ)]=H^{d+1}(G\times SO,\Z) $.  To compute such a set of the
relations,
%
%
we can choose a homomorphism $G\to O$, which will lead to a homomorphism
$H^*(BO,\Z_2)\to H^*(BG,\Z_2)$ as rings.  We know that $H^*(BO,\Z_2)$ is
generated by the Stiefel-Whitney classes $\rw_1,\rw_2,\cdots$.  $\rw_i$ will
map into $\rw_i^G \in H^i(BG,\Z_2)$. 
Then we can treat $\rw_i^G$ as the Stiefel-Whitney classes and use the Wu
formula \eqn{WuF} to compute $Sq^i(\rw_j^G)$.  
The Wu formula and the following defining properties of the Wu classes:
\begin{align}
 Sq^{d-i}(w^G_i)&=u_{d-i} w^G_i, 
\nonumber\\
 Sq^{d-i-j}(w_iw^G_j)&=u_{d-i-j} w_iw^G_j, \cdots
\end{align}
will generate the relations 
(denoted as $ZW^d_\text{top}(A,\Ga)$)
\begin{align}
 & Sq^{d-i}(w^G_i)+u_{d-i} w^G_i, 
\nonumber\\
 & Sq^{d-i-j}(w^G_iw^G_j)+u_{d-i-j} w^G_iw^G_j, \cdots
\end{align}
in $\oplus_{k=1}^{d-1} H^k[BG, H^{d-k}(O,\RZ)]\oplus  \cH^d(O,\RZ) \oplus
\cH^d(G,\RZ)$.  Those relations become the relations in $\oplus_{k=1}^{d-1}
H^k[BG, H^{d-k}(SO,\RZ)]\oplus  \cH^d(SO,\RZ) \oplus
\cH^d(G,\RZ)=H^{d+1}[B(G\times SO),\Z]$ through the natural map $\bt:$
$H^d[B(G\times SO),\Z_2]\to H^{d+1}[B(G\times SO),\Z]$, after  we set
$\rw_1=0$.


$\La^d(G)$ also contain another type of relations: if $a \in \cH^d(G\times
SO,\RZ)$ can be expressed as a mod 2 reduction of $\bar a \in
\text{Free}\cH^d(G\times SO,\Z)$, then $a$ is in $\La^d(G)$.  The reason for
such type of relations is discussed below \eqn{W4c1}.

The relations will generate $\La^d(G)$ which also allow us to compute
$E^d(G)$.  Certainly, the subgroup of $\oplus_{k=1}^{d-1} H^k[BG,
\cH^{d-k}(SO,\RZ)]$, $\oplus_{k=1}^{d-1} H^k(BG, \si\iTOL^{d-k})$, will survive
the quotient by $\La^d(G)$. $E^d(G)$ is the subgroup not in
$\oplus_{k=1}^{d-1} H^k(BG, \si\iTOL^{d-k})$ that also survive the quotient.
Thus $E^d(G)\rtimes \oplus_{k=1}^{d-1} H^k(BG, \si\iTOL^{d-k}) $ describes the
distinct SPT phases.  Next, we will demonstrate the above approach by
computing the pure and mixed SPT states for some simple symmetry groups.

\subsection{$U(1)$ SPT states}

From\cite{CGL1314}
\begin{align}
	H^d(BU(1),\Z) &=0 \text{ if } d=\text{odd};	 
	\nonumber\\
	H^d(BU(1),\Z) &=\Z \text{ if } d=\text{even};	 
\end{align}
we can obtain
\begin{align}
	H^d(BU(1),\Z_2) &=0 \text{ if } d=\text{odd};	 
	\nonumber\\
	H^d(BU(1),\Z_2) &=\Z_2 \text{ if } d=\text{even};	 
\end{align}
using universal coefficient theorem.\cite{W1313,W1447} The ring $H^*[BU(1),
\Z]$ is generated by the first Chern class $c_1$.

This allows us to calculate the $U(1)$ mixed SPT described by
$\oplus_{k=1}^{d-1} H^k[BU(1), \cH^{d-k}(O,\RZ)]$.
We obtain $U(1)$ mixed SPT states in 4+1D described by the group-cohomology
$H^2(BU(1), \si\iTOL^3)=\Z$.
We also obtain mixed SPT states 
in 6+1D described by 
$H^4(BU(1), \si\iTOL^3)\oplus H^2(BU(1), \si\iTOL^5)=\Z\oplus Z_2$.

The well known 2+1D $U(1)$ pure SPT states have the following Chern-Simons
topological invariants
\begin{align}
	W^3_\text{top}(A,\Ga) = \frac{k}{(2\pi)^2} A\dd A,\ \ \ k\in \Z .
\end{align}
where $A$ is the $U(1)$ gauge connection one form. 
Their Hall conductances are given by $\si_{xy}=\frac{2k}{2\pi}$.

The 4+1D $U(1)$ mixed SPT states described by $\cH^2(U(1), \si\iTOL^3)$ has been
discussed in \Ref{WGW1489}. Its gauge-gravity topological invariant is given by
(see a discussion in Appendix \ref{giTO3})
\begin{align}
	W^5_\text{top}(A,\Ga) = \om_3 \frac{\dd A}{2\pi} = \frac{A}{2\pi} p_1,
\end{align}
In 4 spatial dimensions, the $U(1)$ monopole is a 1D loop.  In this SPT state,
such a 1D loop will carry the gapless edge state of 2+1D $(E_8)^3$ bosonic
quantum Hall state.

The 6+1D $U(1)$ mixed SPT states described by $\cH^4(U(1), \si\iTOL^3)=\Z$
have the following topological invariants
\begin{align}
	W^7_\text{top}(A,\Ga) = \frac{k}{(2\pi)^2} \om_3 \dd A\dd A,\ \ \ k\in \Z .
\end{align}
The 6+1D $U(1)$ mixed SPT state described by $\cH^2(U(1), \si\iTOL^5)=\Z_2$
has
\begin{align}
	W^7_\text{top}(A,\Ga) = \frac{1}{2} \rw_2\rw_3 \frac{\dd A}{2\pi}.
\end{align}

To see if there are extra mixed $U(1)$ SPT phases, let us first note that the
ring $H^*[BU(1), \Z_2]$ is generated by $f_2$, which is the mod 2 reduction
(denoted as $\rho_2$) of the first Chern class $c_1$: $f_2=\rho_2c_1$.  If we
choose the natural embedding $U(1) \to O$, we find that
\begin{align}
\rw_2^{U(1)}=f_2 ,\ \ \ \ \rw_i^{U(1)}=0, \ \ \ i=1,\text{ or } i>2.
\end{align}

In 3+1D, the potential extra mixed SPT phases are described by $
H^2[BU(1),\cH^2(SO,\RZ)]=\Z_2 $.  We note that $\cH^2(SO,\RZ)$ is generated by
the $\rw_2$ (see \eqn{HBSOU1}).  Therefore, the extra $U(1)$ SPT phases
described by $H^2[BU(1), \cH^2(SO,\RZ)]=\Z_2$ are generated by $f_2\rw_2$,
In 3+1D, we have a relation
\begin{align}
 Sq^2(\rw_2^{U(1)}) &= u_2 \rw_2^{U(1)},
\nonumber\\
 Sq^2(\rw_2^{U(1)}) &= u^{U(1)}_2 \rw_2^{U(1)},
\end{align}
which gives us
\begin{align}
 \rw_2^{U(1)}\rw_2^{U(1)}&=(\rw_1^2+\rw_2)  \rw_2^{U(1)} ,
\end{align}
In 3+1D, we also have a relation $f_2^2 =\rw_2^{U(1)}\rw_2^{U(1)}=0$ mod 2.
For oriented space-time $\rw_1=0$, so  $f_2\rw_2$ vanishes.
There is no extra 3+1D $U(1)$ SPT phase.

Here is another way to rephrase the above reasoning.  In 3+1D, there is a
potential topological invariant
\begin{align}
\label{W4c1}
W^4_\text{top}(A,\Ga) = \frac{1}{2} \rw_2 \frac{\dd A}{2\pi} = \frac12 \rw_2 \rho_2 c_1,
\end{align}
where the Chern class $c_1=\dd A/2\pi$ and $\rho_2$ is the mod 2 reduction.
Using the relation $Sq^2(\rho_2 c_1)=u_2 \rho_2 c_1$ and $Sq^2(\rho_2 c_1)=(\rho_2 c_1)^2$, we find that
$(\rw_1^2+\rw_2)\rho_2 c_1 = (\rho_2 c_1)^2$. Therefore
on oriented manifold, we have\cite{K1459v2}
\begin{align}
	W^4_\text{top}(A,\Ga) = \frac{1}{2} \rw_2 \frac{\dd A}{2\pi} = 
\frac12 (f_2)^2 =\frac12 c_1^2 = \frac{1}{2(2\pi)^2} \dd A \dd A.
\end{align}
Such a topological invariant is not quantized. It can continuously deform into zero
via $\frac{\th}{(2\pi)^2} \dd A \dd A$ as $\th$ goes from $\pi$ to 0.
This is why there is no $U(1)$ SPT phase in 3+1D.

We note that on space-time $M^4$ with spin structure, $\rw_2=0$.
The above result implies that all the $U(1)$ bundles on such $M^4$ satisfy
\begin{align}
 \int_{M^4} c_1^2=\text{even}.
\end{align}
Or in other words, the $\Z_2$ reduction of $c_1^2$ cannot be probed by any
$M^4$ with spin structure, no mater what $U(1)$-symmetry twists we add.

In 4+1D, the potential extra SPT phases are described by $
H^2[BU(1),\cH^3(SO,\RZ)]\oplus H^4[BU(1),\cH^1(SO,\RZ)] =\Z $. But
$H^2[BU(1),\cH^3(SO,\RZ)]$ is $H^2[BU(1),\si\iTOL^3]$ which has been included
before.  Thus the potential extra mixed SPT phases are described by
$H^4[BU(1),\cH^1(SO,\RZ)] =0$, \ie there is no extra $U(1)$ SPT phase in 4+1D.

It has been pointed out that the following gauge-gravity topological invariant
may exist 
\begin{align}
\label{w3F}
W^5_\text{top}(A,\Ga) = \frac{1}{2} \rw_3 \frac{\dd A}{2\pi} .
\end{align}
It may suggest that there is an extra $U(1)$ SPT phase in 4+1D.  Here we would
like to show that such a term always vanishes. We start with the relation
\eq{SqWu}:
\begin{align}
 Sq^1(\rw_2 \rw_2^{U(1)}) =u_1 \rw_2 \rw_2^{U(1)}.
\end{align}
The left-hand-side gives us
\begin{align}
&\ \ \ \
 Sq^1(\rw_2)\rw_2^{U(1)} +\rw_2 Sq^1(\rw_2^{U(1)})
\nonumber\\
&=(\rw_1\rw_2+\rw_3) \rw_2^{U(1)} +\rw_2 Sq^1(\rw_2^{U(1)}).
\end{align}
Since $\rw_2^{U(1)}$ is a Stiefel-Whitney class of a $O$ vector bundle over
$M^5$ (which is not the tangent bundle that gives rise to Stiefel-Whitney
classes $\rw_i$), we can use the Wu formula \eq{WuF} to calculate
$Sq^1(\rw_2^{U(1)})=\rw_1^{U(1)}\rw_2^{U(1)}+\rw_3^{U(1)}=0$. Thus we have
\begin{align}
&\ \ \ \  Sq^1(\rw_2 \rw_2^{U(1)})=(\rw_1\rw_2+\rw_3) \rw_2^{U(1)} 
\nonumber\\
&
=
u_1\rw_2 \rw_2^{U(1)} = \rw_1\rw_2 \rw_2^{U(1)},
\end{align}
which gives us $\rw_3 f_2= \rw_3 c_1 =0$ mod 2 for any $U(1)$ bundle on
$M^5$ which can even be unorientable. This leads to the vanishing of \eqn{w3F}.

In 5+1D, we may have extra mixed $U(1)$ SPT phases described by $\cH^4[U(1),
\cH^2(SO,\RZ)]\oplus \cH^2[U(1), \cH^4(SO,\RZ)]=2\Z_2$, generated by 
$\rw_2f_2^2,\  \rw_4f_2$.
We have the following relations
\begin{align}
&\ \ \ \ Sq^1(\rw_3 \rw_2^{U(1)})=u_1 \rw_3\rw_2^{U(1)}
\nonumber\\
& \to 0=0
\nonumber\\
&\ \ \ \ Sq^2(\rw_2^{U(1)}\rw_2^{U(1)})=u_2 \rw_2^{U(1)}\rw_2^{U(1)}
\nonumber\\
& \to (\rw_1^2+\rw_2) \rw_2^{U(1)}\rw_2^{U(1)}=0
\nonumber\\
&\ \ \ \ Sq^2(\rw_2 \rw_2^{U(1)})=u_2 \rw_2\rw_2^{U(1)}
\nonumber\\
& \to \rw_2^3+\rw_1^2 \rw_2\rw_2^{U(1)}=0
\end{align}
and $\rw_2^2=\rw_4$.  Since $\rw_1=\rw_2^3=0$ for 6-dimensional orientable
manifold (see Section \ref{lowDiTO}), we only have one relation $\rw_2
f_2^2=0$.  However, $\rw_4f_2=\rw_2^2f_2=p_1f_2$ mod 2 (see Appendix
\ref{PandSW}).  So, the $\Z_2$ class $\rw_4f_2$ is part of an integer class
$p_1c_1$, and the topological invariant from an integer class does not have a
quantized coefficient (see the discussion below \eqn{W4c1}).  So 
the term $\rw_4f_2$ can smoothly connect to zero, and
there is no extra mixed $U(1)$ SPT phases in 5+1D.

In 6+1D, we may have extra $U(1)$ SPT phases described by $H^4[BU(1),
\cH^3(SO,\RZ)]\oplus H^2[BU(1), \cH^5(SO,\RZ)]=\Z\oplus \Z_2$, but they are
discussed before since $H^4[BU(1), \cH^3(SO,\RZ)]\oplus H^2[BU(1),
\cH^5(SO,\RZ)]=H^4[BU(1), \si\iTOL^3]\oplus H^2[BU(1), \si\iTOL^5]$.
So, there is no extra  $U(1)$ SPT phase in 6+1D.


\subsection{$Z_n$ SPT states}

From\cite{CGL1314,W1313,W1447}
\begin{align}
	H^d(BZ_n,\Z) &=0 \text{ if } d=\text{odd};	 
	\nonumber\\
	H^d(BZ_n,\Z) &=\Z_n \text{ if } d=\text{even};	 
\end{align}
we obtain
\begin{align}
	H^d(BZ_n,\Z_2) &=\Z_2 \text{ if } d=0,
\nonumber\\
	H^d(BZ_n,\Z_2) &=\Z_{\<n,2\>}, \text{ if } d>0 ,
\end{align}
where $\<m,n\>$ is the greatest common divisor of $m,,n$.  This allows us to
obtain $Z_n$ mixed SPT states
described by
$\oplus_{k=1}^{d-1} H^k[BZ_n, \cH^{d-k}(O,\RZ)]$.  
There are no such $Z_n$
mixed SPT states in 3+1D.
The  4+1D $Z_n$
mixed SPT states are described by the group-cohomology $H^2(BZ_n,
\si\iTOL^3)=\Z_n$. 
We also obtain mixed SPT states \\
in 5+1D described by $H^1(BZ_n, \si\iTOL^5)=\Z_{\<n,2\>}$,\\
in 6+1D described by $H^4(BZ_n, \si\iTOL^3)\oplus H^2(BZ_n, \si\iTOL^5)=\Z_n\oplus \Z_{\<n,2\>}$,\\
in 7+1D described by $H^3(BZ_n, \si\iTOL^5)=\Z_{\<n,2\>}$.

$H^2(BZ_n, \si\iTOL^3)=\Z_n$ is generated by
$W^5_\text{top}=\bt(A_{Z_n}/2\pi)\om_3$ where $A_{Z_n}/2\pi$ is the
generator of $\cH^1(Z_n,\RZ)=\Z_n$ (or $\oint A_{Z_n}/2\pi=\frac1n $ mod 1.
According the Appendix \ref{giTO3}, $H^2(BZ_n, \si\iTOL^3)=\Z_n$ is generated by
\begin{align}
\label{ZniTO3}
W^5_\text{top}= \frac{A_{Z_n}}{2\pi} p_1. 
\end{align}

The structure of above results also lead to a physical
probe of the corresponding SPT
states
%
%
by dimension reduction.\cite{W1447,WGW1489}  We put the system on a 4D space of a form $S^2\times
D^2$ and put $n$ identical monodromy defects on $S^2$. In the small $S^2$
limit, the effective 2+1D state on $D^2$ will be an $(E_8)^3$ bosonic quantum
Hall state, with gapless excitations on the boundary of $D^2$.  We may also
replace $S^2$ by $\t D^2$ and break the $Z_n$ symmetry on the boundary of $\t
D^2$.  We then create $n$ identical $Z_n$ domain walls on the  boundary of $\t
D^2$.  This will have the same effect as $n$ identical monodromy defects on
$S^2$. We get an $(E_8)^3$ bosonic quantum Hall state on $D^2$ in the small $\t
D^2$ limit.  In fact, all the mixed SPT states described by
$H^2(BG,\si\iTO_L^{d-2})$ and all the $G_1\times G_2$ pure SPT states described by
$H^2[BG_1,\cH^{d-2}(G_2,\RZ)]$ can be probed in this way.

In the following, we will consider if there are extra mixed $Z_n$ SPT phases.
We find that there is no extra mixed $Z_n$ SPT phase if $n=$ odd.  So in
the following, we will assume $n=$ even.  
We first note that the
ring $H^*[BZ_n, \Z_2]$ is generated by $a_1$.
If we choose the natural
embedding $Z_n \to O$ via $Z_n/Z_{n/2}\to O/SO$, we find that
\begin{align}
\rw_1^{Z_n}=a_1 ,\ \ \ \ \rw_i^{Z_n}=0, \ \ \ i>1.
\end{align}

In 2+1D, the potential extra mixed $Z_n$ SPT phases are described by
$H^1[BZ_n, \cH^2(SO,\RZ)]=\Z_{\<n,2\>}$.  We note that $\cH^2(SO,\RZ)$ is
generated by the $\rw_2$ (see \eqn{HBSOU1}).  Therefore, the potential extra
$Z_n$ SPT phases described by $H^1[BZ_n, \cH^2(SO,\RZ)]$ are generated by
$a_1\rw_2$.  In 2+1D, we have the following relations
\begin{align}
 Sq^1[ (\rw_1^{Z_n})^2]=u_1 (\rw_1^{Z_n})^2
& \to \rw_1(\rw_1^{Z_n})^2=0 ;
\nonumber\\
&\ \ \  u_2 =\rw_1^2+\rw_2=0 .
\end{align}
We see that $\rw_2=0$ for orientable 2+1D space-time and $a_1\rw_2$ vanishes.
Thus there is no extra $Z_n$ SPT phase in 2+1D.

In 3+1D, the potential extra mixed $Z_n$ SPT phases are described by
$H^2[BZ_n, \cH^2(SO,\RZ)]=\Z_{\<n,2\>}$, which
are generated by $a_1^2\rw_2$.  In 3+1D, we have the following relations (setting
$\rw_1=0$)
\begin{align}
& \rw_2^2+\rw_4=0 ,
\nonumber\\
& a_1^2 \rw_2 + a_1\rw_3 = a_1^4 = a_1^4 +a_1^2\rw_2 =0.
\end{align}
We see that $a_1^2\rw_2=0$ and there is no extra $Z_n$ SPT phase in 3+1D.

In the above, we also see that $a_1^4 =0$. What is the physical meaning of this
relation? In fact, in 1+1D, we have $a_1^2=0$. Let us discuss this simpler 1+1D
situation. The relation $a_1^2=0$ comes from
\begin{align}
 Sq^1(a_1)=u_1a_1 \to a_1^2=\rw_1a_1 .
\end{align}
We see that on non-orientable $M^2$, $a_1^2$ do not have to be zero.  This
means that $\int_{M^2} a_1^2 $ can be non-zero if  $M^2$ is non-orientable. But
$\int_{M^2} a_1^2 $ must be zero mod 2 if $M^2$ is orientable.  For $Z_n$ SPT
state without time-reversal symmetry, we cannot use the non-orientable $M^2$ to
probe it. So $a_1^2$ cannot produce any measurable topological invariant, and
should be quotient out. This is why $H^2(BZ_n,\RZ)$ is trivial, since its
potential generator $a_1^2$ is not measurable on any orientable space-time for
any symmetry twist.

In 4+1D, the potential extra mixed $Z_n$ SPT phases are described by
$H^3[BZ_n, \cH^2(SO,\RZ)]\oplus H^1[BZ_n, \cH^4(SO,\RZ)]=2\Z_{\<n,2\>}$,
which are  generated by $a_1^3\rw_2,\  a_1\rw_4$.  In 4+1D, we have the follow
relations (setting $\rw_1=0$)
\begin{align}
& \rw_2^2+\rw_4=\rw_5=0 ,
\nonumber\\
& a_1^2 \rw_3 = a_1^5 +a_1^3\rw_2 =0.
\end{align}
We see that $a_1^3\rw_2=a_1^5$
which is already included by $\cH^5(Z_n,\RZ)$.
But nothing restricts $a_1\rw_4$ (except
$\rw_4=\rw_2^2$).  So there is an $Z_n$ SPT phase in 4+1D generated by
$a_1\rw_2^2$ for $n=$ even. Its topological invariant is given by
\begin{align}
	W^5_\text{top}(A,\Ga) = \frac{n}{2} \frac{A_{Z_n}}{2\pi} p_1,
\end{align}
where $A_{Z_n}$ is the flat connection that describe the $Z_n$ twist\cite{WGW1489}
\begin{align}
 \oint A_{Z_n} =0 \text{ mod } 2\pi/n.
\end{align}
But the above topological invariant has been included by
$Z_n$ SPT phases described by $H^2(BZ_n,\si\iTOL^3)$ (see \eqn{ZniTO3}).
So there is no extra $Z_n$ SPT phase in 4+1D.


In 5+1D, the potential extra mixed $Z_n$ SPT phases are described by
$H^4[BZ_n, \cH^2(SO,\RZ)]\oplus H^2[BZ_n, \cH^4(SO,\RZ)]=2\Z_{\<n,2\>}$,
which are generated by $a_1^4\rw_2,\  a_1^2\rw_4$.  In 5+1D, we have the follow
relations (setting $\rw_1=0$)
\begin{align}
& \rw_2^2+\rw_4=\rw_5=\rw_3^2+\rw_2\rw_4=\rw_3^2=\rw_6=0 ,
\nonumber\\
& a_1^2 \rw_4 = a_1^4\rw_2 =a_1^3\rw_3=a_1^6=0 .
\end{align}
We see that $a_1^2\rw_4=a_1^4\rw_2=0$.
So there is no extra $Z_n$ SPT phase in 5+1D.

In 6+1D, the potential extra mixed $Z_n$ SPT phases are described by
$H^5[BZ_n, \cH^2(SO,\RZ)]\oplus H^3[BZ_n, \cH^4(SO,\RZ)]\oplus H^1[BZ_n,
\cH^6(SO,\RZ)]=4\Z_{\<n,2\>}$, which are generated by $a_1^5\rw_2,\  a_1^3\rw_4,
a_1 \rw_6, a_1\rw_2^3$.  In 6+1D, we have the following relations (setting
$\rw_1=0$)
\begin{align}
& \rw_2^2+\rw_4=\rw_5=\rw_3^2+\rw_2\rw_4=\rw_6=0 ,
\nonumber\\
& a_1^2 \rw_2\rw_3+a_1\rw_3^2 = a_1^5\rw_2 =a_1^4\rw_3=0 .
\end{align}
We see that
$a_1^5\rw_2=a_1\rw_6=0$,  but nothing restricts 
$a_1^2\rw_2\rw_3=a_1\rw_3^2$ and $a_1^3\rw_4$. 
However, $a_1^2\rw_2\rw_3$ is already included by
$H^2(BZ_n,\si\iTOL^5)$. So
there is an $Z_n$ SPT phase in 6+1D generated by $a_1^3\rw_2^2$ for $n=$
even.  Its topological invariant is given by
\begin{align}
	W^7_\text{top}(A,\Ga) = \frac{1}{2\pi^3} A_{Z_n}^3 p_1 .
\end{align}
But the above topological invariant has been included by $Z_n$ SPT phases described
by $H^4(BZ_n,\si\iTOL^3)\simeq \cH^3(Z_n,\RZ)$ (see Appendix \ref{giTO3}).  So
there is no extra $Z_n$ SPT phase in 6+1D.

\subsection{$U(1)\rtimes Z_2=O(2)$ SPT states}

In \Ref{CGL1314}, it was shown that
\begin{align}
\label{Z2rU1Z1}
 H^d[ BO_2,\Z] \subset
\begin{cases}
\Z\oplus {\frac d 4}\Z_2,  &  d=0 \text{ mod } 4,\\
{\frac{d-1}{4}}\Z_2,  &  d=1 \text{ mod } 4 ,\\
{\frac{d+2}{4}}\Z_2,  &  d=2 \text{ mod } 4 ,\\
{\frac{d+1}{4}}\Z_2,  &  d=3 \text{ mod } 4 .\\
\end{cases}
\end{align}
In \Ref{R11,B8283}, it was shown that
\begin{align}
H^∗[BO_2,\Z]=\Z[x_2,x_3,x_4]/(2x_2,2x_3,x_3^2-x_2x_4),
\end{align}
where $x_2=\bt\rw^{O_2}_1$, $x_3=\bt\rw^{O_2}_2$, and $x_4=p_1^{O_2}$ is the Pontryagin class.
Here $\bt$ is the natural map $H^d(BG,\Z_2)\to H^{d+1}(G,\Z)$.  In other words,
we have a relation $(\bt\rw^{O_2}_2)^2=\bt\rw^{O_2}_1 p_1^{O_2}$.  We find that
\begin{align}
\label{HO2Z}
 H^0(BO_2,\Z) & = \Z,
\nonumber\\
 H^1(BO_2,\Z) & = 0,
\\
 H^2(BO_2,\Z) & = \Z_2 , \text{ basis } [\bt\rw^{O_2}_1],
\nonumber\\
 H^3(BO_2,\Z) & = \Z_2 , \text{ basis } [\bt\rw^{O_2}_2],
\nonumber\\
 H^4(BO_2,\Z) & = \Z\oplus \Z_2 , \text{ basis } [(\bt\rw^{O_2}_1)^2, p_1^{O_2}],
\nonumber\\
 H^5(BO_2,\Z) & = \Z_2 , \text{ basis } [\bt\rw^{O_2}_1\bt\rw^{O_2}_2],
\nonumber\\
 H^6(BO_2,\Z) & = 2\Z_2 , \text{ basis } [\bt\rw^{O_2}_1 p_1^{O_2}, (\bt\rw^{O_2}_1)^3],
\nonumber\\
 H^7(BO_2,\Z) & = 2\Z_2 , \text{ basis } [\bt\rw^{O_2}_2 p_1^{O_2}, (\bt\rw^{O_2}_1)^2\bt\rw^{O_2}_2],
\nonumber\\
 H^8(BO_2,\Z) & = \Z\oplus 2\Z_2 , \text{ basis } 
[(p_1^{O_2})^2,(\bt\rw^{O_2}_1)^2 p_1^{O_2}, (\bt\rw^{O_2}_1)^4],
\nonumber 
\end{align}
which agrees with
\eqn{Z2rU1Z1} with $\subset$ replaced by $=$. So we actually have
\begin{align}
\label{Z2rU1Z}
 H^d[ BO_2,\Z] =
\begin{cases}
\Z\oplus {\frac d 4}\Z_2,  &  d=0 \text{ mod } 4,\\
{\frac{d-1}{4}}\Z_2,  &  d=1 \text{ mod } 4 ,\\
{\frac{d+2}{4}}\Z_2,  &  d=2 \text{ mod } 4 ,\\
{\frac{d+1}{4}}\Z_2,  &  d=3 \text{ mod } 4 .\\
\end{cases}
\end{align}
which allow us to get\cite{W1313,W1447}
\begin{align}
\label{UZ2Z2}
 H^d[BO_2,\Z_2] =
\begin{cases}
{\frac {d +2}2}\Z_2,  &  d=0 \text{ mod } 2,\\
{\frac{d+1}{2}}\Z_2,  &  d=1 \text{ mod } 2 ,
\end{cases}
\end{align}
We also have
\begin{align}
\label{HO2RZ}
 \cH^0(O_2,\RZ) & = \RZ,
\\
 \cH^1(O_2,\RZ) & = \Z_2 , \text{ basis } [\frac12 \rw^{O_2}_1],
\nonumber\\
 \cH^2(O_2,\RZ) & = \Z_2 , \text{ basis } [\frac12 \rw^{O_2}_2],
\nonumber\\
 \cH^3(O_2,\RZ) & = \Z\oplus \Z_2 , \text{ basis } [\frac12 (\rw^{O_2}_1)^3, \frac{1}{2\pi}A\dd A],
\nonumber\\
 \cH^4(O_2,\RZ) & = \Z_2 , \text{ basis } [\frac12 (\rw^{O_2}_1)^2\rw^{O_2}_2],
\nonumber\\
 \cH^5(O_2,\RZ) & = 2\Z_2 , \text{ basis } [\frac12 \rw^{O_2}_1 (\rw^{O_2}_2)^2, \frac12 (\rw^{O_2}_1)^5],
\nonumber\\
 \cH^6(O_2,\RZ) & = 2\Z_2 , \text{ basis } [\frac12 (\rw^{O_2}_2)^3 , \frac12 (\rw^{O_2}_1)^4\rw^{O_2}_2],
\nonumber\\
 \cH^7(O_2,\RZ) & = \Z\oplus 2\Z_2 , 
\nonumber\\
& \text{ basis } 
[\frac{A(\dd A)^3}{(2\pi)^3}, \frac {(\rw^{O_2}_1)^3 (\rw^{O_2}_2)^2}2 , \frac{(\rw^{O_2}_1)^7}2] ,
\nonumber 
\end{align}
The above basis give rise to the basis in \eqn{HO2Z} after the natural map
$\t\bt$: $\cH^d(G,\RZ)\to H^{d+1}(BG,\Z)$, which becomes the Steenrod square
$Sq^1$ when acting on $\rw^{O_2}_i$'s.  One can use the properties in \eqn{Sq1}
to do the calculation (see Appendix \ref{calgen}).  

This allows us to obtain $O_2$ mixed SPT states which are given by\\ 
in 3+1D:
$H^1(BO_2, \si\iTOL^3)=0$, \\
in 4+1D:
$H^2(BO_2, \si\iTOL^3)=\Z_2$,\\
in 5+1D: 
$H^3(BO_2, \si\iTOL^3)\oplus H^1(BO_2, \si\iTOL^5)=2\Z_2$,\\
in 6+1D: 
$H^4(BO_2, \si\iTOL^3)\oplus H^2(BO_2, \si\iTOL^5)=\Z\oplus 3\Z_2$,\\
in 7+1D: 
$H^5(BO_2, \si\iTOL^3)\oplus H^3(BO_2, \si\iTOL^5)\oplus H^1(BO_2, \si\iTOL^7)=3\Z_2$.

In the following, we will consider if there are extra mixed $O_2$
SPT phases.  We first note that the ring $H^*[BO_2, \Z_2]$ is
generated by $a_1,f_2$.  If we choose the natural embedding $O_2
\to O$ via $O(2) \to O$, we find that
\begin{align}
\rw_1^{O_2}=a_1 ,\ \ \
\rw_2^{O_2}=f_2 ,\ \ \
 \rw_i^{O_2}=0, \  i>2.
\end{align}

In 2+1D, the potential extra mixed $O_2$ SPT phases are described by
$H^1[BO_2, \cH^2(SO,\RZ)]=\Z_2$.  
Therefore, the potential extra
$O_2$ SPT phases are generated by
$a_1\rw_2$.  In 2+1D, we have the following relations
\begin{align}
 Sq^1[ (\rw_1^{O_2})^2]=u_1 (\rw_2^{O_2})^2
& \to \rw_1(\rw_1^{O_2})^2=0 ;
\nonumber\\
&\ \ \  u_2 =\rw_1^2+\rw_2=0 .
\end{align}
We see that $\rw_1=\rw_2=0$ for orientable 2+1D space-time and $a_1\rw_2$
vanishes.  Thus there is no extra $O_2$ SPT phase in 2+1D.

In 3+1D, the potential extra mixed $O_2$ SPT phases are described
by $H^2[BO_2, \cH^2(SO,\RZ)]=2\Z_2$ generated by
$f_2\rw_2,\ a_1^2\rw_2$.  In 3+1D, we have the following relations
(setting $\rw_1=0$)
\begin{align}
& \rw_2^2+\rw_4=0,
\nonumber\\
& a_1^4=\rw_3 a_1 = \rw_2 a_1^2 =f_2^2+\rw_2 f_2=0.
\end{align}
We see that $\rw_2 a_1^2=0$ and
$\rw_2 f_2=f_2^2=c_1^2$ mod 2.
So $\rw_2 f_2$, as part of $c_1^2$, can be smoothly deformed to zero.
Thus there is no extra $O_2$ SPT phase in 3+1D.

In 4+1D, the potential extra mixed $O_2$ SPT phases are described by $H^3[BO_2,
\cH^2(SO,\RZ)]\oplus H^1[BO_2, \cH^4(SO,\RZ)]=3\Z_2$ generated by
$\rw_2a_1f_2,\ \rw_2a_1^3,\ \rw_4a_1$.  In 4+1D, we have the following relations
(setting $\rw_1=0$)
\begin{align}
& \rw_2^2+\rw_4=\rw_5=0,
\nonumber\\
& a_1^3 f_2 = \rw_3 a_1^2 = \rw_3 f_2+\rw_2 a_1f_2
\nonumber\\
&\ \ \ \ \ \ 
=a_1f_2^2+a_1f_2\rw_2
 =0.
\end{align}
We see that $\rw_2 a_1^3=0$,
$w_2 a_1 f_2=\rw_3 f_2=\rw_3 f_2$, and $ a_1f_2^2=a_1f_2\rw_2$.
But $ a_1f_2^2 = \rw_1^{O_2}
(\rw_2^{O_2})^2$
is already included in $H^5(BO_2,\RZ)$
(see \eqn{HO2RZ}).
So $a_1\rw_4=a_1 \rw_2^2=a_1p_1$ is not restricted to zero.
There is an $O_2$ SPT phase in 4+1D with topological invariant
\begin{align}
	W^5_\text{top}(A,\Ga) = \frac{A_{Z_2}}{2\pi} p_1,
\end{align}
where $A_{Z_2}$ is the flat connection that describe the $Z_2$ twist\cite{WGW1489}
\begin{align}
 \oint A_{Z_2} =0 \text{ mod } \pi.
\end{align}
But the above topological invariant has been included by $O_2$ SPT phases described
by $H^2(BO_2,\si\iTOL^3)\simeq \cH^1(O_2,\RZ)$ (see Appendix \ref{giTO3}).  So
there is no extra $O_2$ SPT phase in 4+1D.

In 5+1D, the potential extra mixed $O_2$ SPT phases are described
by $H^4[BO_2, \cH^2(SO,\RZ)]\oplus H^2[BO_2,
\cH^4(SO,\RZ)]=5\Z_2$ generated by 
$
\rw_2a_1^2f_2,\
\rw_2f_2^2,\
\rw_2a_1^4,\
\rw_4a_1^2,\
\rw_4f_2
$.  In 5+1D,
we have the following relations (setting $\rw_1=0$)
\begin{align}
& \rw_2^2+\rw_4=\rw_3^2+\rw_2\rw_4=\rw_3^2=\rw_5=\rw_6=0,
\\
& 
\rw_4a_1^2=
\rw_2 a_1^4=
\rw_2 f_2^2 =
\rw_2a_1^2 f_2+a_1^4f_2
=0.
\nonumber 
\end{align}
We see that $\rw_2 a_1^4=\rw_4a_1^2=\rw_2 f_2^2=0$ and
$ \rw_2a_1^2 f_2 =a_1^4f_2$. 
But $ \rw_2a_1^2 f_2 =a_1^4f_2=
(\rw_1^{O_2})^4
\rw_2^{O_2}
$ is already included in $H^6(BO_2,\RZ)$
(see \eqn{HO2RZ}).
Also $\rw_4 f_2=\rw_2^2 f_2 = p_1c_1$ mod 2 is connected to zero.
There are
no extra $O_2$ SPT phase in 5+1D.

In 6+1D, the potential extra mixed $O_2$ SPT phases are described
by $
H^5[BO_2, \cH^2(SO,\RZ)]\oplus 
H^3[BO_2, \cH^4(SO,\RZ)]\oplus 
H^1[BO_2, \cH^6(SO,\RZ)]
=7\Z_2$ 
generated by 
$
\rw_2a_1^5,\
\rw_2a_1^3f_2,\
\rw_2a_1f_2^2,\
\rw_4a_1^3,\
\rw_4a_1f_2,\
\rw_6a_1,\
\rw_3^2a_1
$.  In 6+1D,
we have the following relations (setting $\rw_1=0$)
\begin{align}
& \rw_2^2+\rw_4=\rw_3^2+\rw_2\rw_4=\rw_5=\rw_6=0,
\\
& 
\rw_2\rw_3 a_1^2 +\rw_3^2 a_1 = 
\rw_4 a_1f_2 = 
\rw_2a_1^5 =
\rw_2 a_1^3f_2 +\rw_2 a_1f_2^2 
\nonumber\\
&=
a_1^3f_2^2+\rw_2 a_1 f_2^2 =
a_1^3f_2^2+\rw_3 a_1^2 f_2 
=
a_1^5 f_2= a_1f_2^3
=0.
\nonumber 
\end{align}
We see that $\rw_2 a_1^5=\rw_4a_1f_2=0$ and $ \rw_2\rw_3 a_1^2  =\rw_3^2 a_1$,
$ \rw_2 a_1 f_2^2  =\rw_2 a_1^3f_2= \rw_3 a_1^2f_2=a_1^3 f_2^2$.  But $ \rw_3^2
a_1 =\rw_2\rw_3 a_1^2 $ is already included in $H^2(BO_2,\si\iTOL^5)$, and $
\rw_2 a_1 f_2^2  =\rw_2 a_1^3f_2= \rw_3 a_1^2f_2=a_1^3 f_2^2$ are  already
included in $H^7(BO_2,\RZ)$ (see \eqn{HO2RZ}).  However, $a_1^3\rw_4
=a_1^3\rw_2^2=a_1^3 p_1$ mod 2 is not restricted.  There is an $O_2$ SPT
phase in 6+1D described by a topological invariant
\begin{align}
	W^7_\text{top}(A,\Ga) = \frac{1}{2\pi^3} A_{Z_2}^3 p_1 .
\end{align}
But the above topological invariant has been included by $O_2$ SPT phases described
by $H^4(BO_2,\si\iTOL^3)\simeq \cH^3(O_2,\RZ)$ (see Appendix \ref{giTO3}).  So
there is no extra $O_2$ SPT phase in 6+1D.

\subsection{$Z_2^T$ SPT states}

Note that\cite{CGL1314,W1313,W1447}
\begin{align}
	H^d(BZ_2^T,\Z) &=0 \text{ if } d=\text{even};	 
	\nonumber\\
	H^d(BZ_2^T,\Z) &=\Z_2 \text{ if } d=\text{odd};	 
\end{align}
\begin{align}
	H^d(BZ_2^T,\Z_2) &=\Z_2 \text{ if } d=0,
\nonumber\\
	H^d(BZ_2^T,\Z_2) &=\Z_2, \text{ if } d>0 ,
\end{align}
where the time-reversal has a non-trivial action $\Z\to -\Z$.  This allows us
to obtain, in 3+1D, $Z_2^T$ pure SPT states described by the
group-cohomology $\cH^4(Z_2^T, \RZ)=\Z_2$, and $Z_2^T$ mixed SPT states
described by the group-cohomology $H^1(BZ_2^T, \si\iTOL^3)=\Z_2$.
We also obtain mixed SPT states\\ 
in 5+1D described by $H^3(BZ_2^T, \si\iTOL^3)\oplus H^1(BZ_2^T, \si\iTOL^5)=2\Z_2$,\\
in 6+1D described by $H^2(BZ_2^T, \si\iTOL^5)=\Z_2$,\\
in 7+1D described by $H^5(BZ_2^T, \si\iTOL^3)\oplus H^2(BZ_2^T, \si\iTOL^5)\oplus 
H^1(BZ_2^T, \si\iTOL^7)=4\Z_2$.

The 3+1D $Z_2^T$ mixed SPT state described by $H^1(BZ_2^T, \si\iTOL^3)$ may be
produced in the following way:\cite{CLV1407,LGW1418,WGW1489} We start with a
system with $Z_2^T$ symmetry whose ground state break the $Z_2^T$ symmetry.
Then, we allow the fluctuations of the domain walls of the $Z_2^T$ order
parameter, to restore a $Z_2^T$ symmetry.  We may bound an 2+1D $(E_8)^3$
bosonic quantum Hall state to such domain wall. In this case, the restored
$Z_2^T$ symmetric state is the mixed SPT state described by $H^1(BZ_2^T,
\si\iTOL^3)$. In fact, all the mixed SPT states described by
$H^1(BG,\si\iTO_L^{d-2})$ and all the $G_1\times G_2$ pure SPT states described
by $H^1[BG_1,\cH^{d-2}(G_2,\RZ)]$ can be constructed in this
way.\cite{CLV1407}

Such a $Z_2^T$ mixed SPT state can be probed by surface symmetry
breaking.\cite{VS1306} The $Z_2^T$ symmetry breaking domain wall on the
boundary will carry the gapless edge state of 2+1D $(E_8)^3$ bosonic quantum
Hall state.  In fact, all the mixed SPT states described by
$H^1(BG,\si\iTO_L^{d-2})$ and all the $G_1\times G_2$ pure SPT states described
by $H^1[BG_1,\cH^{d-2}(G_2,\RZ)]$ can be probed in this way.
 
In the following, we will consider if there are extra mixed $Z_2^T$ SPT phases.
Let $a_1$ be the generator of the ring $H^*(BZ_2^T,\Z_2)$.  Let we choose the
natural embedding $Z_2^T \to O$ via $Z_2^T\to O/SO$, we find that
\begin{align}
\rw_1^{Z_2^T}=a_1 ,\ \ \ \ \rw_i^{Z_2^T}=0, \ \ \ i>1.
\end{align}
However, since the $Z_2^T$ twist can only be implemented by reversing the
space-time orientation, we need to identify 
\begin{align}
\rw_1 \to  \rw_1^{Z_2^T}.
\end{align} 
In 7+1D and below, we did not find any extra mixed $Z_2^T$ SPT phases.

\subsection{$U(1)\times Z_2^T$ SPT states}

In \Ref{CGL1314}, we obtained
\begin{align}
\label{Z2TU1Z}
 H^d[ B(U(1)\times Z_2^T),\Z] =
\begin{cases}
0,  &  d=0 \text{ mod } 2,\\
\frac{d+1}{2}\Z_2,  &  d=1 \text{ mod } 2 ,\\
\end{cases}
\end{align}
which 
allows us to get\cite{W1313,W1447}
\begin{align}
 H^d[ B(U(1)\times Z_2^T),\Z_2] =
\begin{cases}
{\frac {d +2}2}\Z_2,  &  d=0 \text{ mod } 2,\\
{\frac{d+1}{2}}\Z_2,  &  d=1 \text{ mod } 2 ,
\end{cases}
\end{align}
This allows us to obtain $U(1)\times Z_2^T$ mixed SPT states which are given by\\ 
in 3+1D:
$H^1[B(U(1)\times Z_2^T), \si\iTOL^3]=\Z_2$, \\
in 4+1D:
$H^2[B(U(1)\times Z_2^T), \si\iTOL^3]=0$,\\
in 5+1D: 
$H^3[B(U(1)\times Z_2^T), \si\iTOL^3]\oplus H^1[B(U(1)\times Z_2^T), \si\iTOL^5]=3\Z_2$,\\
in 6+1D: 
$H^4[B(U(1)\times Z_2^T), \si\iTOL^3]\oplus H^2[B(U(1)\times Z_2^T], \si\iTOL^5)=2\Z_2$,\\
in 7+1D: 
$H^5[B(U(1)\times Z_2^T), \si\iTOL^3]\oplus H^2[B(U(1)\times Z_2^T), \si\iTOL^5]\oplus H^1[B(U(1)\times Z_2^T), \si\iTOL^7]=6\Z_2$.

In the following, we will consider if there are extra mixed $U(1)\times Z_2^T$
SPT phases.  We first note that the ring $H^*[B(U(1)\times Z_2^T), \Z_2]$ is
generated by $a_1,f_2$ (the same as  $H^*[B(U(1)\times Z_2), \Z_2]$ and
$H^*[B(U(1)\rtimes Z_2), \Z_2]$).  Note that $H^d[ B(U(1)\times Z_2^T),\Z]$ for
$d=$ odd is generated by $a_1c_1^{\frac{d-1}2}$, $a_1^3c_1^{\frac{d-3}2}$,
\etc.  Or $H^d[ B(U(1)\times Z_2^T),\RZ]$ for $d=$ even is generated by
$\frac12 c_1^{\frac{d}2}$, $\frac12 a_1^2c_1^{\frac{d-2}2}$, \etc.

If we choose the natural embedding
$U(1)\times Z_2^T \to O$ via $U(1)\times Z_2 \to SO(3)$ which map $U(1)$ to the
rotation in $x$-$y$ plane and $Z_2$ to the $z\to -z$ reflection, we find that
\begin{align}
\rw_1^{O_2}=a_1 ,\ \ \
\rw_2^{O_2}=f_2 ,\ \ \
 \rw_i^{O_2}=0, \  i>2.
\end{align}
Also since the time-reversal twist is implemented by the orientation reversal,
we need to set $\rw_1=a_1$.

In 2+1D, the potential extra mixed $U(1)\times Z_2^T$ SPT phases are described
by $H^1[B(U(1)\times Z_2^T), \cH^2(SO,\RZ)]=\Z_2$ generated by $a_1\rw_2$.  In
2+1D, we have the following relations (setting $\rw_1=a_1$)
\begin{align}
& \rw_1^2+\rw_2=\rw_1 \rw_2=0,
\end{align}
We see that $\rw_2 \rw_1=0$,
and there is no extra $U(1)\times Z_2^T$ SPT phase in 2+1D.

In 3+1D, the potential extra mixed $U(1)\times Z_2^T$ SPT phases are described
by $H^2[B(U(1)\times Z_2^T), \cH^2(SO,\RZ)]=2\Z_2$ generated by $a_1^2\rw_2,\
f_2 \rw_2$.  In 3+1D, we have the following relations (setting $\rw_1=a_1$)
\begin{align}
& \rw_1 \rw_2=\rw_1\rw_3=\rw_1^4+\rw_2^2+\rw_1\rw_3+\rw_4=0 ,
\nonumber\\
& \rw_1^2 f_2=\rw_1^2f_2+f_2^2+f_2 \rw_2=0 .
\end{align}
We see that $\rw_1^2\rw_2=0$ and $f_2 \rw_2 =f_2^2$.
But $f_2 \rw_2 =f_2^2$ can be deformed to zero smoothly. Thus
there is no extra $U(1)\times Z_2^T$ SPT phase in 3+1D.

In 4+1D, the potential extra mixed $U(1)\times Z_2^T$ SPT phases are described
by $H^3[B(U(1)\times Z_2^T), \cH^2(SO,\RZ)]\oplus H^1[B(U(1)\times Z_2^T),
\cH^4(SO,\RZ)]=3\Z_2$ generated by $\rw_2\rw_1f_2,\ \rw_2\rw_1^3,\ \rw_4\rw_1$.  In
4+1D, we have the following complete set of relations (setting $\rw_1=a_1$)
\begin{align}
\label{UT5d}
& 
\rw_1^5=\rw_1 f_2^2=\rw_1^3 \rw_2=\rw_1 f_2 \rw_2=\rw_1 \rw_2^2=\rw_1^2 \rw_3
\nonumber\\
&=f_2 \rw_3=\rw_1 \rw_4=\rw_5=0
\end{align}
There is no extra $U(1)\times Z_2^T$ SPT phase in 4+1D.


In 5+1D, the potential extra mixed $U(1)\times Z_2^T$ SPT phases are described
by $H^4[B(U(1)\times Z_2^T), \cH^2(SO,\RZ)]\oplus H^2[B(U(1)\times Z_2^T),
\cH^4(SO,\RZ)]=5\Z_2$ generated by 
$
\rw_2a_1^2f_2,\
\rw_2f_2^2,\
\rw_2a_1^4,\
\rw_4a_1^2,\
\rw_4f_2
$.  In 5+1D,
we have the following complete set of relations (setting $\rw_1=a_1$)
\begin{align}
&
\rw_1^4 f_2=\rw_1^6+\rw_1^4 \rw_2=\rw_1^2 f_2^2+\rw_1^2 f_2 \rw_2=f_2^2 \rw_2 
\nonumber\\
&
=\rw_1 f_2 \rw_3=\rw_1 \rw_3 \rw_2+\rw_2^3= \rw_2^3+\rw_3^2 =\rw_1^6+\rw_2 \rw_4 
\nonumber\\
&
=\rw_1^2 \rw_2^2+\rw_1^2 \rw_4 =f_2 \rw_2^2+f_2 \rw_4 =\rw_1^2 \rw_2^2+\rw_1 \rw_5 
\nonumber\\
&
=\rw_1^2 \rw_2^2+\rw_6 = 0
\end{align}
We see that $f_2^2 \rw_2=0$.
Also $\rw_1^4\rw_2=\rw_1^6$ and $\rw_1^2f_2\rw_2=\rw_1^2f_2^2$
are already included in $\cH^6(U(1)\times Z_2^T, \RZ)$.
$f_2\rw_4=f_2\rw_2^2=f_2 p_1$ mod 2 is a part of integer class $c_1 p_1$
and can be smoothly deformed to zero.
However, $\rw_1^2\rw_4 =\rw_1^2\rw_2^2  = \rw_1^2 p_1$ mod 2 is not restricted.
There is an $U(1)\times Z_2^T$ SPT phase in 5+1D described by
\begin{align}
	W^6_\text{top}(A,\Ga) = \frac12 \rw_1^2 p_1 .
\end{align}
But the above topological invariant has been included by $U(1)\times Z_2^T$ SPT
phases described by $H^3[B(U(1)\times Z_2^T),\si\iTOL^3]\simeq \cH^2(U(1)\times
Z_2^T,\RZ)$ (see Appendix \ref{giTO3}).  So there is no extra $U(1)\times
Z_2^T$ SPT phase in 5+1D.

In 6+1D, the potential extra mixed $U(1)\times Z_2^T$ SPT phases are described
by $H^5[B(U(1)\times Z_2^T), \cH^2(SO,\RZ)]\oplus H^3[B(U(1)\times Z_2^T),
\cH^4(SO,\RZ)]\oplus H^1[B(U(1)\times Z_2^T),
\cH^6(SO,\RZ)]=7\Z_2$ 
generated by 
$
\rw_2a_1^5,\
\rw_2a_1^3f_2,\
\rw_2a_1f_2^2,\
\rw_4a_1^3,\
\rw_4a_1f_2,\
\rw_6a_1,\
\rw_3^2a_1
$.  In 6+1D,
we have the following complete set of relations (setting $\rw_1=a_1$)
\begin{align}
&
\rw_1^7=\rw_1^3 f_2^2=\rw_1^5 \rw_2=\rw_1^5 f_2+\rw_1^3 f_2 \rw_2=\rw_1 f_2^2 \rw_2
\nonumber\\
&
=\rw_1^3 \rw_2^2=\rw_1 \rw_2^3=\rw_1^4 \rw_3=\rw_1^5 f_2+\rw_1^2 f_2 \rw_3=f_2^2 \rw_3
\nonumber\\
&
=\rw_2^2 \rw_3=\rw_1 \rw_3^2=\rw_1^3 \rw_4=\rw_1 f_2 \rw_2^2+\rw_1 f_2 \rw_4
\nonumber\\
&
=\rw_1^2 \rw_2 \rw_3+\rw_1 \rw_2 \rw_4=\rw_3 \rw_4 =\rw_1 f_2 \rw_2^2+f_2 \rw_5
\nonumber\\
&
=\rw_1^2 \rw_5 =\rw_1^2 \rw_2 \rw_3+\rw_2 \rw_5=\rw_1 \rw_6=\rw_7 =0.
\end{align}
We see that $\rw_2 \rw_1^5=\rw_1f_2^2\rw_2=\rw_4\rw_1^3=\rw_1\rw_6=\rw_1\rw_3^2 =0$ and
$\rw_1^3f_2\rw_2=\rw_1^5f_2$, $\rw_1f_2\rw_4=\rw_1f_2\rw_2^2=f_2\rw_5$.  But
$\rw_1^3f_2\rw_2=\rw_1^5f_2$ is already included in $\cH^7(U(1)\times Z_2^T,\RZ)$.
Only $\rw_1f_2\rw_2^2$ is not restricted.  Thus there is an extra $U(1)\times
Z_2^T$ SPT phase in 6+1D described by a topological invariant
\begin{align}
	W^7_\text{top}(A,\Ga) = \frac{1}{2} \rw_1 p_1 \frac{\dd A}{2\pi}.
\end{align}

\subsection{$U(1)\rtimes Z_2^T$ SPT states}

In \Ref{CGL1314}, we also obtained
\begin{align}
\label{Z2TrU1Z1}
 H^d[ B(U(1)\rtimes Z_2^T),\Z] \subset
\begin{cases}
{\frac d 4}\Z_2,  &  d=0 \text{ mod } 4,\\
{\frac{d+3}{4}}\Z_2,  &  d=1 \text{ mod } 4 ,\\
\Z\oplus {\frac{d-2}{4}}\Z_2,  &  d=2 \text{ mod } 4 ,\\
{\frac{d+1}{4}}\Z_2,  &  d=3 \text{ mod } 4 ,\\
\end{cases}
\end{align}
If we assume
\begin{align}
\label{Z2TrU1Z}
 H^d[ B(U(1)\rtimes Z_2^T),\Z] =
\begin{cases}
{\frac d 4}\Z_2,  &  d=0 \text{ mod } 4,\\
{\frac{d+3}{4}}\Z_2,  &  d=1 \text{ mod } 4 ,\\
\Z\oplus {\frac{d-2}{4}}\Z_2,  &  d=2 \text{ mod } 4 ,\\
{\frac{d+1}{4}}\Z_2,  &  d=3 \text{ mod } 4 ,\\
\end{cases}
\end{align}
then, we will obtain\cite{W1313,W1447}
\begin{align}
\label{UrTZ2Z2}
 H^d[ B(U(1)\rtimes Z_2^T),\Z_2] =
\begin{cases}
{\frac {d +2}2}\Z_2,  &  d=0 \text{ mod } 2,\\
{\frac{d+1}{2}}\Z_2,  &  d=1 \text{ mod } 2 ,
\end{cases}
\end{align}
which should agree with \eqn{UZ2Z2}. Indeed, it agrees, implying that
\eqn{Z2TrU1Z} is correct.

Eqns.  (\ref{Z2TrU1Z}) and (\ref{UrTZ2Z2}) allow us to obtain $U(1)\rtimes Z_2^T$ mixed SPT states which are given by\\ 
in 3+1D:
$H^1[B(U(1)\rtimes Z_2^T), \si\iTOL^3]=\Z_2$, \\
in 4+1D:
$H^2[B(U(1)\rtimes Z_2^T), \si\iTOL^3]=\Z$,\\
in 5+1D: 
$H^3[B(U(1)\rtimes Z_2^T), \si\iTOL^3]\oplus H^1[B(U(1)\rtimes Z_2^T), \si\iTOL^5]=2\Z_2$,\\
in 6+1D: 
$H^4[B(U(1)\rtimes Z_2^T), \si\iTOL^3]\oplus H^2[B(U(1)\rtimes Z_2^T), \si\iTOL^5]=3\Z_2$,\\
in 7+1D: 
$H^5[B(U(1)\rtimes Z_2^T), \si\iTOL^3]\oplus H^2[B(U(1)\rtimes Z_2^T), \si\iTOL^5]\oplus H^1[B(U(1)\rtimes Z_2^T), \si\iTOL^7]=6\Z_2$.

In the following, we will consider if there are extra mixed $U(1)\rtimes Z_2^T$
SPT phases.  We first note that the ring $H^*[B(U(1)\rtimes Z_2^T), \Z_2]$ is
generated by $a_1,f_2$ (the same as  $H^*[B(U(1)\times Z_2), \Z_2]$ and
$H^*[B(U(1)\rtimes Z_2), \Z_2]$), where $f_2$ is $c_1$ mod 2.  As discussed in
Appendix \ref{LHS}, $\cH^d[U(1)\rtimes Z_2^T,\RZ]$ is generated by the subgroup
of the factor group of $\cH^k[Z_2^T,\cH^{d-k}(U(1), \RZ)]$.  We find that in
our case here, $\cH^d[U(1)\rtimes Z_2^T,\RZ]$ is generated by the full
$\cH^k[Z_2^T,\cH^{d-k}(U(1), \RZ)]$.  $\cH^{n}(U(1), \RZ)=\Z$ is generated by
the Chern-Simons term $ac_1^{\frac{n-1}2}$. $Z_2^T$ acts on $\cH^{n}(U(1),
\RZ)$ by $\cH^{n}(U(1), \RZ) \to \cH^{n}(U(1), \RZ)$ if $\frac{n-1}2 =$ even,
and by $\cH^{n}(U(1), \RZ) \to -\cH^{n}(U(1), \RZ)$ if $\frac{n-1}2 =$ odd.  So
$\cH^d[U(1)\rtimes Z_2^T,\RZ]$ is generated by $a_1^m a c_1^n$, with $m+2n+1=d$
and $(m,n)$ = (even,even) or $(m,n)$ = (odd,odd).  As discussed in Appendix
\ref{giTO3}, $a_1^m a c_1^n$ can be viewed as $a_1^{m-1} c_1^{n+1}$. This
allows us to obtain the generators of $\cH^d[U(1)\rtimes Z_2^T,\RZ]$.

If we choose the natural embedding
$U(1)\rtimes Z_2^T \to O$ via $U(1)\rtimes Z_2 =O(2) \to O$, we find that
\begin{align}
\rw_1^{O_2}=a_1 ,\ \ \
\rw_2^{O_2}=f_2 ,\ \ \
 \rw_i^{O_2}=0, \  i>2.
\end{align}
Since the time-reversal twist is implemented by the orientation reversal, we
need to set $\rw_1=a_1$.  We also note that $H^k[B(U(1)\rtimes Z_2^T),
\Z_2]=H^k[B(U(1)\times Z_2^T), \Z_2]$, since when the coefficient is $\Z_2$,
there is no distinction between $U(1)\rtimes Z_2^T$ and $U(1)\times Z_2^T$.
The above results implies that the extra mixed $U(1)\rtimes Z_2^T$ SPT phases
are the same as the extra mixed $U(1)\times Z_2^T$ SPT phases, So we can used
the results from the last section.


\section{SPT states protected by mirror reflection symmetry}

In this section, we are going to consider SPT state protected by mirror
reflection symmetry $Z_2^M$, which can be probed by the fixed-point partition
function on space-time $M^d$ with symmetry twist.  However, here, the symmetry
twists make the space-time unoriented.\cite{K1467,HMC1402,K1459v2,KTT1429}  So the $Z_2^M$
SPT states are described by the gravitational topological invariants
$W^d_\text{top}(\Ga)$, which take non-trivial values for unoriented
space-times.

Here we would like to remark that the symmetry twists of the time reversal
$Z_2^T$ can be implemented by unoriented space-time, since the
action amplitude for cells with opposite orientation differ by a complex
conjugation (see \eqn{lattS}).  This suggest that the L-type $Z_2^T$ SPT states
and the L-type $Z_2^M$ SPT states are the same.

To study the potential $Z_2^M$ SPT states, we note that
the ring of the cobordism group $\Om^{O}_d$
of closed unoriented smooth manifolds
is\cite{UOCob} 
\begin{align}
 \Om^O=\sum_d \Om^{O}_d = \Z_2[\{x_d\}],\ \ \ d>1,\ d\neq 2^i-1,
\end{align}
where $\M[\{x_d\}]$ is the polynomial ring generated by $x_d$'s with $\M$
as coefficient. Also $x_{2i}=\R P^{2i}$.
In lower dimensions, we have\\
$\Om^{O}_{1}=0$, since circles bound disks.\\
$\Om^{O}_{2}=\Z_2$, generated by $x_2$.\\
$\Om^{O}_{3}=0$.\\
$\Om^{O}_{4}=2\Z_2$, generated by $x_4$ and $x_2^2$.\\
$\Om^{O}_{5}=\Z_2$, generated by $x_5=H_{2,4}$. \\
$\Om^{O}_{6}=3\Z_2$,  generated by $x_6$, $x_2x_4$, and $x_2^3$. \\
$\Om^{O}_{7}=\Z_2$, generated by $x_2x_5=H_{2,4}\times \R P^2$.\\
$\Om^{O}_{8}=5\Z_2$, generated by $x_8$, $x_2x_6$, $x_2^4x_4$, $x_4^2$, $x_2^4$.\\
$H_{m,n}$ is a manifold of dimension $m + n - 1$ defined as the subset of $\R
P^m \times \R P^n$ of points satisfying the homogeneous equation $x_0
y_0+\cdots + x_m y_m = 0$.

The potential gravitational topological invariants for $Z_2^M$ SPT phases have been
obtained in \Ref{K1467,K1459v2}. However, their realizations have not been
discussed systematically. In this paper, we show that all the $Z_2^M$ potential
gravitational topological invariants are realizable by the $O(\infty)$ NL$\si$Ms.
This is because the unoriented  cobordism group has no free parts. Thus there
is no Chern-Simon potential gravitational topological invariants.  As discussed in
Section \ref{Pggterm}, the locally-null potential gravitational topological
invariants are all realizable.  

In the following, we will calculate the corresponding locally-null
gravitational topological invariants for those $Z_2^M$ SPT phases. Many
results have been obtained in \Ref{K1467,K1459v2}.

In 1+1D, the gravitational topological invariants $W^d_\text{top}(\Ga)$ are
generated by $\pi \rw_1^2=\pi \rw_2$, since the condition $u_2=0$ requires
$\rw_1^2=\rw_2$.  In 3+1D, the gravitational topological invariants
$W^d_\text{top}(\Ga)$ are generated by $\pi \rw_1^3$ and $\pi \rw_3$, since the
condition $u_3=0$ requires $\rw_1\rw_2=0$.  

In 4+1D, there are seven Stiefel-Whitney classes
$\rw_1^5$, $\rw_1^3\rw_2$, $\rw_1^2\rw_3$, $\rw_1\rw_2^2$, $\rw_1\rw_4$, $\rw_2\rw_3$, $\rw_5$.  The Wu classes
$u_3=u_4=u_5=0$ give us
\begin{align}
&\ \ \ \
\rw_1\rw_2=\rw_1^4+\rw_2^2+\rw_1\rw_3+\rw_4
\nonumber\\
&=\rw_1^3\rw_2+\rw_1\rw_2^2+\rw_1^2\rw_3+\rw_1\rw_4 =0
\end{align}
Other relations can be obtained by applying the Steenrod squares:
\begin{align}
 Sq^1(u_3)&= \rw_1\rw_3 =0.
\nonumber\\
 Sq^1(u_4)&=\rw_1 \rw_4+\rw_5=0.
\nonumber\\
 Sq^2(u_3)&=\rw_1^2\rw_2+\rw_1\rw_2^2 +\rw_1^2\rw_3 =0.
\end{align}
Additional relations can be obtained from \eqn{SqWu}
\begin{align}
Sq^1(\rw_1^4)+u_1\rw_1^4 &= \rw_1^5= 0
\nonumber\\
Sq^1(\rw_4)+u_1\rw_4 &= \rw_5= 0
	\\
Sq^2(\rw_3)+u_2\rw_3 &= \rw_1\rw_4+\rw_5+\rw_1^2\rw_3=0.
	\nonumber 
\end{align}
We find that $\rw_1^5=\rw_1\rw_2=\rw_1\rw_3=\rw_1\rw_4=\rw_5=0$.  So we have an realizable
gauge-gravity topological invariant in 4+1D:
\begin{align}
	W^5_\text{top}(\Ga)= \frac12 \rw_2\rw_3 .
\end{align}
But such a topological invariant can exist even if we break the time-reversal
symmetry (see \eq{w2w3}). So it actually describes a topologically ordered
phase.  There is no L-type time-reversal SPT in 4+1D.  In general, the L-type
realizable $Z_2^M$ SPT phases in $d$-dimensional space-time are not described
by $\Om^{O}_{d}$, but by a quotient of 
$\Om^{O}_{d}$
\begin{align}
\PSPT^d_{Z_2^M}=\Om^{O}_d/\bar\Om^{SO}_d, 
\end{align}
where $\bar\Om^{SO}_d$ is the orientation invariant subgroup of $\Om^{SO}_d$
(\ie the manifold $M^d$ and its orientation reversal $-M^d$ belong to the same
oriented cobordism class).

\section{Summary}

In this paper, we use $G\times SO(\infty)$ non-linear NL$\si$Ms to construct
pure SPT and mixed SPT states, as well as iTO states.  We find that those
topological states are classified by a quotient of $\cH^d(G\times SO,\RZ)$.
For example, the quotient of $\cH^d(SO,\RZ)$ give rise to iTO phases:
$\cH^d(SO,\RZ)/\La^d=\si\iTOL^d$.  Writing $\cH^d(G\times SO,\RZ)$ as
$\cH^d(G,\RZ)\oplus \cH^d( SO,\RZ)\oplus \oplus_{k=1}^{d-1} H^k[BG,
\cH^{d-k}(SO,\RZ)] $ and use the quotient to reduce $\cH^d(SO,\RZ)$ to
$\si\iTOL^d$, we find that L-type realizable $G$ SPT phases are classified by
$E^d(G)\rtimes \oplus_{k=1}^{d-1} H^k(BG, \si\iTOL^{d-k})\oplus \cH^d(G,\RZ) $.
This classification include both the pure states [classified by $\cH^d(G,\RZ)$]
and the mixed SPT states  [classified by $E^d(G)\rtimes \oplus_{k=1}^{d-1} H^k(BG,
\si\iTOL^{d-k})$].  (Some of the mixed SPT states were also referred
as the beyond-group-cohomology SPT states. In this paper, we see that those
beyond-group-cohomology SPT states are actually within another type of group
cohomology classification.)

More general SPT states exist, which cannot be obtained from $G\times
SO(\infty)$ non-linear NL$\si$Ms.  Those  SPT states are described by
$E^d(G)\rtimes \oplus_{k=1}^{d-1} H^k(BG, \iTOL^{d-k})$.  We note that, as
Abelian groups, $\si\iTOL^{d}$ is isomorphic to $\iTOL^{d}$, although
$\si\iTOL^{d} \subset \iTOL^{d}$ (see Table \ref{invTop}).  As a result,
$E^d(G)\rtimes \oplus_{k=1}^{d-1} H^k(BG, \iTOL^{d-k})$  is isomorphic to
$E^d(G)\rtimes \oplus_{k=1}^{d-1} H^k(BG, \si\iTOL^{d-k})$, as Abelian groups.

XGW would like to thank Anton Kapustin, Dan Freed, Zheng-Cheng Gu, and Cenke Xu
for many helpful discussions.  This research is supported by NSF Grant No.
DMR-1005541 and NSFC 11274192.  He is also supported by the BMO Financial Group
and the John Templeton Foundation Grant No. 39901.  Research at Perimeter
Institute is supported by the Government of Canada through Industry Canada and
by the Province of Ontario through the Ministry of Research.

\appendix

\section{Group cohomology theory}
\label{gcoh}

\subsection{Homogeneous group cocycle}

In this section, we will briefly introduce group cohomology.  The group
cohomology class $\cH^d(G,\M)$ is an Abelian group constructed from a group $G$
and an Abelian group $\M$.   We will use ``+'' to represent the multiplication
of the Abelian groups.  Each elements of $G$ also induce a mapping $\M\to \M$,
which is denoted as
\begin{eqnarray}
g\cdot m = m', \ \ \ g\in G,\ m,m'\in \M.
\end{eqnarray}
The map $g\cdot$ is a group homomorphism:
\begin{eqnarray}
g\cdot (m_1+m_2)= g\cdot m_1 +g \cdot m_2.
\end{eqnarray}
The  Abelian group $\M$ with such a $G$-group homomorphism, is call a
$G$-module.

A homogeneous $d$-cochain
is a function $\nu_d: G^{d+1}\to \M$, that satisfies
\begin{align}
\label{scond}
\nu_d(g_0,\cdots,g_d)
=g\cdot \nu_d(gg_0,\cdots,gg_d), \ \ \ \ g,g_i \in G.
\end{align}
We denote the set of $d$-cochains as $\cC^d(G,\M)$. Clearly $\cC^d(G,\M)$ is an
Abelian group.
homogeneous group cocycle

Let us define a mapping $\dd$ (group homomorphism) from
$\cC^d(G,\M)$ to $\cC^{d+1}(G,\M)$:
\begin{align}
  (\dd \nu_d)( g_0,\cdots, g_{d+1})=
  \sum_{i=0}^{d+1} (-)^i \nu_d( g_0,\cdots, \hat g_i
  ,\cdots,g_{d+1})
\end{align}
where $g_0,\cdots, \hat g_i ,\cdots,g_{d+1}$ is the sequence $g_0,\cdots, g_i
,\cdots,g_{d+1}$ with $g_i$ removed.
One can check that $\dd^2=0$.
The homogeneous $d$-cocycles are then
the homogeneous $d$-cochains that also satisfy the cocycle condition
\begin{eqnarray}
\label{cccond}
 \dd \nu_d =0.
\end{eqnarray}
We denote the set of $d$-cocycles as
$\cZ^d(G,\M)$. Clearly $\cZ^d(G,\M)$ is an Abelian subgroup of $\cC^d(G,\M)$.

Let us denote  $\cB^d(G,\M)$ as the image of the map $\dd: \cC^{d-1}(G,\M) \to
\cC^d(G,\M)$ and $\cB^0(G,\M)=\{0\}$.
The elements in $\cB^d(G,\M)$ are called $d$-coboundary.
Since $\dd^2=0$, $\cB^d(G,\M)$  is a subgroup of $\cZ^d(G,\M)$:
\begin{eqnarray}
\label{cb}
\cB^d(G,\M) =
\{\dd \nu_{d-1}| \nu_{d-1}\in \cC^{d-1}(G,\M)\} \subset \cZ^d(G,\M).
\end{eqnarray}
The group cohomology class $\cH^d(G,\M)$ is then defined as
\begin{eqnarray}
\cH^d(G,\M) =  \cZ^d(G,\M)/ \cB^d(G,\M) .
\end{eqnarray}
We note that the $\dd$ operator and the cochains $\cC^d(G,\M)$ (for all values
of $d$) form a so called cochain complex,
\begin{align}
\cdots
\stackrel{\dd}{\to}
\cC^d(G,\M)
\stackrel{\dd}{\to}
\cC^{d+1}(G,\M)
\stackrel{\dd}{\to}
\cdots
\end{align}
which is denoted as $C(G,\M)$.  So we may also write the group cohomology
$\cH^d(G,\M)$ as the standard cohomology of the cochain complex $H^d[C(G,\M)]$.

\subsection{Inhomogeneous group cocycle}
\label{inhomo}

The above definition of group cohomology class can be rewritten in terms of
inhomogeneous group cochains/cocycles.  An inhomogeneous group $d$-cochain is
a function $\om_d: G^d \to M$. All $\om_d(g_1,\cdots,g_d)$ form $\cC^d(G,\M)$.
The inhomogeneous group cochains and the homogeneous group cochains are
related as
\begin{eqnarray}
\nu_d(g_0,g_1,\cdots,g_d)=
\om_d( g_{01},\cdots, g_{d-1,d}),
\end{eqnarray}
with
\begin{align}
g_0=1,\ \
g_1=g_0 g_{01}, \ \
g_2=g_1 g_{12}, \ \ \cdots \ \
g_d=g_{d-1} g_{d-1,d}.
\end{align}
Now the $\dd$ map has a form on $\om_d$:
\begin{align}
&
(\dd\om_d)(g_{01},\cdots, g_{d,d+1})=
 g_{01}\cdot \om_d( g_{12},\cdots,g_{d,d+1})
\nonumber\\
& \ \ \
+\sum_{i=1}^d (-)^i \om_d(g_{01},\cdots, g_{i-1,i}g_{i,i+1},\cdots, g_{d,d+1})
\nonumber\\
& \ \ \
+(-)^{d+1}\om_d(g_{01},\cdots,\t g_{d-1,d})
\end{align}
This allows us to define the inhomogeneous group $d$-cocycles which satisfy
$\dd \om_d=0$ and  the inhomogeneous group $d$-coboundaries which have a form
$\om_d = \dd \mu_{d-1}$.  In the following, we are going to use inhomogeneous
group cocycles to study group cohomology.  Geometrically, we may view $g_i$ as
living on the vertex $i$, while $g_{ij}$ as living on the edge connecting the
two vertices $i$ to $j$.

\section{L-type potential gauge topological invariants} 
\label{agterm}

In Section \ref{lattG}, we introduced  the  gauge topological invariant
$W^d_\text{top}( A)$.  In fact, the gauge invariance \eq{gaugeinv} put a strong
constrain on the quantized class of the  gauge topological invariant
$W^d_\text{top}( A)$.  In this section, we will solve those self consistent
conditions and obtain the \emph{potential gauge topological invariants} directly
without going through the NL$\si$M (\ie we do not concern about if a gauge
topological invariant can be generated/realized by a well defined local bosonic
model or not).  

First, it appears that all gauge topological invariants are trivial, since we can
always rescale them $W^d_\text{top}(A) \in \R$ to  $\t W^d_\text{top}(A)=\la
W^d_\text{top}(A)$ and send $\la \to 0$.  The new rescaled topological invariant $
\t W^d_\text{top}(A)$ will vanish. This way, we showed that there is no
non-trivial gauge topological invariant that does not smoothly connect to
zero.

There are two related ways 
to see the mistake in the above argument. First, we note that
gauge topological invariants $W^d_\text{top}(A)$ can be gauge invariant only up
to a $2\pi$ phase. If we scale $W^d_\text{top}(A)$ by an arbitrary real number,
it will not be gauge invariant.  

So different non-trivial gauge topological invariants that do not smoothly connect
to zero are classified by their quantized changes under gauge transformations:
\begin{align}
\label{AgA}
\int_{M^d} W^d_\text{top}(A^g) - \int_{M^d} W^d_\text{top}(A) = 0 \text{ mod } 1.
\end{align} 
We note that the change of the gauge topological invariant, $\int_{{M^d}}
[W^d_\text{top}(A+\del A)-W^d_\text{top}(A)]$, 
can be expressed as\cite{DW9093}
\begin{align}
\label{diffWA}
\int_{M^d} [W^d_\text{top}(A+\del A)-W^d_\text{top}(A)]
& =\int_{\t N^{d+1}} P_\text{top}(F^N)
\nonumber\\
P_\text{top} (F^N) &= \dd  W^d_\text{top}(A), 
\end{align}
where $M^d$ is closed $\prt M^d=\emptyset$, ${\t N^{d+1}}={M^d}\times I$ and the gauge connection $A^N$ on ${\t
N^{d+1}}$ satisfies that on one boundary of ${\t N^{d+1}}$ $A^N=A$ and on the
other boundary $A^N=A+\del A$.  We call $A^N$ an extension of $A,A+\del A$ on
the boundary ${M^d}\cup (-{M^d})=\prt {\t N^{d+1}}$ to ${\t N^{d+1}}$.
Therefore, \eqn{AgA} can be rewritten as
\begin{align}
\label{Pquan1a}
 \int_{N^{d+1}} P_\text{top} (F^N) &= 0 \text{ mod } 1, \ \ \ 
{N^{d+1}}={M^d}\times S^1,
\end{align}
where $G$-bundle on ${N^{d+1}}={M^d}\times S^1$ has a twist generate by $g$ around $S^1$.
We note that $ P_\text{top} (F^N)$ is a closed form (or a cocycle)
$\dd P_\text{top} (F^N)=0$.
Its change under a gauge transformation on ${N^{d+1}}$ is given by
\begin{align}
  \int_{{N^{d+1}}\times S^1} \dd P_\text{top}  = 0 .
\end{align}
Thus $ \int_{N^{d+1}} P_\text{top} (F^N)$ is gauge invariant. So we can express
$P_\text{top} (F^N)$ as a function of the field strength.  Also, a smooth
change of the local bosonic Lagrangian will change 
$W^d_\text{top}(A)$ by an gauge invariant term
$\Del W(F^N)$ and change
$P_\text{top} (F^N)$ by an
exact form $P_\text{top} (F^N)\to P'_\text{top} (F^N)=P_\text{top} (F^N)+\dd
\Del W(F^N)$.

Also, when $g$ is trivial on ${M^d}$ or when the $G$-bundle on ${N^{d+1}}={M^d}\times S^1$ can
be reduced to a  $G$-bundle on ${M^d}$, we have
\begin{align}
 \int_{N^{d+1}} P_\text{top} (F^N) &= 0, \ \ \ {N^{d+1}}={M^d}\times S^1,
\end{align}
In other words,  when the $G$-bundle on ${N^{d+1}}={M^d}\times S^1$ can be
extended to a  $G$-bundle on $\t D^{d+2}={M^d}\times D^2$, where $D^2$ is a
disk, we have 
\begin{align}
\label{Pz1}
 \int_{\prt ({M^d}\times D^2)} P_\text{top} (F^N) &= 0 .
\end{align}
The above also implies that
\begin{align}
 \int_{\prt D^{d+2}} P_\text{top} (F^N) &= 0 ,
\end{align}
where $D^{d+2}$ is a $(d+2)$-dimensional disk.

To see the second  mistake, we note that $W^d_\text{top}(A)$ is only required
to be well defined when $A$ is deformable to $A=0$. In general, only the
difference $\int_{M^d} [W^d_\text{top}(\t A)-W^d_\text{top}(A)]$ is well
defined, and only up to a $2\pi$ phase. If we scale $W^d_\text{top}(A)$ by an
arbitrary real number, it will not be well defined.  In this case, we need to
use \eqn{diffWA} to define the difference.  More generally, if we want to define
the difference of the topological invariant on spaces with different geometry, we
need to generalize \eqn{diffWA} to
\begin{align}
\label{diffWAG}
\int_{\t M^d} W^d_\text{top}(\t A)- \int_{M^d} W^d_\text{top}(A)
& =\int_{N^{d+1}} P_\text{top}(F^N)
\nonumber\\
P_\text{top} (F^N) &= \dd  W^d_\text{top}(A), 
\end{align}
where $\prt N^{d+1}=\t
M^d \cup (-M^d)$ and the gauge connection $A^N$ on $ N^{d+1}$ satisfies that
on one boundary $-M^d$, $A^N=A$, and on the other boundary $\t
M^d$, $A^N=\t A$. 
In order for the above difference to be well defined, we require that
\begin{align}
\label{Pquan1}
 \int_{N^{d+1}} P_\text{top} (F^N) &= 0 \text{ mod } 1, 
\text{ for any closed } N^{d+1}.
\end{align}
where is a stronger quantization condition on $P_\text{top} (F^N)$.

Now, we would like to retell the above story in terms of classifying space, and
following \Ref{DW9093}, try to understand the different quantized topological
invariants $P_\text{top}(F)$ from the  classifying space point of view.\footnote{For
a simple introduction on classifying space, see the Wiki article ``Classifying
space''. For a continuous group $G$, the classifying space $BG$ in this paper
is defined with real manifold topology on $G$.}  We first note that all the
gauge configurations on ${N^{d+1}}$ can be understood through classifying space
$BG$ and universal bundles $EG$ (with a connection): all $G$-bundles on
${N^{d+1}}$ with all the possible connections can be obtained by choosing a
suitable map of ${N^{d+1}}$ into $BG$, $\gamma: {N^{d+1}} \to BG$.\cite{DW9093}
$BG$ is a very large space, often infinite dimensional.  If we pick a
connection in the universal bundle $EG$, even the different connections in the
same $G$-bundle on ${M^d}$ can  be obtained by different maps $\gamma$.
Therefore, we can express $P_\text{top}(F)$ as
\begin{align}
\int_{N^{d+1}} P_\text{top}(F)=Q^{d+1}_\text{top}(\ga)
\end{align}
We will further assume that we can express $P_\text{top}(F)$ as
\begin{align}
\int_{N^{d+1}} P_\text{top}(F)=Q^{d+1}_\text{top}(N^{d+1}_\ga)
\end{align}
where $N^{d+1}_\ga$ is the image of ${N^{d+1}}$ in the classifying space $BG$ under the map
$\ga$.  We will come back to this point later.  Here we use the superscript
$d+1$ to stress that $Q^{d+1}_\text{top}(\cdot)$ is function of
$(d+1)$-dimensional manifolds.

We see that once we specify a connection on $BG$, every map $\ga: {M^d}\to BG$
will define a connection $F$ on ${N^{d+1}}$. Thus we can view the function of
$N^{d+1}_\ga$, $Q^{d+1}_\text{top}(N^{d+1}_\ga)$, as a function of the
connection, $P_\text{top}(F)$.  Therefore, we can study the properties (such
the quantization condition) of gauge topological invariant $P_\text{top}(F)$ via the
function $Q^{d+1}_\text{top}(N^{d+1}_\ga)$ in the classifying space $BG$.

The function $Q^{d+1}_\text{top}({N^{d+1}})$ has the following defining
properties (see \eqn{Pz1}):
\begin{align}
& Q^{d+1}_\text{top}({N^{d+1}}) \in \R,
\nonumber\\
&
 Q^{d+1}_\text{top}(N^{d+1}_1)
+ Q^{d+1}_\text{top}(N^{d+1}_2)
=Q^{d+1}_\text{top}(N^{d+1}_1\cup N^{d+1}_2),
\nonumber\\
& Q^{d+1}_\text{top}({N^{d+1}}) = 0 , \text{ if } {N^{d+1}}= \prt (M^d\times D^2).
\end{align}
Here $N^{d+1}_1\cup N^{d+1}_2$ is an algebraic union of $N^{d+1}_1$ and
$N^{d+1}_2$.  For example, if $N^{d+1}_2$ is $N^{d+1}_1$ with an opposite
orientation, then $N^{d+1}_1\cup N^{d+1}_2 = \emptyset$.  (More precisely,
$N^{d+1}_1$ and $N^{d+1}_2$ should be viewed as chains, and $N^{d+1}_1\cup
N^{d+1}_2$ as the addition of chains in homological theory.) The above also
implies that
\begin{align}
Q^{d+1}_\text{top}({N^{d+1}}) = 0 , \text{ if } {N^{d+1}}= \prt (D^{d+2}).
\end{align}
Then using the additive property of $Q^{d+1}_\text{top}$, we can show that
\begin{align}
\label{Padd}
& Q^{d+1}_\text{top}({N^{d+1}}) \in \R,
\nonumber\\
&
 Q^{d+1}_\text{top}(N^{d+1}_1)
+ Q^{d+1}_\text{top}(N^{d+1}_2)
=Q^{d+1}_\text{top}(N^{d+1}_1\cup N^{d+1}_2),
\nonumber\\
& Q^{d+1}_\text{top}({N^{d+1}}) = 0 , \text{ if } {N^{d+1}}= \prt O^{d+2},
\end{align}
Also, from \eqn{Pquan1}, we obtain
\begin{align}
\label{Pquan}
& Q^{d+1}_\text{top}(N^{d+1}) = 0 \text{ mod } 1, \text{ if } \prt N^{d+1}= \emptyset.
\end{align}
From the condition \eqn{Padd}, we see that the function $Q^{d+1}_\text{top}({N^{d+1}})$
can be described by a cocycle $\om_{d+1} \in Z^{d+1}(BG,\R)$, where
$Z^{d+1}(BG,\R)$ is the space of all cocycles on the classifying space $BG$
with coefficient $\R$:
\begin{align}
 Q^{d+1}_\text{top}({N^{d+1}}) =\<\om_{d+1},{N^{d+1}}\>.
\end{align}
Certainly not every cocycle in
$C^{d+1}(BG,\R)$ satisfies the quantization condition \eqn{Pquan}.  Let us 
use $Z_\Z^{d+1}(BG,\R)$ to denote the set of cocycles that satisfy the
quantization condition  \eqn{Pquan},  
and use $B^{d+1}(BG,\R)$ to denote the set coboundaries.
Since the  coboundaries are all connected and represent local smooth changes of
the bosonic Lagrangian,
$Z_\Z^{d+1}(BG,\R)/B^{d+1}(BG,\R)$
describes the quantized topological invariants,
which are not smoothly connect to each other by the
local smooth changes of
the bosonic Lagrangian.
It turns out that
\begin{align}
\text{Free}[H^{d+1}(BG,\Z)]
 \equiv Z_\Z^{d+1}(BG,\R)/B^{d+1}(BG,\R).
\end{align}
Thus $\text{Free}[H^{d+1}(BG,\Z)]$ describes a set of the quantized potential
topological invariants.

But $\text{Free}[H^{d+1}(BG,\Z)]$ does not describe all the potential
topological invariants.  $\text{Free}[H^{d+1}(BG,\Z)]$ only describe a type of
topological invariants that change their value under a smooth change of the gauge
configuration $\int_{M^d} [W^d_\text{top}(A+\del A)-W^d_\text{top}(A)] \neq 0$.
We will call such type of topological invariants as Chern-Simons topological invariants.
However, there are another type of topological invariants that do not change under a
smooth change of the gauge configuration $\int_{M^d} [W^d_\text{top}(A+\del
A)-W^d_\text{top}(A)] = 0$.  We will call such type of topological invariants as
locally-null  topological invariants. The locally-null topological invariants correspond
to $P(F^N)=0$. So it is missed by our discussion above.  In the classifying
space approach, the locally-null topological invariants $\ee^{\int_{M^d}
W^d_\text{top}(A)}$ is described by cocycles in
$H^d(BG,\RZ)$.\cite{HW1267,HW1227,W1313}
However, $H^d(BG,\RZ)$ may contain continuous part, such as $\RZ$.
So the quantized potential locally-null topological invariants
are described by $\text{Dis}[H^d(BG,\RZ)]$, the discrete part of
$H^d(BG,\RZ)$.

This way, we show that \frm{the potential gauge topological invariants that cannot
connect to zero and cannot connect to each other are described by
$\text{Free}[H^{d+1}(BG,\Z)]\oplus \text{Dis}[H^d(BG,\RZ)]$.  Since
$\text{Dis}[H^d(BG,\RZ)]=\text{Tor}[ H^{d+1}(BG,\Z)]$, we may say that the
potential gauge topological invariants  are described by $H^{d+1}(BG,\Z)$.}

Since the different gauge transformation properties
\begin{align}
2\pi	\int_{M^d} W^d_\text{top}(A^g) - W^d_\text{top}(A) = 
\Big|_{A=0} \int_{M^d} L^d_\text{top}(g^{-1}\prt g)
\end{align}
are classified by group cohomology $\cH^d(G,\RZ)$ (where $ L^d_\text{top}(g)$ is
a cocycle in $\cH^d(G,\RZ)$) and since $H^{d+1}(BG,\Z)=\cH^d_B(G,\R)$ (see, for
example, \Ref{W1313}), we find that the L-type potential gauge topological
invariants coincide with the realizable L-type gauge topological invariants produced by
the NL$\si$M.  This means that \frm{the L-type potential gauge
topological invariants described by $H^{d+1}(BG,\Z)$ can all be produced by L-type
local bosonic models (\ie the NL$\si$Ms with fields in $G$), if
we ``gauge'' the symmetry $G$. In other words, all the L-type potential gauge
topological invariants described by $H^{d+1}(BG,\Z)$ are realizable by L-type local bosonic systems.}


For example, when $G=U(1)$, $H^{4}(BU(1),\Z)=\Z$, whose generator is $c_1^2$
with $c_1=\frac{1}{2\pi}F$  and $c_1^2 = \frac{1}{4\pi^2} F F=P_\text{top}(F)$.
The corresponding gauge topological invariant is 
$W^3_\text{top}(A)=\frac{1}{(2\pi)^2}A F$, where $F$ is the curvature two-form of the $U(1)$ connection one-form $A$.
Such a gauge topological invariant describes a $U(1)$ SPT state in $\cH^3[U(1),\RZ]$
in 2+1D.

\section{L-type potential gauge-gravity topological invariants}
\label{Pggterm}

In Section \ref{Rggterm}, we introduced L-type realizable gauge-gravity
topological invariants $W^d_\text{top}(A,\Ga)$.  In this section, we will discuss
the L-type potential gauge-gravity topological invariants $W^d_\text{top}(A,\Ga)$,
by repeating the discussion in Appendix \ref{agterm}.
We can use a $(d+1)$-form  $P_\text{top}$ to define difference of the potential
gauge-gravity topological invariant $W^d_\text{top}(A,\Ga)$:\cite{DW9093}
\begin{align}
&\ \ \ \
	\int_{\t M^d} W^d_\text{top}(\t A,\Ga)
-
\int_{ M^d} W^d_\text{top}( A,\Ga)
\\
&
= \int_{N^{d+1}} P_\text{top}(F^N,R^N),
\  \text{ with } \prt N^{d+1}=\t M^d \cup (-M^d).
\nonumber 
\end{align}
In the classifying space approach,
$P_\text{top}(F^N,R^N)$ is expressed as
\begin{align}
\int_{N^{d+1}_\ga} P_\text{top}(F^N,R^N) = Q^{d+1}_\text{top}(N^{d+1}_\ga)
\end{align}
where
$N^{d+1}_\ga$ is the image of the map $\ga: N^{d+1}\to B(G\times SO)$.
We find that $Q^{d+1}_\text{top}(N^{d+1}_\ga)$ satisfies
\begin{align}
\label{PSO}
& Q^{d+1}_\text{top}(N^{d+1}_\ga) \in \R,
\nonumber\\
&
 Q^{d+1}_\text{top}(N^{d+1}_\ga)
+ Q^{d+1}_\text{top}(\t N^{d+1}_\ga)
=Q^{d+1}_\text{top}(N^{d+1}_\ga\cup \t N^{d+1}_\ga),
\nonumber\\
& Q^{d+1}_\text{top}(N^{d+1}_\ga) = 0 , \text{ if } {N^{d+1}_\ga}= \prt O.
\nonumber \\
& Q^{d+1}_\text{top}(N^{d+1}_\ga) = 0 \text{ mod } 1, 
\ \ \ \text{ if } 
\prt N^{d+1}_\ga= \emptyset.
\end{align}
However, the quantization condition $Q^{d+1}_\text{top}(N^{d+1}_\ga) = 0 \text{
mod } 1$ is required only for a subset of cycles $N^{d+1}_\ga$ in $B(SO\times
G)$.  This is because for a closed $N^{d+1}$ with a given topology, a generic
map $\ga:N^{d+1}\to B(G\times SO)$ can give rise to an arbitrary $G\times SO$
principle bundle over $N^{d+1}$ (whose fiber is the $G\times SO$ group). The
corresponding $SO$ vector bundle (whose fiber is the vector space that forms
the fundamental representation of $SO$) may not be the tangent bundle over
$N^{d+1}$. Such a map is not allowed.  The quantization condition
$Q^{d+1}_\text{top}(N^{d+1}_\ga) = 0 \text{ mod } 1$ is required only for the
maps $\ga$ that give rise to the tangent bundle over $N^{d+1}$.  

Let $Z_G^{d+1}[B(G\times SO),\R]$ be the space of quantized cocycles,
and let $B_{d+1}[B(G\times SO),\R]$ be the space of coboundaries.
Then the potential gauge-gravity topological invariants are described by
\begin{align}
&\ \ \ \
H_G^{d+1}[B(G\times SO),\R]
\nonumber\\
&\equiv Z_G^{d+1}[B(G\times SO),\R]/B_{d+1}[B(G\times SO),\R]
.
\end{align}
Since the quantization condition is enforced only one a subset of
$(d+1)$-cycles $N^{d+1}_\ga$, $H_G^{d+1}[B(G\times SO),\R]$ may contain
unquantized continuous part $\R$.  It may also contain quantized discrete part
$\Z$. In other words,
\begin{align}
 H_G^{d+1}[B(G\times SO),\R]=
(\oplus_{i=1}^{n_\R^G}\R)
\oplus
(\oplus_{i=1}^{n_\Z^G}\Z)
\end{align}

We note that the cocycles in $H^{d+1}(BG,\Z)$ also satisfy all the 
conditions in \eqn{PSO}, thus we have a group homomorphism
\begin{align}
\label{HPi}
 \text{Free}[H^{d+1}[B(G\times SO),\Z]] \to H_G^{d+1}[B(G\times SO),\R].
\end{align}
The image of the map is formed by realizable gauge-gravity topological invariants.
We also note that there is another  group homomorphism
(an exact sequence)
\begin{align}
0\to \text{Dis}[H_G^{d+1}(BG,\R)]
 \to \text{Free}[H^{d+1}(B(G\times SO),\Z(\frac1n))]
\end{align}
for a certain $n$, where $\Z(\frac1n)$ is the fractional integer $\{0,\pm
\frac1n,\pm \frac2n,\cdots\}$.  This is because all unquantized cocycles
are dropped, and a quantized cocycle corresponds an element of
$H^{d+1}(B(G\times SO),\Z(\frac1n))$.  Also different quantized cocycles
correspond different elements of $H^{d+1}(B(G\times SO),\Z(\frac1n))$.  
If we write
\begin{align}
 \text{Free}[H^{d+1}(B(G\times SO),\Z)]=
(\oplus_{i=1}^{n_\Z}\Z),
\end{align}
we have 
\begin{align}
 n_\Z =n_\R^G+n_\Z^G.
\end{align}

Note that $H_G^{d+1}(BG,\R)$ only describe Chern-Simons gauge-gravity
topological invariants.  The locally-null  gauge-gravity topological invariants are
described by
\begin{align}
 H_G^d [B(G\times SO),\RZ] \equiv  \text{Dis}\Big(H^d [B(G\times SO),\RZ]/\La_G^d\Big),
\end{align}
where $\La_G^d$ is a subgroup of $H^d [B(G\times SO),\RZ]$ form by cocycles
$\om^d$ that satisfy
\begin{align}
 \<\om^d,N_\ga^d\> = 0,
\end{align}
where $N_\ga^d$ is all the close $d$-manifolds in $B(G\times SO)$ such that the
$SO$ bundle on $N_\ga^d$ is smoothly connected to the tangent bundle of
$N_\ga^d$. Since $\text{Dis}(H^d [B(G\times SO),\RZ])\simeq \text{Tor}(H^{d+1}
[B(G\times SO),\Z])$, we have
\begin{align}
H_G^d [B(G\times SO),\RZ] \subset
\text{Tor}( H^{d+1} [B(G\times SO),\Z]), 
\end{align}
that describes the locally-null potential gauge-gravity topological invariants.
Those locally-null gauge-gravity topological invariants
are all realizable.
We also have
\begin{align}
\text{Dis}(H_G^{d+1} [B(G\times SO),\R]) \subset
\text{Free}( H^{d+1} [B(G\times SO),\Z(\frac1n)]), 
\end{align}
that describes the Chern-Simons potential gauge-gravity topological invariants.  A
subset of those Chern-Simons gauge-gravity topological invariants that are also in
$\text{Free}( H^{d+1} [B(G\times SO),\Z])$ are realizable.


We like to remark that, in general, the image of the map \eq{HPi} is not the
whole $H_G^{d+1}(B(G\times SO),\R)$.  This means that some potential
gauge-gravity topological invariants cannot be generated from NL$\si$M
construction discussed in Section \ref{gterm}.  However, it is not clear if
there are some other bosonic path integrals that can generate the missing
potential topological invariants.

\section{The ring of $H^*(BSO,\Z)$}
\label{HBSOZ}

According to \Ref{B8283}, the ring $H^{*}(BSO,\Z)$ is a polynomial ring
generated by $p_i$ and $\bt(\rw_{2i_1}\rw_{2i_2}\cdots)$, $0<i_1<i_2<\cdots$,
with integer coefficients. Here $p_i\in H^{4i}(BSO,\Z)$ is the Pontryagin
class of dimension $4i$ and $\rw_i\in H^{i}(BSO,\Z_2)$ is the Stiefel-Whitney
class of dimension $i$.  Since
$\text{Tor}H^d(BG,\RZ)=\text{Tor}H^{d+1}(BG,\Z)$ (see, for example,
\Ref{W1313}), the natural map $H^d(BG,\Z_2)\to \text{Tor}H^d(BG,\RZ)$ induces
a natural map 
$H^d(BG,\Z_2)\to H^{d+1}(BG,\Z)$: $\bt: H^{i}(BSO,\Z_2)\to H^{i+1}(BSO,\Z)$.
Therefore, $\bt(\rw_{2i_1}\rw_{2i_2}\cdots)$ has a dimension
$1+2i_1+2i_2+\cdots$.  

More precisely, to obtain the ring $H^{*}(BSO,\Z)$ from a polynomial ring
generated by $p_i$ and $\bt(\rw_{2i_1}\rw_{2i_2}\cdots)$, we need to quotient
out certain equivalence relations:
\begin{align}
 H^{*}(BSO,\Z)=\Z[\{p_i\}, \{\bt (\rw_{2i_1}\rw_{2i_2}\cdots)\}]/ \sim ,
\end{align}
where the  equivalence relations $\sim$ contain
\begin{align}
& 2 \bt (\rw_{2i_1}\rw_{2i_2}\cdots) =0 ,
\\
& \bt  w(I) \bt  w(J) =
\nonumber\\
&\sum_{k\in I} \bt  \rw_{2k} \bt  \rw[(I-k)\cup J-(I-k)\cap J]p[(I-k)\cap J] ,
\nonumber 
\end{align}
where $I=\{i_1,i_2,\cdots\}$, $\rw(I)=\rw_{2i_1}\rw_{2i_2}\cdots$.  and
$p(I)=p_{i_1}p_{i_2}\cdots$.  Here we list all the second kind of the
equivalence relations for low dimensions
\begin{align}
 \bt  \rw_2 \bt  \rw_2 &= \bt  \rw_2 \bt  \rw_2,
\nonumber\\
 \bt  \rw_2 \bt  \rw_4 &= \bt  \rw_2 \bt  \rw_4,
\nonumber\\
 \bt  (\rw_2\rw_4) \bt  \rw_2 &= \bt  \rw_2 \bt  (\rw_2 \rw_4).
\end{align}
We see that those relations are identities (mod 2), and thus there are no
effective equivalence relations of the second kind for dimensions less than
12.  So for low dimensions, 
\begin{align}
\label{HBSOZs}
 H^1(BSO,\Z) &=0,
\nonumber \\
 H^2(BSO,\Z) &=0,
\nonumber\\
 H^3(BSO,\Z) &=\Z_2=\{m \bt  \rw_2\},
\nonumber\\
 H^4(BSO,\Z) &=\Z=\{np_1\},
\nonumber\\
 H^5(BSO,\Z) &=\Z_2=\{m \bt  \rw_4\},
\\
 H^6(BSO,\Z) &=\Z_2=\{m \bt  \rw_2\bt  \rw_2\},
\nonumber\\
 H^7(BSO,\Z) &=2\Z_2=\{ m_1\bt  \rw_6 + m_2 \bt  \rw_2 p_1 \},
\nonumber\\
 H^8(BSO,\Z) &=2\Z\oplus \Z_2=\{n_1p_1^2+n_2p_2+m \bt  \rw_2 \bt  \rw_4\},
\nonumber 
\end{align}
where $m$'s are in $Z_2$ and $n$'s in $\Z$.


Also, according to \Ref{B8283}, the ring $H^{*}(BO_n,\Z_2)$ 
is given by
\begin{align}
 H^{*}(BO_n,\Z_2)=\Z_2[\rw_1,\rw_2,\cdots].
\end{align}
For low dimensions, we find that
\begin{align}
 H^1(BO,\Z_2) &=\Z_2=\{m  \rw_1\},
\nonumber\\
 H^2(BO,\Z_2) &=2\Z_2=\{m_1  \rw_2+m_2\rw_1^2\},
\\
 H^3(BO,\Z_2) &=3\Z_2=\{m_1 \rw_3+m_2\rw_1\rw_2+m_3\rw_1^3\},
\nonumber\\
 H^4(BO,\Z_2) &=5\Z_2,
\nonumber\\
 H^5(BO,\Z_2) &=7\Z_2,
\nonumber\\
 H^6(BO,\Z_2) &=11\Z_2,
\nonumber 
\end{align}
where $m$'s are in $Z_2$.

\section{Calculate the generators in \eqn{HBSOU1} and \eqn{HO2RZ} from
\eqn{HBSOZ1} and \eqn{HO2Z}}
\label{calgen}

The basis in \eqn{HBSOU1} and \eqn{HO2RZ} give rise to the basis in
\eqn{HBSOZ1} and \eqn{HO2Z} after the one-to-one natural map $\t\bt$:
$\cH^d(G,\RZ)\to H^{d+1}(BG,\Z)$.
We also have a natural map
$\pi: \cH^d(G,\Z_2)\to \cH^d(G,\RZ)$, such as
$\pi \rw_i =\frac12 \rw_i$.
The Bockstein homomorphism 
$\bt: \cH^d(G,\Z_2)=H^d(BG,\Z_2)\to H^{d+1}(BG,\Z)$ is given by
$\bt=\t\bt\pi$, which is equal to the Steenrod
square $Sq^1$.  One can use the properties (see 
Section \ref{Rswc})
\begin{align}
\label{Sq1}
&
 Sq^1Sq^1=0, \ \ \
Sq^1(xy)=Sq^1(x)y+xSq^1(y), 
\nonumber\\
&
Sq^1(w_i)=\rw_1\rw_i+(i+1)\rw_{i+1}, \ \ \ 
Sq^1(w_2^2)=0;
\end{align}
for $x,y \in H^*(X,\Z_2)$ to compute the action of $Sq^1$.  


Let us first calculate the generators in \eqn{HBSOU1} from those in
\eqn{HBSOZ1}.  In 2-dimensional space-time $\cH^2(SO,\RZ)=H^3(BSO,\Z)=\Z_2$.
$H^3(BSO,\Z)$ is generated by the promoted 3-dimensional topological invariant
$K^3(\Ga) = Sq^1(\rw_2)=\rw_1\rw_2+\rw_3$.  $\cH^2(SO,\RZ)$ is generated by
$W^2_\text{top}(\Ga)$ which is the pull back of $K^3(\Ga) =\rw_1\rw_2+\rw_3$ by
the natural map $\t\bt: \cH^2(SO,\RZ) \to H^3(BSO,\Z)$:
\begin{align}
	\t\bt[W^d_\text{top}(\Ga)]=K^{d+1}(\Ga).
\end{align}
Using $\t\bt=Sq^1 \pi^{-1}$, we find that $W^2_\text{top}=\frac12 \rw_2$, since
$\pi^{-1} \frac12 \rw_2=\rw_2=$ and $Sq^1(\rw_2)=\rw_1\rw_2+\rw_3$.  
In 2+1D space-time, the corresponding $H^4(BSO,\Z)=\Z$ is generated by
$K^4(\Ga)=p_1$.  The pull back of the promoted generator $p_1$ by the natural
map $\t\bt: \cH^3(SO,\RZ) \to H^4(BSO,\Z)$ is the gauge-gravity topological
invariant $W^3_\text{top}=\om_3$.  
In 3+1D space-time, the corresponding $\cH^4(SO,\RZ)=\Z_2$ is generated by
the gauge-gravity topological invariant
$W^4_\text{top}=\frac12 \rw_4$, since
$\t\bt \frac12 \rw_4=Sq^1\pi^{-1} \frac12 \rw_4=Sq^1\rw_4=\bt\rw_4$.
In 4+1D space-time, the corresponding $\cH^5(SO,\RZ)=\Z_2$ is generated by the
gauge-gravity topological invariant $W^4_\text{top}=\frac12
\rw_2(\rw_1\rw_2+\rw_3)$.  This is because $\t\bt\frac12
\rw_2(\rw_1\rw_2+\rw_3)=Sq^1\pi^{-1}\frac12 \rw_2(\rw_1\rw_2+\rw_3)=Sq^1
\rw_2(\rw_1\rw_2+\rw_3)=Sq^1\rw_2Sq^1\rw_2=\bt\rw_2\bt\rw_2$, where we have
used \eqn{Sq1}.
In 5+1D space-time, again, using $\t\bt=Sq^1\pi^{-1}$ and \eqn{Sq1}, we can
show that the corresponding $\cH^6(SO,\RZ)=2\Z_2$ is generated by the
gauge-gravity topological invariant $W^4_\text{top}=\frac12 \rw_6, \frac12
\rw_2^3 $.  
Similarly, we can show that,
in 6+1D space-time, the corresponding $\cH^8(SO,\RZ)=2\Z\oplus \Z_2$ is
generated by the gauge-gravity topological invariant
$W^4_\text{top}=\om_7^{p_1^2}, \om_7^{p_2}, \frac12 (\rw_1\rw_2+\rw_3)\rw_4$.

We would like to remark that $\t\bt$ maps both $\frac12
(\rw_1\rw_2+\rw_3)\rw_4$ and $\frac12 \rw_2 (\rw_1\rw_4+\rw_5)$ in
$\cH^7(SO,\RZ)$ to the same $\bt\rw_2\bt\rw_4$ in $H^8(BSO,\Z)$, since
$\bt=Sq^1$ maps both $(\rw_1\rw_2+\rw_3)\rw_4$ and $\rw_2 (\rw_1\rw_4+\rw_5)$
in $\cH^7(SO,\Z_2)$ to the same $\bt\rw_2\bt\rw_4$.  Since both $\bt$ and $\pi$
are many-to-one maps, the above fact does not contradict with the facts that
$\t\bt$ is an one-to-one map and $\bt=\t\bt\pi$.  Although
$(\rw_1\rw_2+\rw_3)\rw_4$ and $\rw_2 (\rw_1\rw_4+\rw_5)$ are different cocycles
in $\cH^7(SO,\Z_2)$, their images under $\pi$, $\frac12
(\rw_1\rw_2+\rw_3)\rw_4$ and $\frac12 \rw_2 (\rw_1\rw_4+\rw_5)$, belong to the
same cocycle in $\cH^7(SO,\RZ)$ (\ie differ only by a coboundary).

Using a similar approach, we can calculate the generators in \eqn{HO2RZ} from
those in \eqn{HO2Z}. For example, we can show $\t\bt \frac12
(\rw_1^{O_2})^3(\rw_2^{O_2})^2=(\bt\rw_1^{O_2})^2p_1^{O_2}$.  This is because
$\t\bt \frac12 (\rw_1^{O_2})^3(\rw_2^{O_2})^2=Sq^1[
(\rw_1^{O_2})^3(\rw_2^{O_2})^2]=Sq^1[ (\rw_1^{O_2})^3](\rw_2^{O_2})^2=Sq^1[
(\rw_1^{O_2})^3]p_1^{O_2}$, where we have used $(\rw_2^{O_2})^2=p_1^{O_2}$ mod
2.  Then using $Sq^1[(\rw_1^{O_2})^3]=(Sq^1\rw_1^{O_2})^2=(\bt\rw_1^{O_2})^2$,
we find  $\t\bt \frac12
(\rw_1^{O_2})^3(\rw_2^{O_2})^2=(\bt\rw_1^{O_2})^2p_1^{O_2}$, where we have used
$Sq^1\rw_1^{O_2}=(\rw_1^{O_2})^2$.

\section{Relation between Pontryagin classes and Stiefel-Whitney classes}
\label{PandSW}


There is result due to Wu that relate  Pontryagin classes and Stiefel-Whitney classes (see \Ref{T6067} Theorem C):\\
Let $B$ be a vector bundle over a manifold $X$, $\rw_i$ be its Stiefel-Whitney
classes and $p_i$ its Pontryagin classes. Let $\rho_4$ be the reduction modulo
4 and $\theta_2$ be the embedding of $\mathbb{Z}_2$ into $\mathbb{Z}_4$ (as
well as their induced actions on cohomology groups). Then
\begin{align}
\cP_2(\rw_{2i}) = \rho_4(p_i) + \theta_2 \Big( \rw_1 Sq^{2i-1} \rw_{2i} + \sum_{j = 0}^{i-1} \rw_{2j} \rw_{4i-2j} \Big),
\end{align}
where $\cP_2$ is the Pontryagin square,\cite{PS} which maps $x\in
H^{2n}(X,\Z_2)$ to $\cP_2(x) \in H^{4n}(X,\Z_4)$.  Let $\rho_2$ be the
reduction modulo 2. The Pontryagin square has a property that $\rho_2
\cP_2(x)=x^2$.
Therefore
\begin{align}
\rho_2 \cP_2(\rw_{2i})=\rw_{2i}^2 = \rho_2(p_i).
\end{align}

\section{The K\"unneth formula} \label{HBGRZ}

The K\"unneth formula is a very helpful formula that allows us to calculate the
cohomology of chain complex $X\times X'$ in terms of  the cohomology of chain
complex $X$ and chain complex $X'$.
The K\"unneth formula is expressed in terms of
the tensor-product operation $\otimes_R$ and the torsion-product operation
$\boxtimes_R\equiv \text{Tor}_1^R$, which have the following properties:
\begin{align}
\label{tnprd}
& \M \otimes_\Z \M' \simeq \M' \otimes_\Z \M ,
\nonumber\\
& \Z \otimes_\Z \M \simeq \M \otimes_\Z \Z =\M ,
\nonumber\\
& \Z_n \otimes_\Z \M \simeq \M \otimes_\Z \Z_n = \M/n\M ,
\nonumber\\
& \Z_n \otimes_\Z \RZ \simeq \RZ \otimes_\Z \Z_n = 0,
\nonumber\\
& \Z_m \otimes_\Z \Z_n  =\Z_{\<m,n\>} ,
\nonumber\\
& \RZ \otimes_\Z \RZ = 0,
\nonumber\\
& \R \otimes_\Z \RZ = 0,
\nonumber\\
& \R \otimes_\Z \R = \R,
\nonumber\\
&  (\M'\oplus \M'')\otimes_R \M = (\M' \otimes_R \M)\oplus (\M'' \otimes_R \M)   ,
\nonumber\\
& \M \otimes_R (\M'\oplus \M'') = (\M \otimes_R \M')\oplus (\M \otimes_R \M'')   ;
\end{align}
and
\begin{align}
\label{trprd}
& \text{Tor}_1^R(\M,\M') \equiv \M\boxtimes_R \M'  ,
\nonumber\\
& \M\boxtimes_R \M' \simeq \M'\boxtimes_R \M  ,
\nonumber\\
& \Z\boxtimes_\Z  \M = \M\boxtimes_\Z  \Z = 0,
\nonumber\\
& \Z_n\boxtimes_\Z \M = \{m\in \M| nm=0\},
\nonumber\\
& \Z_n\boxtimes_\Z \RZ = \Z_n,
\nonumber\\
& \Z_m\boxtimes_\Z \Z_n = \Z_{\<m,n\>} ,
\nonumber\\
& \M'\oplus \M''\boxtimes_R\M = \M'\boxtimes_R \M\oplus\M''\boxtimes_R \M,
\nonumber\\
& \M\boxtimes_R\M'\oplus \M'' = \M\boxtimes_R\M'\oplus\M\boxtimes_RB
,
\end{align}
where $\<m,n\>$ is the greatest common divisor of $m$ and $n$.  These
expressions allow us to compute the tensor-product $\otimes_R$ and  the
torsion-product $\boxtimes_R$.  Here $R$ is a ring and $\M,\M',\M''$ are
$R$-modules.  A $R$-module is like a vector space over $R$ (\ie we can
``multiply'' a vector by an element of $R$.)

The K\"unneth formula itself is given by (see
\Ref{Spa66} page 247)
\begin{align}
\label{kunn}
&\ \ \ \ H^d(X\times X',\M\otimes_R \M')
\nonumber\\
&\simeq \Big[\oplus_{k=0}^d H^k(X,\M)\otimes_R H^{d-k}(X',\M')\Big]\oplus
\nonumber\\
&\ \ \ \ \ \
\Big[\oplus_{k=0}^{d+1}
H^k(X,\M)\boxtimes_R H^{d-k+1}(X',\M')\Big]  .
\end{align}
Here $R$ is a principle ideal domain and $\M,\M'$ are $R$-modules such that
$\M\boxtimes_R \M'=0$.  
We also require either\\
(1) $H_d(X,\Z)$ and  $H_d(X',\Z)$ are finitely generated, or\\
(2) $\M'$ and $H_d(X',\Z)$ are
finitely generated.\\
For example, $\M'=\Z\oplus \cdots \oplus \Z\oplus \Z_n\oplus\cdots\oplus\Z_m$.

For more details on principal ideal domain and $R$-module, see the
corresponding Wiki articles.  Note that $\Z$ and $\R$ are principal ideal
domains, while $\R/\Z$ is not.  Also, $\R$ and $\R/\Z$ are not finitely
generate $R$-modules if $R=\Z$.  The K\"unneth formula works for topological
cohomology where $X$ and $X'$ are treated as topological spaces.  But, it does
not work for group cohomology by treating  $H^d$ as $\cH^d$ and $X$ and $X'$ as
groups, $X=G$ and $X'=G'$, as demonstrated by the example $\M=\M'=\RZ$ and
$X=X'=Z_n$.  However, since $\cH^d(G,\Z)=H^d(BG,\Z)$,
the above K\"unneth formula works for group cohomology when $\M=\M'=\Z$.  The
above K\"unneth formula also works for group cohomology when $G,G'$ are finite
or when $G'$ is finite and $\M'$ is finitely generate (such as $\M'$ is $\Z$ or
$\Z_n$).



As the first application of K\"unneth formula, we like to use it to calculate
$H^*(X',\M)$ from $H^*(X',\Z)$,  by choosing $R=\M'=\Z$. In this case, the
condition $\M\boxtimes_R\M'=\M\boxtimes_\Z \Z=0$ is always
satisfied. $\M$ can be $\R/\Z$, $\Z$, $\Z_n$ \etc. So we have
\begin{align}
\label{kunnZ}
&\ \ \ \ H^d(X\times X',\M)
\nonumber\\
&\simeq \Big[\oplus_{k=0}^d H^k(X,\M)\otimes_{\Z} H^{d-k}(X',\Z)\Big]\oplus
\nonumber\\
&\ \ \ \ \ \
\Big[\oplus_{k=0}^{d+1}
H^k(X,{\M})\boxtimes_\Z H^{d-k+1}(X',\Z)\Big]  .
\end{align}
The above is valid for topological cohomology.
It is also valid for group  cohomology:
\begin{align}
\label{kunnZG1}
&\ \ \ \ \cH^d(G\times G',\M)
\nonumber\\
&\simeq \Big[\oplus_{k=0}^d \cH^k(G,\M)\otimes_{\Z} \cH^{d-k}(G',\Z)\Big]\oplus
\nonumber\\
&\ \ \ \ \ \
\Big[\oplus_{k=0}^{d+1}
\cH^k(G,\M)\boxtimes_\Z \cH^{d-k+1}(G',\Z)\Big]  .
\end{align}
provided that $G'$ is a finite group.
Using \eqn{HdR}, we can rewrite the above as
\begin{align}
\label{kunnZG2}
&\ \ \ \ \cH^d(G\times G',\M)
\simeq
\cH^d(G,\M)\oplus
\nonumber\\
&\ \ \ \ \ \
\Big[\oplus_{k=0}^{d-2} \cH^k(G,\M)\otimes_{\Z} \cH^{d-k-1}(G',\RZ)\Big]\oplus
\nonumber\\
&\ \ \ \ \ \
\Big[\oplus_{k=0}^{d-1}
\cH^k(G,\M)\boxtimes_\Z \cH^{d-k}(G',\RZ)\Big]  ,
\end{align}
where we have used
\begin{align}
 \cH^1(G',\Z)=0.
\end{align}
If we further choose $\M=\RZ$, we obtain
\begin{align}
\label{kunnZG}
&\ \ \ \ \cH^d(G\times G',\RZ)
\nonumber\\
&\simeq
\cH^d(G,\RZ)\oplus
\cH^d(G',\RZ)\oplus
\nonumber\\
&\ \ \ \ \ \
\Big[\oplus_{k=1}^{d-2} \cH^k(G,\RZ)\otimes_{\Z} \cH^{d-k-1}(G',\RZ)\Big]\oplus
\nonumber\\
&\ \ \ \ \ \
\Big[\oplus_{k=1}^{d-1}
\cH^k(G,\RZ)\boxtimes_\Z \cH^{d-k}(G',\RZ)\Big]  ,
\end{align}
where $G'$ is a finite group.

We can further choose $X$ to be the space of one point (or the trivial group) in \eqn{kunnZ},
and use
\begin{align}
H^{d}(X,\M))=
\begin{cases}
\M, & \text{ if } d=0,\\
0, & \text{ if } d>0,
\end{cases}
\end{align}
to reduce \eqn{kunnZ} to
\begin{align}
\label{ucf}
 H^d(X,\M)
&\simeq  \M \otimes_{\Z} H^d(X,\Z)
\oplus
\M\boxtimes_\Z H^{d+1}(X,\Z)  .
\end{align}
where $X'$ is renamed as $X$.  The above is a form of the universal coefficient
theorem which can be used to calculate $H^*(X,\M)$ from $H^*(X,\Z)$ and the
module $\M$.  The  universal coefficient theorem works for topological
cohomology where $X$ is a topological space.  The  universal coefficient
theorem also works for group cohomology when $X$ is a finite group.

Using the universal
coefficient theorem, we can rewrite \eqn{kunnZ} as
\begin{align}
\label{kunnHX}
H^d(X\times X',\M) \simeq \oplus_{k=0}^d H^k[X, H^{d-k}(X',\M)] .
\end{align}
The above is valid for topological cohomology.
It is also valid for group  cohomology:
\begin{align}
\label{kunnG}
\cH^d(G\times G',\M) \simeq \oplus_{k=0}^d \cH^k[G, \cH^{d-k}(G',\M)] ,
\end{align}
provided that both $G$ and $G'$ are finite groups.

We may apply the above to the classifying
spaces of group $G$ and $G'$. Using
$B(G\times G')=BG\times BG'$, we find
\begin{align*}
H^d[B(G\times G'),\M] \simeq \oplus_{k=0}^d H^k[BG, H^{d-k}(BG',\M)] .
\end{align*}
Choosing $\M=\RZ$ and using 
\begin{align}
\label{HdR}
& \cH^d(G,\R/\Z)=\cH^{d+1}(G,\Z)=H^{d+1}(BG,\Z),
\end{align}
we have
\begin{align}
\label{kunnU}
&\ \ \ \
 \cH_B^d(G\times G',\RZ)=
 H^{d+1}[B(G\times G'),\Z]
\nonumber\\
&= \oplus_{k=0}^{d+1} H^k[BG, H^{d+1-k}(BG',\Z)]
\nonumber\\
&=\cH_B^d(G,\RZ)\oplus \cH_B^d(G',\RZ)\oplus 
\nonumber\\
&\ \ \ \ \ \ \ \
\oplus_{k=1}^{d-1} H^k[BG, \cH_B^{d-k}(G',\RZ)]
\end{align}
where we have used $H^0(BG',\Z)=\Z$, 
$\cH^1_B(G',\Z)=H^1(BG',\Z)$,
and $\cH^1_B(G',\Z)=0$ if $G'$ is compact (or finite).
Eqn.  \ref{kunnU} is valid for any groups $G$ and $G'$.
If $G$ also satisfies (for example when $G$ is finite)
\begin{align}
\label{HdcHdZ}
 H^d(BG,\Z)=\cH_B^d(G,\Z),\ \ \ \
 H^d(BG,\Z_n)=\cH_B^d(G,\Z_n),
\end{align}
we can rewrite the above as
\begin{align}
\label{kunnUfinite}
\cH^d(G  \times G',\RZ) 
 = \oplus_{k=0}^{d} \cH^{k}[G,\cH^{d-k}(G',\RZ)]
.
\end{align}
Such a result is consistent with \eqn{Lyndon} for arbitrary $G,G'$.

Choosing $X=BG$, $\M=\Z_n$, \eqn{ucf} becomes
\begin{align}
\label{ucfG}
 \cH^d(G,\Z_n)
&\simeq  \Z_n \otimes_{\Z} \cH^d(G,\Z)
\oplus
\Z_n\boxtimes_\Z \cH^{d+1}(G,\Z)  ,
\end{align}
where we have used \eqn{HdcHdZ}.
Using \eqn{ucfG}, we find that
\begin{align}
\label{ucfGG}
 \cH^d[G,&\cH^{d'}(G',\R/\Z)]
\simeq  \cH^{d'}(G',\R/\Z) \otimes_{\Z} \cH^{d-1}(G,\R/\Z)
\oplus
\nonumber\\
&
\cH^{d'}(G',\R/\Z)\boxtimes_\Z \cH^{d}(G,\R/\Z)  ,
\end{align}

\section{Lyndon-Hochschild-Serre spectral sequence}
\label{LHS}

The Lyndon-Hochschild-Serre spectral sequence (see \Ref{L4871} page 280,291,
and \Ref{HS5310}) allows us to understand the structure of $\cH^d(G,\RZ)$ to a certain degree. (Here $G$ is a group extension
of $SG$ by $GG$: $SG=G/GG$.) We find that $\cH^d(G,\M)$, when
viewed as an Abelian group, contains a chain of subgroups
\begin{align}
\label{Lyndon}
\{0\}=H_{d+1}
\subset H_d
\subset \cdots
\subset H_0
=
 \cH^d(G,\M)
\end{align}
such that $H_k/H_{k+1}$ is a subgroup of a factor
group of $\cH^k[SG,\cH^{d-k}(GG,\M)_{SG}]$,
\ie $\cH^k[SG,\cH^{d-k}(GG,\M)_{SG}]$
contains a   subgroup $\Ga^k$, such that
\begin{align}
 H_k/H_{k+1} &\subset \cH^k[SG,\cH^{d-k}(GG,\RZ)_{SG}]/\Ga^k,
\nonumber\\
k&=0,\cdots,d.
\end{align}
Note that $G$  may have a non-trivial action on $\M$ and $SG$ may have a
non-trivial action on $\cH^{d-k}(GG,\M)$ as determined by the structure $1\to
GG \to GG\gext SG \to SG \to 1$.  We add the subscript $SG$ to
$\cH^{d-k}(GG,\RZ)$ to stress this point.  We also have
\begin{align}
 H_0/H_{1} &\subset \cH^0[SG,\cH^{d}(GG,\RZ)_{SG}],
\nonumber\\
 H_d/H_{d+1}&=H_d = \cH^d(SG,\RZ)/\Ga^d.
\end{align}
In other words, all the elements in $\cH^d(GG\gext SG,\RZ)$ can be one-to-one
labeled by $(x_0,x_1,\cdots,x_d)$ with
\begin{align}
 x_k\in H_k/H_{k+1} \subset \cH^k[SG,\cH^{d-k}(GG,\RZ)_{SG}]/\Ga^k.
\end{align}

Note that here $\M$ can be $\Z,\Z_n,\R,\R/\Z$ \etc.  Let $x_{k,\al}$,
$\al=1,2,\cdots$, be the generators of $H^k/H^{k+1}$. Then we say $x_{k,\al}$
for all $k,\al$ are the generators of $\cH^d(G,\M)$.  We also call
$H_k/H_{k+1}$, $k=0,\cdots,d$, the generating sub-factor groups of
$\cH^d(G,\M)$.


The above result implies that we can use $(m_0,m_1,\cdots,m_d)$ with
$ m_k\in \cH^k[SG,\cH^{d-k}(GG,\RZ)_{SG}] $
to  label all the elements in $\cH^d(G,\RZ)$. However, such a labeling scheme
may not be one-to-one, and it may happen that only some of
$(m_0,m_1,\cdots,m_d)$ correspond to  the  elements in $\cH^d(G,\RZ)$.  But, on
the other hand, for every element in $\cH^d(G,\RZ)$, we can find a
$(m_0,m_1,\cdots,m_d)$ that corresponds to it.


\section{Generators of $H^k(BG,\si\iTO_L^3)$.}
\label{giTO3}

The Abelian group $H^k(BG,\si\iTO_L^3)$ is generated by
$W^k_\text{top}(A,\Ga)/2\pi = x \om_3$ where $x$ are the generators of
$H^k(BG,\Z)$.  Since $\si\iTO_L^3 =\cH^3(SO,\RZ)=H^4(BSO,\Z)$ and since
$H^4(BSO,\Z)$ is generated by the first Pontryagin class $p_1$, we may also say
that $H^k(BG,\si\iTO_L^3)$ is generated by $x p_1$ in $H^k[BG,H^4(BSO,\Z)]$.  We
also know that $H^k(BG,\Z) \simeq \cH^{k-1}(G,\RZ)$, thus we can further say
that \frm{$H^k(BG,\si\iTO_L^3)$ is generated by $ap_1$ in $H^{k+3}[B(G\times
SO),\RZ]$ where $a$ are the generators of $\cH^{k-1}(G,\RZ)$ and
$\bt(ap_1)=xp_1$ under the natural map $\bt: H^{k+3}[B(G\times SO),\RZ]\to
H^{k+4}[B(G\times SO),\Z]$.} For example, when $k=2$ and $G=U(1)$,
$H^2[BU(1),\si\iTO_L^3]$ is generated by $W^5_\text{top}(A,\Ga)=c_1\om_3$.  W
can also say that it is generated by $W^5_\text{top}(A,\Ga)=a p_1$ where
$\dd a =c_1$ and $a$ generates $\cH^1[U(1),\RZ]$.
Thus
\begin{align}
 c_1\om_3 = \bt(a) \om_3=-a  \dd \om_3 = - a p_1.
\end{align}
We see that the natural map $\bt: \cH^k(G,\RZ)\to \cH^{k+1}(BG,\Z)$ behave like
a derivative $\dd$.  Similarly
\begin{align}
 \bt(a_1)\om_3&=-\frac 12 a_1 \bt(\om_3) = -\frac 12 a_1 p_1.
\end{align}
Note that when acting on the cocycles with $\Z_2$ coefficient, $\bt = Sq^1$.
\\[-4mm]

\bibliography{../../bib/wencross,../../bib/all,../../bib/publst,./local} 

\begin{thebibliography}{104}%
\makeatletter
\providecommand \@ifxundefined [1]{%
 \@ifx{#1\undefined}
}%
\providecommand \@ifnum [1]{%
 \ifnum #1\expandafter \@firstoftwo
 \else \expandafter \@secondoftwo
 \fi
}%
\providecommand \@ifx [1]{%
 \ifx #1\expandafter \@firstoftwo
 \else \expandafter \@secondoftwo
 \fi
}%
\providecommand \natexlab [1]{#1}%
\providecommand \enquote  [1]{``#1''}%
\providecommand \bibnamefont  [1]{#1}%
\providecommand \bibfnamefont [1]{#1}%
\providecommand \citenamefont [1]{#1}%
\providecommand \href@noop [0]{\@secondoftwo}%
\providecommand \href [0]{\begingroup \@sanitize@url \@href}%
\providecommand \@href[1]{\@@startlink{#1}\@@href}%
\providecommand \@@href[1]{\endgroup#1\@@endlink}%
\providecommand \@sanitize@url [0]{\catcode `\\12\catcode `\$12\catcode
  `\&12\catcode `\#12\catcode `\^12\catcode `\_12\catcode `\%12\relax}%
\providecommand \@@startlink[1]{}%
\providecommand \@@endlink[0]{}%
\providecommand \url  [0]{\begingroup\@sanitize@url \@url }%
\providecommand \@url [1]{\endgroup\@href {#1}{\urlprefix }}%
\providecommand \urlprefix  [0]{URL }%
\providecommand \Eprint [0]{\href }%
\providecommand \doibase [0]{http://dx.doi.org/}%
\providecommand \selectlanguage [0]{\@gobble}%
\providecommand \bibinfo  [0]{\@secondoftwo}%
\providecommand \bibfield  [0]{\@secondoftwo}%
\providecommand \translation [1]{[#1]}%
\providecommand \BibitemOpen [0]{}%
\providecommand \bibitemStop [0]{}%
\providecommand \bibitemNoStop [0]{.\EOS\space}%
\providecommand \EOS [0]{\spacefactor3000\relax}%
\providecommand \BibitemShut  [1]{\csname bibitem#1\endcsname}%
\let\auto@bib@innerbib\@empty
\bibitem [{\citenamefont {Haldane}(1983)}]{H8364}%
  \BibitemOpen
  \bibfield  {author} {\bibinfo {author} {\bibfnamefont {F.~D.~M.}\
  \bibnamefont {Haldane}},\ }\href@noop {} {\bibfield  {journal} {\bibinfo
  {journal} {Physics Letters A}\ }\textbf {\bibinfo {volume} {93}},\ \bibinfo
  {pages} {464} (\bibinfo {year} {1983})}\BibitemShut {NoStop}%
\bibitem [{\citenamefont {Gu}\ and\ \citenamefont {Wen}(2009)}]{GW0931}%
  \BibitemOpen
  \bibfield  {author} {\bibinfo {author} {\bibfnamefont {Z.-C.}\ \bibnamefont
  {Gu}}\ and\ \bibinfo {author} {\bibfnamefont {X.-G.}\ \bibnamefont {Wen}},\
  }\href@noop {} {\bibfield  {journal} {\bibinfo  {journal} {Phys. Rev. B}\
  }\textbf {\bibinfo {volume} {80}},\ \bibinfo {pages} {155131} (\bibinfo
  {year} {2009})},\ \Eprint {http://arxiv.org/abs/arXiv:0903.1069}
  {arXiv:0903.1069} \BibitemShut {NoStop}%
\bibitem [{\citenamefont {Chen}\ \emph {et~al.}(2010)\citenamefont {Chen},
  \citenamefont {Gu},\ and\ \citenamefont {Wen}}]{CGW1038}%
  \BibitemOpen
  \bibfield  {author} {\bibinfo {author} {\bibfnamefont {X.}~\bibnamefont
  {Chen}}, \bibinfo {author} {\bibfnamefont {Z.-C.}\ \bibnamefont {Gu}}, \ and\
  \bibinfo {author} {\bibfnamefont {X.-G.}\ \bibnamefont {Wen}},\ }\href@noop
  {} {\bibfield  {journal} {\bibinfo  {journal} {Phys. Rev. B}\ }\textbf
  {\bibinfo {volume} {82}},\ \bibinfo {pages} {155138} (\bibinfo {year}
  {2010})},\ \Eprint {http://arxiv.org/abs/arXiv:1004.3835} {arXiv:1004.3835}
  \BibitemShut {NoStop}%
\bibitem [{\citenamefont {Chen}\ \emph
  {et~al.}(2011{\natexlab{a}})\citenamefont {Chen}, \citenamefont {Gu},\ and\
  \citenamefont {Wen}}]{CGW1107}%
  \BibitemOpen
  \bibfield  {author} {\bibinfo {author} {\bibfnamefont {X.}~\bibnamefont
  {Chen}}, \bibinfo {author} {\bibfnamefont {Z.-C.}\ \bibnamefont {Gu}}, \ and\
  \bibinfo {author} {\bibfnamefont {X.-G.}\ \bibnamefont {Wen}},\ }\href@noop
  {} {\bibfield  {journal} {\bibinfo  {journal} {Phys. Rev. B}\ }\textbf
  {\bibinfo {volume} {83}},\ \bibinfo {pages} {035107} (\bibinfo {year}
  {2011}{\natexlab{a}})},\ \Eprint {http://arxiv.org/abs/arXiv:1008.3745}
  {arXiv:1008.3745} \BibitemShut {NoStop}%
\bibitem [{\citenamefont {Chen}\ \emph
  {et~al.}(2011{\natexlab{b}})\citenamefont {Chen}, \citenamefont {Gu},\ and\
  \citenamefont {Wen}}]{CGW1128}%
  \BibitemOpen
  \bibfield  {author} {\bibinfo {author} {\bibfnamefont {X.}~\bibnamefont
  {Chen}}, \bibinfo {author} {\bibfnamefont {Z.-C.}\ \bibnamefont {Gu}}, \ and\
  \bibinfo {author} {\bibfnamefont {X.-G.}\ \bibnamefont {Wen}},\ }\href@noop
  {} {\bibfield  {journal} {\bibinfo  {journal} {Phys. Rev. B}\ }\textbf
  {\bibinfo {volume} {84}},\ \bibinfo {pages} {235128} (\bibinfo {year}
  {2011}{\natexlab{b}})},\ \Eprint {http://arxiv.org/abs/arXiv:1103.3323}
  {arXiv:1103.3323} \BibitemShut {NoStop}%
\bibitem [{\citenamefont {Schuch}\ \emph {et~al.}(2011)\citenamefont {Schuch},
  \citenamefont {Perez-Garcia},\ and\ \citenamefont {Cirac}}]{SPC1139}%
  \BibitemOpen
  \bibfield  {author} {\bibinfo {author} {\bibfnamefont {N.}~\bibnamefont
  {Schuch}}, \bibinfo {author} {\bibfnamefont {D.}~\bibnamefont
  {Perez-Garcia}}, \ and\ \bibinfo {author} {\bibfnamefont {I.}~\bibnamefont
  {Cirac}},\ }\href@noop {} {\bibfield  {journal} {\bibinfo  {journal} {Phys.
  Rev. B}\ }\textbf {\bibinfo {volume} {84}},\ \bibinfo {pages} {165139}
  (\bibinfo {year} {2011})},\ \Eprint {http://arxiv.org/abs/arXiv:1010.3732}
  {arXiv:1010.3732} \BibitemShut {NoStop}%
\bibitem [{\citenamefont {Pollmann}\ \emph {et~al.}(2010)\citenamefont
  {Pollmann}, \citenamefont {Berg}, \citenamefont {Turner},\ and\ \citenamefont
  {Oshikawa}}]{PBT1039}%
  \BibitemOpen
  \bibfield  {author} {\bibinfo {author} {\bibfnamefont {F.}~\bibnamefont
  {Pollmann}}, \bibinfo {author} {\bibfnamefont {E.}~\bibnamefont {Berg}},
  \bibinfo {author} {\bibfnamefont {A.~M.}\ \bibnamefont {Turner}}, \ and\
  \bibinfo {author} {\bibfnamefont {M.}~\bibnamefont {Oshikawa}},\ }\href
  {\doibase 10.1103/PhysRevB.81.064439} {\bibfield  {journal} {\bibinfo
  {journal} {Phys. Rev. B}\ }\textbf {\bibinfo {volume} {81}},\ \bibinfo
  {pages} {064439} (\bibinfo {year} {2010})},\ \Eprint
  {http://arxiv.org/abs/arXiv:0910.1811} {arXiv:0910.1811} \BibitemShut
  {NoStop}%
\bibitem [{\citenamefont {Fidkowski}\ and\ \citenamefont
  {Kitaev}(2010)}]{FK1009}%
  \BibitemOpen
  \bibfield  {author} {\bibinfo {author} {\bibfnamefont {L.}~\bibnamefont
  {Fidkowski}}\ and\ \bibinfo {author} {\bibfnamefont {A.}~\bibnamefont
  {Kitaev}},\ }\href@noop {} {\bibfield  {journal} {\bibinfo  {journal} {Phys.
  Rev. B}\ }\textbf {\bibinfo {volume} {81}},\ \bibinfo {pages} {134509}
  (\bibinfo {year} {2010})},\ \Eprint {http://arxiv.org/abs/arXiv:0904.2197}
  {arXiv:0904.2197} \BibitemShut {NoStop}%
\bibitem [{\citenamefont {Turner}\ \emph {et~al.}(2011)\citenamefont {Turner},
  \citenamefont {Pollmann},\ and\ \citenamefont {Berg}}]{TPB1102}%
  \BibitemOpen
  \bibfield  {author} {\bibinfo {author} {\bibfnamefont {A.~M.}\ \bibnamefont
  {Turner}}, \bibinfo {author} {\bibfnamefont {F.}~\bibnamefont {Pollmann}}, \
  and\ \bibinfo {author} {\bibfnamefont {E.}~\bibnamefont {Berg}},\ }\href@noop
  {} {\bibfield  {journal} {\bibinfo  {journal} {Phys. Rev. B}\ }\textbf
  {\bibinfo {volume} {83}},\ \bibinfo {pages} {075102} (\bibinfo {year}
  {2011})},\ \Eprint {http://arxiv.org/abs/arXiv:1008.4346} {arXiv:1008.4346}
  \BibitemShut {NoStop}%
\bibitem [{\citenamefont {Fidkowski}\ and\ \citenamefont
  {Kitaev}(2011)}]{FK1103}%
  \BibitemOpen
  \bibfield  {author} {\bibinfo {author} {\bibfnamefont {L.}~\bibnamefont
  {Fidkowski}}\ and\ \bibinfo {author} {\bibfnamefont {A.}~\bibnamefont
  {Kitaev}},\ }\href@noop {} {\bibfield  {journal} {\bibinfo  {journal} {Phys.
  Rev. B}\ }\textbf {\bibinfo {volume} {83}},\ \bibinfo {pages} {075103}
  (\bibinfo {year} {2011})},\ \Eprint {http://arxiv.org/abs/arXiv:1008.4138}
  {arXiv:1008.4138} \BibitemShut {NoStop}%
\bibitem [{\citenamefont {Pollmann}\ \emph {et~al.}(2012)\citenamefont
  {Pollmann}, \citenamefont {Berg}, \citenamefont {Turner},\ and\ \citenamefont
  {Oshikawa}}]{PBT1225}%
  \BibitemOpen
  \bibfield  {author} {\bibinfo {author} {\bibfnamefont {F.}~\bibnamefont
  {Pollmann}}, \bibinfo {author} {\bibfnamefont {E.}~\bibnamefont {Berg}},
  \bibinfo {author} {\bibfnamefont {A.~M.}\ \bibnamefont {Turner}}, \ and\
  \bibinfo {author} {\bibfnamefont {M.}~\bibnamefont {Oshikawa}},\ }\href
  {\doibase 10.1103/PhysRevB.85.075125} {\bibfield  {journal} {\bibinfo
  {journal} {Phys. Rev. B}\ }\textbf {\bibinfo {volume} {85}},\ \bibinfo
  {pages} {075125} (\bibinfo {year} {2012})},\ \Eprint
  {http://arxiv.org/abs/arXiv:0909.4059} {arXiv:0909.4059} \BibitemShut
  {NoStop}%
\bibitem [{\citenamefont {Chen}\ \emph
  {et~al.}(2011{\natexlab{c}})\citenamefont {Chen}, \citenamefont {Liu},\ and\
  \citenamefont {Wen}}]{CLW1141}%
  \BibitemOpen
  \bibfield  {author} {\bibinfo {author} {\bibfnamefont {X.}~\bibnamefont
  {Chen}}, \bibinfo {author} {\bibfnamefont {Z.-X.}\ \bibnamefont {Liu}}, \
  and\ \bibinfo {author} {\bibfnamefont {X.-G.}\ \bibnamefont {Wen}},\
  }\href@noop {} {\bibfield  {journal} {\bibinfo  {journal} {Phys. Rev. B}\
  }\textbf {\bibinfo {volume} {84}},\ \bibinfo {pages} {235141} (\bibinfo
  {year} {2011}{\natexlab{c}})},\ \Eprint
  {http://arxiv.org/abs/arXiv:1106.4752} {arXiv:1106.4752} \BibitemShut
  {NoStop}%
\bibitem [{\citenamefont {Chen}\ \emph {et~al.}(2013)\citenamefont {Chen},
  \citenamefont {Gu}, \citenamefont {Liu},\ and\ \citenamefont
  {Wen}}]{CGL1314}%
  \BibitemOpen
  \bibfield  {author} {\bibinfo {author} {\bibfnamefont {X.}~\bibnamefont
  {Chen}}, \bibinfo {author} {\bibfnamefont {Z.-C.}\ \bibnamefont {Gu}},
  \bibinfo {author} {\bibfnamefont {Z.-X.}\ \bibnamefont {Liu}}, \ and\
  \bibinfo {author} {\bibfnamefont {X.-G.}\ \bibnamefont {Wen}},\ }\href@noop
  {} {\bibfield  {journal} {\bibinfo  {journal} {Phys. Rev. B}\ }\textbf
  {\bibinfo {volume} {87}},\ \bibinfo {pages} {155114} (\bibinfo {year}
  {2013})},\ \Eprint {http://arxiv.org/abs/arXiv:1106.4772} {arXiv:1106.4772}
  \BibitemShut {NoStop}%
\bibitem [{\citenamefont {Chen}\ \emph {et~al.}(2012)\citenamefont {Chen},
  \citenamefont {Gu}, \citenamefont {Liu},\ and\ \citenamefont
  {Wen}}]{CGL1204}%
  \BibitemOpen
  \bibfield  {author} {\bibinfo {author} {\bibfnamefont {X.}~\bibnamefont
  {Chen}}, \bibinfo {author} {\bibfnamefont {Z.-C.}\ \bibnamefont {Gu}},
  \bibinfo {author} {\bibfnamefont {Z.-X.}\ \bibnamefont {Liu}}, \ and\
  \bibinfo {author} {\bibfnamefont {X.-G.}\ \bibnamefont {Wen}},\ }\href@noop
  {} {\bibfield  {journal} {\bibinfo  {journal} {Science}\ }\textbf {\bibinfo
  {volume} {338}},\ \bibinfo {pages} {1604} (\bibinfo {year} {2012})},\ \Eprint
  {http://arxiv.org/abs/arXiv:1301.0861} {arXiv:1301.0861} \BibitemShut
  {NoStop}%
\bibitem [{\citenamefont {Liu}\ and\ \citenamefont {Wen}(2013)}]{LW1305}%
  \BibitemOpen
  \bibfield  {author} {\bibinfo {author} {\bibfnamefont {Z.-X.}\ \bibnamefont
  {Liu}}\ and\ \bibinfo {author} {\bibfnamefont {X.-G.}\ \bibnamefont {Wen}},\
  }\href@noop {} {\bibfield  {journal} {\bibinfo  {journal} {Phys. Rev. Lett.}\
  }\textbf {\bibinfo {volume} {110}},\ \bibinfo {pages} {067205} (\bibinfo
  {year} {2013})},\ \Eprint {http://arxiv.org/abs/arXiv:1205.7024}
  {arXiv:1205.7024} \BibitemShut {NoStop}%
\bibitem [{\citenamefont {Vishwanath}\ and\ \citenamefont
  {Senthil}(2013)}]{VS1306}%
  \BibitemOpen
  \bibfield  {author} {\bibinfo {author} {\bibfnamefont {A.}~\bibnamefont
  {Vishwanath}}\ and\ \bibinfo {author} {\bibfnamefont {T.}~\bibnamefont
  {Senthil}},\ }\href@noop {} {\bibfield  {journal} {\bibinfo  {journal} {Phys.
  Rev. X}\ }\textbf {\bibinfo {volume} {3}},\ \bibinfo {pages} {011016}
  (\bibinfo {year} {2013})},\ \Eprint {http://arxiv.org/abs/arXiv:1209.3058}
  {arXiv:1209.3058} \BibitemShut {NoStop}%
\bibitem [{\citenamefont {Wang}\ and\ \citenamefont {Senthil}(2013)}]{WS1334}%
  \BibitemOpen
  \bibfield  {author} {\bibinfo {author} {\bibfnamefont {C.}~\bibnamefont
  {Wang}}\ and\ \bibinfo {author} {\bibfnamefont {T.}~\bibnamefont {Senthil}},\
  }\href@noop {} {\bibfield  {journal} {\bibinfo  {journal} {Phys. Rev. B}\
  }\textbf {\bibinfo {volume} {87}},\ \bibinfo {pages} {235122} (\bibinfo
  {year} {2013})},\ \Eprint {http://arxiv.org/abs/arXiv:1302.6234}
  {arXiv:1302.6234} \BibitemShut {NoStop}%
\bibitem [{\citenamefont {Burnell}\ \emph {et~al.}(2013)\citenamefont
  {Burnell}, \citenamefont {Chen}, \citenamefont {Fidkowski},\ and\
  \citenamefont {Vishwanath}}]{BCF1372}%
  \BibitemOpen
  \bibfield  {author} {\bibinfo {author} {\bibfnamefont {F.~J.}\ \bibnamefont
  {Burnell}}, \bibinfo {author} {\bibfnamefont {X.}~\bibnamefont {Chen}},
  \bibinfo {author} {\bibfnamefont {L.}~\bibnamefont {Fidkowski}}, \ and\
  \bibinfo {author} {\bibfnamefont {A.}~\bibnamefont {Vishwanath}},\
  }\href@noop {} {\  (\bibinfo {year} {2013})},\ \Eprint
  {http://arxiv.org/abs/arXiv:1302.7072} {arXiv:1302.7072} \BibitemShut
  {NoStop}%
\bibitem [{\citenamefont {{Lu}}\ and\ \citenamefont
  {{Vishwanath}}(2012)}]{LV1219}%
  \BibitemOpen
  \bibfield  {author} {\bibinfo {author} {\bibfnamefont {Y.-M.}\ \bibnamefont
  {{Lu}}}\ and\ \bibinfo {author} {\bibfnamefont {A.}~\bibnamefont
  {{Vishwanath}}},\ }\href {\doibase 10.1103/PhysRevB.86.125119} {\bibfield
  {journal} {\bibinfo  {journal} {Phys. Rev. B}\ }\textbf {\bibinfo {volume}
  {86}},\ \bibinfo {pages} {125119} (\bibinfo {year} {2012})},\ \Eprint
  {http://arxiv.org/abs/1205.3156} {arXiv:1205.3156} \BibitemShut {NoStop}%
\bibitem [{\citenamefont {Senthil}\ and\ \citenamefont {Levin}(2013)}]{SL1301}%
  \BibitemOpen
  \bibfield  {author} {\bibinfo {author} {\bibfnamefont {T.}~\bibnamefont
  {Senthil}}\ and\ \bibinfo {author} {\bibfnamefont {M.}~\bibnamefont
  {Levin}},\ }\href@noop {} {\bibfield  {journal} {\bibinfo  {journal} {Phys.
  Rev. Lett.}\ }\textbf {\bibinfo {volume} {110}},\ \bibinfo {pages} {046801}
  (\bibinfo {year} {2013})},\ \Eprint {http://arxiv.org/abs/arXiv:1206.1604}
  {arXiv:1206.1604} \BibitemShut {NoStop}%
\bibitem [{\citenamefont {{Xu}}(2013)}]{X1321}%
  \BibitemOpen
  \bibfield  {author} {\bibinfo {author} {\bibfnamefont {C.}~\bibnamefont
  {{Xu}}},\ }\href {\doibase 10.1103/PhysRevB.87.144421} {\bibfield  {journal}
  {\bibinfo  {journal} {Phys. rev. B}\ }\textbf {\bibinfo {volume} {87}},\
  \bibinfo {pages} {144421} (\bibinfo {year} {2013})},\ \Eprint
  {http://arxiv.org/abs/arXiv:1209.4399} {arXiv:1209.4399} \BibitemShut
  {NoStop}%
\bibitem [{\citenamefont {Oon}\ \emph {et~al.}(2013)\citenamefont {Oon},
  \citenamefont {Cho},\ and\ \citenamefont {Xu}}]{OCX1325}%
  \BibitemOpen
  \bibfield  {author} {\bibinfo {author} {\bibfnamefont {J.}~\bibnamefont
  {Oon}}, \bibinfo {author} {\bibfnamefont {G.~Y.}\ \bibnamefont {Cho}}, \ and\
  \bibinfo {author} {\bibfnamefont {C.}~\bibnamefont {Xu}},\ }\href {\doibase
  10.1103/PhysRevB.88.014425} {\bibfield  {journal} {\bibinfo  {journal} {Phys.
  Rev. B}\ }\textbf {\bibinfo {volume} {88}},\ \bibinfo {pages} {014425}
  (\bibinfo {year} {2013})},\ \Eprint {http://arxiv.org/abs/1212.1726}
  {arXiv:1212.1726} \BibitemShut {NoStop}%
\bibitem [{\citenamefont {Xu}\ and\ \citenamefont {Senthil}(2013)}]{XS1372}%
  \BibitemOpen
  \bibfield  {author} {\bibinfo {author} {\bibfnamefont {C.}~\bibnamefont
  {Xu}}\ and\ \bibinfo {author} {\bibfnamefont {T.}~\bibnamefont {Senthil}},\
  }\href@noop {} {\bibfield  {journal} {\bibinfo  {journal} {Phys. Rev. B}\
  }\textbf {\bibinfo {volume} {87}},\ \bibinfo {pages} {174412} (\bibinfo
  {year} {2013})},\ \Eprint {http://arxiv.org/abs/arXiv:1301.6172}
  {arXiv:1301.6172} \BibitemShut {NoStop}%
\bibitem [{\citenamefont {Bi}\ \emph {et~al.}(2013)\citenamefont {Bi},
  \citenamefont {Rasmussen},\ and\ \citenamefont {Xu}}]{BRX1315}%
  \BibitemOpen
  \bibfield  {author} {\bibinfo {author} {\bibfnamefont {Z.}~\bibnamefont
  {Bi}}, \bibinfo {author} {\bibfnamefont {A.}~\bibnamefont {Rasmussen}}, \
  and\ \bibinfo {author} {\bibfnamefont {C.}~\bibnamefont {Xu}},\ }\href@noop
  {} {\  (\bibinfo {year} {2013})},\ \Eprint
  {http://arxiv.org/abs/arXiv:1309.0515} {arXiv:1309.0515} \BibitemShut
  {NoStop}%
\bibitem [{\citenamefont {{You}}\ and\ \citenamefont {{Xu}}(2014)}]{YX1420}%
  \BibitemOpen
  \bibfield  {author} {\bibinfo {author} {\bibfnamefont {Y.-Z.}\ \bibnamefont
  {{You}}}\ and\ \bibinfo {author} {\bibfnamefont {C.}~\bibnamefont {{Xu}}},\
  }\href {\doibase 10.1103/PhysRevB.90.245120} {\bibfield  {journal} {\bibinfo
  {journal} {\prb}\ }\textbf {\bibinfo {volume} {90}},\ \bibinfo {pages}
  {245120} (\bibinfo {year} {2014})},\ \Eprint {http://arxiv.org/abs/1409.0168}
  {arXiv:1409.0168} \BibitemShut {NoStop}%
\bibitem [{\citenamefont {Ye}\ and\ \citenamefont {Wen}(2013)}]{YW1328}%
  \BibitemOpen
  \bibfield  {author} {\bibinfo {author} {\bibfnamefont {P.}~\bibnamefont
  {Ye}}\ and\ \bibinfo {author} {\bibfnamefont {X.-G.}\ \bibnamefont {Wen}},\
  }\href {\doibase 10.1103/PhysRevB.87.195128} {\bibfield  {journal} {\bibinfo
  {journal} {Phys. Rev. B}\ }\textbf {\bibinfo {volume} {87}},\ \bibinfo
  {pages} {195128} (\bibinfo {year} {2013})},\ \Eprint
  {http://arxiv.org/abs/arXiv:1212.2121} {arXiv:1212.2121} \BibitemShut
  {NoStop}%
\bibitem [{\citenamefont {Mei}\ and\ \citenamefont {Wen}(2014)}]{MW1469}%
  \BibitemOpen
  \bibfield  {author} {\bibinfo {author} {\bibfnamefont {J.-W.}\ \bibnamefont
  {Mei}}\ and\ \bibinfo {author} {\bibfnamefont {X.-G.}\ \bibnamefont {Wen}},\
  }\href@noop {} {\  (\bibinfo {year} {2014})},\ \Eprint
  {http://arxiv.org/abs/arXiv:1407.0869} {arXiv:1407.0869} \BibitemShut
  {NoStop}%
\bibitem [{\citenamefont {Liu}\ \emph {et~al.}(2014{\natexlab{a}})\citenamefont
  {Liu}, \citenamefont {Mei}, \citenamefont {Ye},\ and\ \citenamefont
  {Wen}}]{LMY1476}%
  \BibitemOpen
  \bibfield  {author} {\bibinfo {author} {\bibfnamefont {Z.-X.}\ \bibnamefont
  {Liu}}, \bibinfo {author} {\bibfnamefont {J.-W.}\ \bibnamefont {Mei}},
  \bibinfo {author} {\bibfnamefont {P.}~\bibnamefont {Ye}}, \ and\ \bibinfo
  {author} {\bibfnamefont {X.-G.}\ \bibnamefont {Wen}},\ }\href@noop {} {\
  (\bibinfo {year} {2014}{\natexlab{a}})},\ \Eprint
  {http://arxiv.org/abs/arXiv:1408.1676} {arXiv:1408.1676} \BibitemShut
  {NoStop}%
\bibitem [{\citenamefont {{Chen}}\ \emph {et~al.}(2014)\citenamefont {{Chen}},
  \citenamefont {{Lu}},\ and\ \citenamefont {{Vishwanath}}}]{CLV1407}%
  \BibitemOpen
  \bibfield  {author} {\bibinfo {author} {\bibfnamefont {X.}~\bibnamefont
  {{Chen}}}, \bibinfo {author} {\bibfnamefont {Y.-M.}\ \bibnamefont {{Lu}}}, \
  and\ \bibinfo {author} {\bibfnamefont {A.}~\bibnamefont {{Vishwanath}}},\
  }\href {\doibase 10.1038/ncomms4507} {\bibfield  {journal} {\bibinfo
  {journal} {Nature Communications}\ }\textbf {\bibinfo {volume} {5}},\
  \bibinfo {pages} {3507} (\bibinfo {year} {2014})},\ \Eprint
  {http://arxiv.org/abs/arXiv:1303.4301} {arXiv:1303.4301} \BibitemShut
  {NoStop}%
\bibitem [{\citenamefont {{Ye}}\ and\ \citenamefont {{Gu}}(2014)}]{YG1494}%
  \BibitemOpen
  \bibfield  {author} {\bibinfo {author} {\bibfnamefont {P.}~\bibnamefont
  {{Ye}}}\ and\ \bibinfo {author} {\bibfnamefont {Z.-C.}\ \bibnamefont
  {{Gu}}},\ }\href@noop {} {\  (\bibinfo {year} {2014})},\ \Eprint
  {http://arxiv.org/abs/1410.2594} {arXiv:1410.2594} \BibitemShut {NoStop}%
\bibitem [{\citenamefont {{Xu}}\ and\ \citenamefont {{You}}(2014)}]{XY1486}%
  \BibitemOpen
  \bibfield  {author} {\bibinfo {author} {\bibfnamefont {C.}~\bibnamefont
  {{Xu}}}\ and\ \bibinfo {author} {\bibfnamefont {Y.-Z.}\ \bibnamefont
  {{You}}},\ }\href@noop {} {\  (\bibinfo {year} {2014})},\ \Eprint
  {http://arxiv.org/abs/1410.6486} {arXiv:1410.6486} \BibitemShut {NoStop}%
\bibitem [{\citenamefont {{Bi}}\ and\ \citenamefont {{Xu}}(2015)}]{BX1571}%
  \BibitemOpen
  \bibfield  {author} {\bibinfo {author} {\bibfnamefont {Z.}~\bibnamefont
  {{Bi}}}\ and\ \bibinfo {author} {\bibfnamefont {C.}~\bibnamefont {{Xu}}},\
  }\href@noop {} {\  (\bibinfo {year} {2015})},\ \Eprint
  {http://arxiv.org/abs/1501.02271} {arXiv:1501.02271} \BibitemShut {NoStop}%
\bibitem [{\citenamefont {Zeng}\ and\ \citenamefont {Wen}(2014)}]{ZW1490}%
  \BibitemOpen
  \bibfield  {author} {\bibinfo {author} {\bibfnamefont {B.}~\bibnamefont
  {Zeng}}\ and\ \bibinfo {author} {\bibfnamefont {X.-G.}\ \bibnamefont {Wen}},\
  }\href@noop {} {\  (\bibinfo {year} {2014})},\ \Eprint
  {http://arxiv.org/abs/arXiv:1406.5090} {arXiv:1406.5090} \BibitemShut
  {NoStop}%
\bibitem [{\citenamefont {Lan}\ and\ \citenamefont {Wen}(2013)}]{LW1384}%
  \BibitemOpen
  \bibfield  {author} {\bibinfo {author} {\bibfnamefont {T.}~\bibnamefont
  {Lan}}\ and\ \bibinfo {author} {\bibfnamefont {X.-G.}\ \bibnamefont {Wen}},\
  }\href@noop {} {\  (\bibinfo {year} {2013})},\ \Eprint
  {http://arxiv.org/abs/arXiv:1311.1784} {arXiv:1311.1784} \BibitemShut
  {NoStop}%
\bibitem [{\citenamefont {Kong}\ and\ \citenamefont {Wen}(2014)}]{KW1458}%
  \BibitemOpen
  \bibfield  {author} {\bibinfo {author} {\bibfnamefont {L.}~\bibnamefont
  {Kong}}\ and\ \bibinfo {author} {\bibfnamefont {X.-G.}\ \bibnamefont {Wen}},\
  }\href@noop {} {\  (\bibinfo {year} {2014})},\ \Eprint
  {http://arxiv.org/abs/arXiv:1405.5858} {arXiv:1405.5858} \BibitemShut
  {NoStop}%
\bibitem [{\citenamefont {Wen}(1989)}]{Wtop}%
  \BibitemOpen
  \bibfield  {author} {\bibinfo {author} {\bibfnamefont {X.-G.}\ \bibnamefont
  {Wen}},\ }\href@noop {} {\bibfield  {journal} {\bibinfo  {journal} {Phys.
  Rev. B}\ }\textbf {\bibinfo {volume} {40}},\ \bibinfo {pages} {7387}
  (\bibinfo {year} {1989})}\BibitemShut {NoStop}%
\bibitem [{\citenamefont {Wen}\ and\ \citenamefont {Niu}(1990)}]{WNtop}%
  \BibitemOpen
  \bibfield  {author} {\bibinfo {author} {\bibfnamefont {X.-G.}\ \bibnamefont
  {Wen}}\ and\ \bibinfo {author} {\bibfnamefont {Q.}~\bibnamefont {Niu}},\
  }\href@noop {} {\bibfield  {journal} {\bibinfo  {journal} {Phys. Rev. B}\
  }\textbf {\bibinfo {volume} {41}},\ \bibinfo {pages} {9377} (\bibinfo {year}
  {1990})}\BibitemShut {NoStop}%
\bibitem [{\citenamefont {Wen}(1990)}]{Wrig}%
  \BibitemOpen
  \bibfield  {author} {\bibinfo {author} {\bibfnamefont {X.-G.}\ \bibnamefont
  {Wen}},\ }\href@noop {} {\bibfield  {journal} {\bibinfo  {journal} {Int. J.
  Mod. Phys. B}\ }\textbf {\bibinfo {volume} {4}},\ \bibinfo {pages} {239}
  (\bibinfo {year} {1990})}\BibitemShut {NoStop}%
\bibitem [{\citenamefont {Keski-Vakkuri}\ and\ \citenamefont
  {Wen}(1993)}]{KW9327}%
  \BibitemOpen
  \bibfield  {author} {\bibinfo {author} {\bibfnamefont {E.}~\bibnamefont
  {Keski-Vakkuri}}\ and\ \bibinfo {author} {\bibfnamefont {X.-G.}\ \bibnamefont
  {Wen}},\ }\href@noop {} {\bibfield  {journal} {\bibinfo  {journal} {Int. J.
  Mod. Phys. B}\ }\textbf {\bibinfo {volume} {7}},\ \bibinfo {pages} {4227}
  (\bibinfo {year} {1993})}\BibitemShut {NoStop}%
\bibitem [{\citenamefont {Plamadeala}\ \emph {et~al.}(2013)\citenamefont
  {Plamadeala}, \citenamefont {Mulligan},\ and\ \citenamefont
  {Nayak}}]{PMN1372}%
  \BibitemOpen
  \bibfield  {author} {\bibinfo {author} {\bibfnamefont {E.}~\bibnamefont
  {Plamadeala}}, \bibinfo {author} {\bibfnamefont {M.}~\bibnamefont
  {Mulligan}}, \ and\ \bibinfo {author} {\bibfnamefont {C.}~\bibnamefont
  {Nayak}},\ }\href@noop {} {\  (\bibinfo {year} {2013})},\ \Eprint
  {http://arxiv.org/abs/arXiv:1304.0772} {arXiv:1304.0772} \BibitemShut
  {NoStop}%
\bibitem [{\citenamefont {Blok}\ and\ \citenamefont {Wen}(1990)}]{BW9045}%
  \BibitemOpen
  \bibfield  {author} {\bibinfo {author} {\bibfnamefont {B.}~\bibnamefont
  {Blok}}\ and\ \bibinfo {author} {\bibfnamefont {X.-G.}\ \bibnamefont {Wen}},\
  }\href@noop {} {\bibfield  {journal} {\bibinfo  {journal} {Phys. Rev. B}\
  }\textbf {\bibinfo {volume} {42}},\ \bibinfo {pages} {8145} (\bibinfo {year}
  {1990})}\BibitemShut {NoStop}%
\bibitem [{\citenamefont {Read}(1990)}]{R9002}%
  \BibitemOpen
  \bibfield  {author} {\bibinfo {author} {\bibfnamefont {N.}~\bibnamefont
  {Read}},\ }\href@noop {} {\bibfield  {journal} {\bibinfo  {journal} {Phys.
  Rev. Lett.}\ }\textbf {\bibinfo {volume} {65}},\ \bibinfo {pages} {1502}
  (\bibinfo {year} {1990})}\BibitemShut {NoStop}%
\bibitem [{\citenamefont {Fr{\"o}hlich}\ and\ \citenamefont
  {Zee}(1991)}]{FZ9117}%
  \BibitemOpen
  \bibfield  {author} {\bibinfo {author} {\bibfnamefont {J.}~\bibnamefont
  {Fr{\"o}hlich}}\ and\ \bibinfo {author} {\bibfnamefont {A.}~\bibnamefont
  {Zee}},\ }\href@noop {} {\bibfield  {journal} {\bibinfo  {journal} {Nucl.
  Phys. B}\ }\textbf {\bibinfo {volume} {364}},\ \bibinfo {pages} {517}
  (\bibinfo {year} {1991})}\BibitemShut {NoStop}%
\bibitem [{\citenamefont {Fr{\"o}hlich}\ and\ \citenamefont
  {Kerler}(1991)}]{FK9169}%
  \BibitemOpen
  \bibfield  {author} {\bibinfo {author} {\bibfnamefont {J.}~\bibnamefont
  {Fr{\"o}hlich}}\ and\ \bibinfo {author} {\bibfnamefont {T.}~\bibnamefont
  {Kerler}},\ }\href@noop {} {\bibfield  {journal} {\bibinfo  {journal} {Nucl.
  Phys. B}\ }\textbf {\bibinfo {volume} {354}},\ \bibinfo {pages} {369}
  (\bibinfo {year} {1991})}\BibitemShut {NoStop}%
\bibitem [{\citenamefont {Wen}\ and\ \citenamefont {Zee}(1992)}]{WZ9290}%
  \BibitemOpen
  \bibfield  {author} {\bibinfo {author} {\bibfnamefont {X.-G.}\ \bibnamefont
  {Wen}}\ and\ \bibinfo {author} {\bibfnamefont {A.}~\bibnamefont {Zee}},\
  }\href@noop {} {\bibfield  {journal} {\bibinfo  {journal} {Phys. Rev. B}\
  }\textbf {\bibinfo {volume} {46}},\ \bibinfo {pages} {2290} (\bibinfo {year}
  {1992})}\BibitemShut {NoStop}%
\bibitem [{\citenamefont {Fr{\"o}hlich}\ and\ \citenamefont
  {Studer}(1993)}]{FS9333}%
  \BibitemOpen
  \bibfield  {author} {\bibinfo {author} {\bibfnamefont {J.}~\bibnamefont
  {Fr{\"o}hlich}}\ and\ \bibinfo {author} {\bibfnamefont {U.~M.}\ \bibnamefont
  {Studer}},\ }\href@noop {} {\bibfield  {journal} {\bibinfo  {journal} {Rev.
  of Mod. Phys.}\ }\textbf {\bibinfo {volume} {65}},\ \bibinfo {pages} {733}
  (\bibinfo {year} {1993})}\BibitemShut {NoStop}%
\bibitem [{\citenamefont {Wen}(1995)}]{W9505}%
  \BibitemOpen
  \bibfield  {author} {\bibinfo {author} {\bibfnamefont {X.-G.}\ \bibnamefont
  {Wen}},\ }\href@noop {} {\bibfield  {journal} {\bibinfo  {journal} {Advances
  in Physics}\ }\textbf {\bibinfo {volume} {44}},\ \bibinfo {pages} {405}
  (\bibinfo {year} {1995})}\BibitemShut {NoStop}%
\bibitem [{\citenamefont {{Freed}}(2014)}]{F1478}%
  \BibitemOpen
  \bibfield  {author} {\bibinfo {author} {\bibfnamefont {D.~S.}\ \bibnamefont
  {{Freed}}},\ }\href@noop {} {\  (\bibinfo {year} {2014})},\ \Eprint
  {http://arxiv.org/abs/1406.7278} {arXiv:1406.7278} \BibitemShut {NoStop}%
\bibitem [{\citenamefont {Wen}(2014)}]{W1447}%
  \BibitemOpen
  \bibfield  {author} {\bibinfo {author} {\bibfnamefont {X.-G.}\ \bibnamefont
  {Wen}},\ }\href {\doibase 10.1103/PhysRevB.89.035147} {\bibfield  {journal}
  {\bibinfo  {journal} {Phys. Rev. B}\ }\textbf {\bibinfo {volume} {89}},\
  \bibinfo {pages} {035147} (\bibinfo {year} {2014})},\ \Eprint
  {http://arxiv.org/abs/arXiv:1301.7675} {arXiv:1301.7675} \BibitemShut
  {NoStop}%
\bibitem [{\citenamefont {Hung}\ and\ \citenamefont
  {Wen}(2013{\natexlab{a}})}]{HW1339}%
  \BibitemOpen
  \bibfield  {author} {\bibinfo {author} {\bibfnamefont {L.-Y.}\ \bibnamefont
  {Hung}}\ and\ \bibinfo {author} {\bibfnamefont {X.-G.}\ \bibnamefont {Wen}},\
  }\href@noop {} {\  (\bibinfo {year} {2013}{\natexlab{a}})},\ \Eprint
  {http://arxiv.org/abs/arXiv:1311.5539} {arXiv:1311.5539} \BibitemShut
  {NoStop}%
\bibitem [{\citenamefont {Levin}\ and\ \citenamefont {Gu}(2012)}]{LG1220}%
  \BibitemOpen
  \bibfield  {author} {\bibinfo {author} {\bibfnamefont {M.}~\bibnamefont
  {Levin}}\ and\ \bibinfo {author} {\bibfnamefont {Z.-C.}\ \bibnamefont {Gu}},\
  }\href@noop {} {\bibfield  {journal} {\bibinfo  {journal} {Phys. Rev. B}\
  }\textbf {\bibinfo {volume} {86}},\ \bibinfo {pages} {115109} (\bibinfo
  {year} {2012})},\ \Eprint {http://arxiv.org/abs/arXiv:1202.3120}
  {arXiv:1202.3120} \BibitemShut {NoStop}%
\bibitem [{\citenamefont {Ryu}\ and\ \citenamefont {Zhang}(2012)}]{RZ1232}%
  \BibitemOpen
  \bibfield  {author} {\bibinfo {author} {\bibfnamefont {S.}~\bibnamefont
  {Ryu}}\ and\ \bibinfo {author} {\bibfnamefont {S.-C.}\ \bibnamefont
  {Zhang}},\ }\href {\doibase 10.1103/PhysRevB.85.245132} {\bibfield  {journal}
  {\bibinfo  {journal} {Phys. Rev. B}\ }\textbf {\bibinfo {volume} {85}},\
  \bibinfo {pages} {245132} (\bibinfo {year} {2012})}\BibitemShut {NoStop}%
\bibitem [{\citenamefont {{Sule}}\ \emph {et~al.}(2013)\citenamefont {{Sule}},
  \citenamefont {{Chen}},\ and\ \citenamefont {{Ryu}}}]{SCR1325}%
  \BibitemOpen
  \bibfield  {author} {\bibinfo {author} {\bibfnamefont {O.~M.}\ \bibnamefont
  {{Sule}}}, \bibinfo {author} {\bibfnamefont {X.}~\bibnamefont {{Chen}}}, \
  and\ \bibinfo {author} {\bibfnamefont {S.}~\bibnamefont {{Ryu}}},\ }\href
  {\doibase 10.1103/PhysRevB.88.075125} {\bibfield  {journal} {\bibinfo
  {journal} {\prb}\ }\textbf {\bibinfo {volume} {88}},\ \bibinfo {pages}
  {075125} (\bibinfo {year} {2013})},\ \Eprint {http://arxiv.org/abs/1305.0700}
  {arXiv:1305.0700} \BibitemShut {NoStop}%
\bibitem [{\citenamefont {Metlitski}\ \emph {et~al.}(2013)\citenamefont
  {Metlitski}, \citenamefont {Kane},\ and\ \citenamefont {Fisher}}]{MKF1331}%
  \BibitemOpen
  \bibfield  {author} {\bibinfo {author} {\bibfnamefont {M.~A.}\ \bibnamefont
  {Metlitski}}, \bibinfo {author} {\bibfnamefont {C.~L.}\ \bibnamefont {Kane}},
  \ and\ \bibinfo {author} {\bibfnamefont {M.~P.~A.}\ \bibnamefont {Fisher}},\
  }\href@noop {} {\bibfield  {journal} {\bibinfo  {journal} {Phys. Rev. B}\
  }\textbf {\bibinfo {volume} {88}},\ \bibinfo {pages} {035131} (\bibinfo
  {year} {2013})},\ \Eprint {http://arxiv.org/abs/arXiv:1302.6535}
  {arXiv:1302.6535} \BibitemShut {NoStop}%
\bibitem [{\citenamefont {Ye}\ and\ \citenamefont {Wen}(2014)}]{YW1427}%
  \BibitemOpen
  \bibfield  {author} {\bibinfo {author} {\bibfnamefont {P.}~\bibnamefont
  {Ye}}\ and\ \bibinfo {author} {\bibfnamefont {X.-G.}\ \bibnamefont {Wen}},\
  }\href@noop {} {\bibfield  {journal} {\bibinfo  {journal} {Phys. Rev. B}\
  }\textbf {\bibinfo {volume} {89}},\ \bibinfo {pages} {045127} (\bibinfo
  {year} {2014})},\ \Eprint {http://arxiv.org/abs/arXiv:1303.3572}
  {arXiv:1303.3572} \BibitemShut {NoStop}%
\bibitem [{\citenamefont {{Bi}}\ \emph {et~al.}(2014)\citenamefont {{Bi}},
  \citenamefont {{Rasmussen}},\ and\ \citenamefont {{Xu}}}]{BRX1424}%
  \BibitemOpen
  \bibfield  {author} {\bibinfo {author} {\bibfnamefont {Z.}~\bibnamefont
  {{Bi}}}, \bibinfo {author} {\bibfnamefont {A.}~\bibnamefont {{Rasmussen}}}, \
  and\ \bibinfo {author} {\bibfnamefont {C.}~\bibnamefont {{Xu}}},\ }\href
  {\doibase 10.1103/PhysRevB.89.184424} {\bibfield  {journal} {\bibinfo
  {journal} {Phys. Rev. B}\ }\textbf {\bibinfo {volume} {89}},\ \bibinfo
  {pages} {184424} (\bibinfo {year} {2014})},\ \Eprint
  {http://arxiv.org/abs/arXiv:1304.7272} {arXiv:1304.7272} \BibitemShut
  {NoStop}%
\bibitem [{\citenamefont {Wang}\ \emph
  {et~al.}(2014{\natexlab{a}})\citenamefont {Wang}, \citenamefont {Santos},\
  and\ \citenamefont {Wen}}]{WSW1456}%
  \BibitemOpen
  \bibfield  {author} {\bibinfo {author} {\bibfnamefont {J.}~\bibnamefont
  {Wang}}, \bibinfo {author} {\bibfnamefont {L.~H.}\ \bibnamefont {Santos}}, \
  and\ \bibinfo {author} {\bibfnamefont {X.-G.}\ \bibnamefont {Wen}},\
  }\href@noop {} {\  (\bibinfo {year} {2014}{\natexlab{a}})},\ \Eprint
  {http://arxiv.org/abs/arXiv:1403.5256} {arXiv:1403.5256} \BibitemShut
  {NoStop}%
\bibitem [{\citenamefont {Chen}\ and\ \citenamefont {Wen}(2012)}]{CW1235}%
  \BibitemOpen
  \bibfield  {author} {\bibinfo {author} {\bibfnamefont {X.}~\bibnamefont
  {Chen}}\ and\ \bibinfo {author} {\bibfnamefont {X.-G.}\ \bibnamefont {Wen}},\
  }\href@noop {} {\bibfield  {journal} {\bibinfo  {journal} {Phys. Rev. B}\
  }\textbf {\bibinfo {volume} {86}},\ \bibinfo {pages} {235135} (\bibinfo
  {year} {2012})},\ \Eprint {http://arxiv.org/abs/arXiv:1206.3117}
  {arXiv:1206.3117} \BibitemShut {NoStop}%
\bibitem [{\citenamefont {Wang}\ \emph
  {et~al.}(2014{\natexlab{b}})\citenamefont {Wang}, \citenamefont {Gu},\ and\
  \citenamefont {Wen}}]{WGW1489}%
  \BibitemOpen
  \bibfield  {author} {\bibinfo {author} {\bibfnamefont {J.}~\bibnamefont
  {Wang}}, \bibinfo {author} {\bibfnamefont {Z.-C.}\ \bibnamefont {Gu}}, \ and\
  \bibinfo {author} {\bibfnamefont {X.-G.}\ \bibnamefont {Wen}},\ }\href@noop
  {} {\  (\bibinfo {year} {2014}{\natexlab{b}})},\ \Eprint
  {http://arxiv.org/abs/arXiv:1405.7689} {arXiv:1405.7689} \BibitemShut
  {NoStop}%
\bibitem [{\citenamefont {Fidkowski}\ \emph {et~al.}(2006)\citenamefont
  {Fidkowski}, \citenamefont {Freedman}, \citenamefont {Nayak}, \citenamefont
  {Walker},\ and\ \citenamefont {Wang}}]{FFN0683}%
  \BibitemOpen
  \bibfield  {author} {\bibinfo {author} {\bibfnamefont {L.}~\bibnamefont
  {Fidkowski}}, \bibinfo {author} {\bibfnamefont {M.}~\bibnamefont {Freedman}},
  \bibinfo {author} {\bibfnamefont {C.}~\bibnamefont {Nayak}}, \bibinfo
  {author} {\bibfnamefont {K.}~\bibnamefont {Walker}}, \ and\ \bibinfo {author}
  {\bibfnamefont {Z.}~\bibnamefont {Wang}},\ }\href@noop {} {\  (\bibinfo
  {year} {2006})},\ \Eprint {http://arxiv.org/abs/arXiv:cond-mat/0610583}
  {arXiv:cond-mat/0610583} \BibitemShut {NoStop}%
\bibitem [{\citenamefont {Wang}(2010)}]{Wang10}%
  \BibitemOpen
  \bibfield  {author} {\bibinfo {author} {\bibfnamefont {Z.}~\bibnamefont
  {Wang}},\ }\href@noop {} {\emph {\bibinfo {title} {Topological Quantum
  Computation}}}\ (\bibinfo  {publisher} {CBMS Regional Conference Series in
  Mathematics},\ \bibinfo {year} {2010})\BibitemShut {NoStop}%
\bibitem [{\citenamefont {Kapustin}(bove)}]{K1459v2}%
  \BibitemOpen
  \bibfield  {author} {\bibinfo {author} {\bibfnamefont {A.}~\bibnamefont
  {Kapustin}},\ }\href@noop {} {\  (\bibinfo {year} {version v2 and above})},\
  \Eprint {http://arxiv.org/abs/arXiv:1404.6659} {arXiv:1404.6659} \BibitemShut
  {NoStop}%
\bibitem [{\citenamefont {Kapustin}(2014)}]{K1467}%
  \BibitemOpen
  \bibfield  {author} {\bibinfo {author} {\bibfnamefont {A.}~\bibnamefont
  {Kapustin}},\ }\href@noop {} {\  (\bibinfo {year} {2014})},\ \Eprint
  {http://arxiv.org/abs/arXiv:1403.1467} {arXiv:1403.1467} \BibitemShut
  {NoStop}%
\bibitem [{\citenamefont {{Kapustin}}\ \emph {et~al.}(2014)\citenamefont
  {{Kapustin}}, \citenamefont {{Thorngren}}, \citenamefont {{Turzillo}},\ and\
  \citenamefont {{Wang}}}]{KTT1429}%
  \BibitemOpen
  \bibfield  {author} {\bibinfo {author} {\bibfnamefont {A.}~\bibnamefont
  {{Kapustin}}}, \bibinfo {author} {\bibfnamefont {R.}~\bibnamefont
  {{Thorngren}}}, \bibinfo {author} {\bibfnamefont {A.}~\bibnamefont
  {{Turzillo}}}, \ and\ \bibinfo {author} {\bibfnamefont {Z.}~\bibnamefont
  {{Wang}}},\ }\href@noop {} {\  (\bibinfo {year} {2014})},\ \Eprint
  {http://arxiv.org/abs/1406.7329} {arXiv:1406.7329} \BibitemShut {NoStop}%
\bibitem [{\citenamefont {Kane}\ and\ \citenamefont {Fisher}(1997)}]{KF9732}%
  \BibitemOpen
  \bibfield  {author} {\bibinfo {author} {\bibfnamefont {C.~L.}\ \bibnamefont
  {Kane}}\ and\ \bibinfo {author} {\bibfnamefont {M.~P.~A.}\ \bibnamefont
  {Fisher}},\ }\href {\doibase 10.1103/PhysRevB.55.15832} {\bibfield  {journal}
  {\bibinfo  {journal} {Phys. Rev. B}\ }\textbf {\bibinfo {volume} {55}},\
  \bibinfo {pages} {15832} (\bibinfo {year} {1997})},\ \Eprint
  {http://arxiv.org/abs/cond-mat/9603118} {cond-mat/9603118} \BibitemShut
  {NoStop}%
\bibitem [{\citenamefont {Hughes}\ \emph {et~al.}(2012)\citenamefont {Hughes},
  \citenamefont {Leigh},\ and\ \citenamefont {Parrikar}}]{HLP1242}%
  \BibitemOpen
  \bibfield  {author} {\bibinfo {author} {\bibfnamefont {T.~L.}\ \bibnamefont
  {Hughes}}, \bibinfo {author} {\bibfnamefont {R.~G.}\ \bibnamefont {Leigh}}, \
  and\ \bibinfo {author} {\bibfnamefont {O.}~\bibnamefont {Parrikar}},\
  }\href@noop {} {\  (\bibinfo {year} {2012})},\ \Eprint
  {http://arxiv.org/abs/arXiv:1211.6442} {arXiv:1211.6442} \BibitemShut
  {NoStop}%
\bibitem [{\citenamefont {Hung}\ and\ \citenamefont {Wen}(2012)}]{HW1267}%
  \BibitemOpen
  \bibfield  {author} {\bibinfo {author} {\bibfnamefont {L.-Y.}\ \bibnamefont
  {Hung}}\ and\ \bibinfo {author} {\bibfnamefont {X.-G.}\ \bibnamefont {Wen}},\
  }\href@noop {} {\  (\bibinfo {year} {2012})},\ \Eprint
  {http://arxiv.org/abs/arXiv:1211.2767} {arXiv:1211.2767} \BibitemShut
  {NoStop}%
\bibitem [{\citenamefont {Hung}\ and\ \citenamefont
  {Wen}(2013{\natexlab{b}})}]{HW1227}%
  \BibitemOpen
  \bibfield  {author} {\bibinfo {author} {\bibfnamefont {L.-Y.}\ \bibnamefont
  {Hung}}\ and\ \bibinfo {author} {\bibfnamefont {X.-G.}\ \bibnamefont {Wen}},\
  }\href@noop {} {\bibfield  {journal} {\bibinfo  {journal} {Phys. Rev. B}\
  }\textbf {\bibinfo {volume} {87}},\ \bibinfo {pages} {165107} (\bibinfo
  {year} {2013}{\natexlab{b}})},\ \Eprint
  {http://arxiv.org/abs/arXiv:1212.1827} {arXiv:1212.1827} \BibitemShut
  {NoStop}%
\bibitem [{\citenamefont {Wen}(2013)}]{W1313}%
  \BibitemOpen
  \bibfield  {author} {\bibinfo {author} {\bibfnamefont {X.-G.}\ \bibnamefont
  {Wen}},\ }\href@noop {} {\bibfield  {journal} {\bibinfo  {journal} {Phys.
  Rev. B}\ }\textbf {\bibinfo {volume} {88}},\ \bibinfo {pages} {045013}
  (\bibinfo {year} {2013})},\ \Eprint {http://arxiv.org/abs/arXiv:1303.1803}
  {arXiv:1303.1803} \BibitemShut {NoStop}%
\bibitem [{\citenamefont {Wu}(1985)}]{W8570}%
  \BibitemOpen
  \bibfield  {author} {\bibinfo {author} {\bibfnamefont {Y.-S.}\ \bibnamefont
  {Wu}},\ }\href {\doibase 10.1016/0370-2693(85)91444-3} {\bibfield  {journal}
  {\bibinfo  {journal} {Physics Letters B}\ }\textbf {\bibinfo {volume}
  {153}},\ \bibinfo {pages} {70} (\bibinfo {year} {1985})}\BibitemShut
  {NoStop}%
\bibitem [{\citenamefont {Wen}\ and\ \citenamefont
  {Wang}(2008{\natexlab{a}})}]{WW0808}%
  \BibitemOpen
  \bibfield  {author} {\bibinfo {author} {\bibfnamefont {X.-G.}\ \bibnamefont
  {Wen}}\ and\ \bibinfo {author} {\bibfnamefont {Z.}~\bibnamefont {Wang}},\
  }\href@noop {} {\bibfield  {journal} {\bibinfo  {journal} {Phys. Rev. B}\
  }\textbf {\bibinfo {volume} {77}},\ \bibinfo {pages} {235108} (\bibinfo
  {year} {2008}{\natexlab{a}})},\ \Eprint
  {http://arxiv.org/abs/arXiv:0801.3291} {arXiv:0801.3291} \BibitemShut
  {NoStop}%
\bibitem [{\citenamefont {Wen}\ and\ \citenamefont
  {Wang}(2008{\natexlab{b}})}]{WW0809}%
  \BibitemOpen
  \bibfield  {author} {\bibinfo {author} {\bibfnamefont {X.-G.}\ \bibnamefont
  {Wen}}\ and\ \bibinfo {author} {\bibfnamefont {Z.}~\bibnamefont {Wang}},\
  }\href@noop {} {\bibfield  {journal} {\bibinfo  {journal} {Phys. Rev. B}\
  }\textbf {\bibinfo {volume} {78}},\ \bibinfo {pages} {155109} (\bibinfo
  {year} {2008}{\natexlab{b}})},\ \Eprint
  {http://arxiv.org/abs/arXiv:0803.1016} {arXiv:0803.1016} \BibitemShut
  {NoStop}%
\bibitem [{\citenamefont {Kapustin}\ and\ \citenamefont
  {Thorngren}(2014)}]{KT1430}%
  \BibitemOpen
  \bibfield  {author} {\bibinfo {author} {\bibfnamefont {A.}~\bibnamefont
  {Kapustin}}\ and\ \bibinfo {author} {\bibfnamefont {R.}~\bibnamefont
  {Thorngren}},\ }\href@noop {} {\  (\bibinfo {year} {2014})},\ \Eprint
  {http://arxiv.org/abs/arXiv:1404.3230} {arXiv:1404.3230} \BibitemShut
  {NoStop}%
\bibitem [{\citenamefont {{Kapustin}}\ and\ \citenamefont
  {{Thorngren}}(2014)}]{KT1417}%
  \BibitemOpen
  \bibfield  {author} {\bibinfo {author} {\bibfnamefont {A.}~\bibnamefont
  {{Kapustin}}}\ and\ \bibinfo {author} {\bibfnamefont {R.}~\bibnamefont
  {{Thorngren}}},\ }\href {\doibase 10.1103/PhysRevLett.112.231602} {\bibfield
  {journal} {\bibinfo  {journal} {Physical Review Letters}\ }\textbf {\bibinfo
  {volume} {112}},\ \bibinfo {eid} {231602} (\bibinfo {year} {2014})},\ \Eprint
  {http://arxiv.org/abs/1403.0617} {arXiv:1403.0617} \BibitemShut {NoStop}%
\bibitem [{Note1()}]{Note1}%
  \BibitemOpen
  \bibinfo {note} {For example, the pure 2+1D gravitational anomalies described
  by \protect \emph {unquantized} thermal Hall conductivity are not classified
  by topologically ordered states.}\BibitemShut {Stop}%
\bibitem [{\citenamefont {Ryu}\ \emph {et~al.}(2009)\citenamefont {Ryu},
  \citenamefont {Schnyder}, \citenamefont {Furusaki},\ and\ \citenamefont
  {Ludwig}}]{RSF0957}%
  \BibitemOpen
  \bibfield  {author} {\bibinfo {author} {\bibfnamefont {S.}~\bibnamefont
  {Ryu}}, \bibinfo {author} {\bibfnamefont {A.}~\bibnamefont {Schnyder}},
  \bibinfo {author} {\bibfnamefont {A.}~\bibnamefont {Furusaki}}, \ and\
  \bibinfo {author} {\bibfnamefont {A.}~\bibnamefont {Ludwig}},\ }\href@noop {}
  {\  (\bibinfo {year} {2009})},\ \Eprint
  {http://arxiv.org/abs/arXiv:0912.2157} {arXiv:0912.2157} \BibitemShut
  {NoStop}%
\bibitem [{\citenamefont {Ryu}\ \emph {et~al.}(2012)\citenamefont {Ryu},
  \citenamefont {Moore},\ and\ \citenamefont {Ludwig}}]{RML1204}%
  \BibitemOpen
  \bibfield  {author} {\bibinfo {author} {\bibfnamefont {S.}~\bibnamefont
  {Ryu}}, \bibinfo {author} {\bibfnamefont {J.~E.}\ \bibnamefont {Moore}}, \
  and\ \bibinfo {author} {\bibfnamefont {A.~W.~W.}\ \bibnamefont {Ludwig}},\
  }\href {\doibase 10.1103/PhysRevB.85.045104} {\bibfield  {journal} {\bibinfo
  {journal} {Phys. Rev. B}\ }\textbf {\bibinfo {volume} {85}},\ \bibinfo
  {pages} {045104} (\bibinfo {year} {2012})},\ \Eprint
  {http://arxiv.org/abs/arXiv:1010.0936} {arXiv:1010.0936} \BibitemShut
  {NoStop}%
\bibitem [{\citenamefont {Liu}\ \emph {et~al.}(2014{\natexlab{b}})\citenamefont
  {Liu}, \citenamefont {Gu},\ and\ \citenamefont {Wen}}]{LGW1418}%
  \BibitemOpen
  \bibfield  {author} {\bibinfo {author} {\bibfnamefont {Z.-X.}\ \bibnamefont
  {Liu}}, \bibinfo {author} {\bibfnamefont {Z.-C.}\ \bibnamefont {Gu}}, \ and\
  \bibinfo {author} {\bibfnamefont {X.-G.}\ \bibnamefont {Wen}},\ }\href@noop
  {} {\  (\bibinfo {year} {2014}{\natexlab{b}})},\ \Eprint
  {http://arxiv.org/abs/arXiv:1404.2818} {arXiv:1404.2818} \BibitemShut
  {NoStop}%
\bibitem [{\citenamefont {{Gu}}\ \emph {et~al.}(2015)\citenamefont {{Gu}},
  \citenamefont {{Wang}},\ and\ \citenamefont {{Wen}}}]{GWW1568}%
  \BibitemOpen
  \bibfield  {author} {\bibinfo {author} {\bibfnamefont {Z.-C.}\ \bibnamefont
  {{Gu}}}, \bibinfo {author} {\bibfnamefont {J.~C.}\ \bibnamefont {{Wang}}}, \
  and\ \bibinfo {author} {\bibfnamefont {X.-G.}\ \bibnamefont {{Wen}}},\
  }\href@noop {} {\  (\bibinfo {year} {2015})},\ \Eprint
  {http://arxiv.org/abs/1503.01768} {arXiv:1503.01768} \BibitemShut {NoStop}%
\bibitem [{\citenamefont {Gu}\ and\ \citenamefont {Wen}(2014)}]{GW14}%
  \BibitemOpen
  \bibfield  {author} {\bibinfo {author} {\bibfnamefont {Z.-C.}\ \bibnamefont
  {Gu}}\ and\ \bibinfo {author} {\bibfnamefont {X.-G.}\ \bibnamefont {Wen}},\
  }\href@noop {} {\bibfield  {journal} {\bibinfo  {journal} {to appear}\ }
  (\bibinfo {year} {2014})}\BibitemShut {NoStop}%
\bibitem [{\citenamefont {Gu}\ and\ \citenamefont {Wen}(2012)}]{GW1248}%
  \BibitemOpen
  \bibfield  {author} {\bibinfo {author} {\bibfnamefont {Z.-C.}\ \bibnamefont
  {Gu}}\ and\ \bibinfo {author} {\bibfnamefont {X.-G.}\ \bibnamefont {Wen}},\
  }\href@noop {} {\  (\bibinfo {year} {2012})},\ \Eprint
  {http://arxiv.org/abs/arXiv:1201.2648} {arXiv:1201.2648} \BibitemShut
  {NoStop}%
\bibitem [{\citenamefont {{Hsieh}}\ \emph {et~al.}(2014)\citenamefont
  {{Hsieh}}, \citenamefont {{Mayodele Sule}}, \citenamefont {{Cho}},
  \citenamefont {{Ryu}},\ and\ \citenamefont {{Leigh}}}]{HMC1402}%
  \BibitemOpen
  \bibfield  {author} {\bibinfo {author} {\bibfnamefont {C.-T.}\ \bibnamefont
  {{Hsieh}}}, \bibinfo {author} {\bibfnamefont {O.}~\bibnamefont {{Mayodele
  Sule}}}, \bibinfo {author} {\bibfnamefont {G.~Y.}\ \bibnamefont {{Cho}}},
  \bibinfo {author} {\bibfnamefont {S.}~\bibnamefont {{Ryu}}}, \ and\ \bibinfo
  {author} {\bibfnamefont {R.~G.}\ \bibnamefont {{Leigh}}},\ }\href@noop {}
  {\bibfield  {journal} {\bibinfo  {journal} {ArXiv e-prints}\ } (\bibinfo
  {year} {2014})},\ \Eprint {http://arxiv.org/abs/1403.6902} {arXiv:1403.6902}
  \BibitemShut {NoStop}%
\bibitem [{\citenamefont {Freedman}\ \emph {et~al.}(2004)\citenamefont
  {Freedman}, \citenamefont {Nayak}, \citenamefont {Shtengel}, \citenamefont
  {Walker},\ and\ \citenamefont {Wang}}]{FNS0428}%
  \BibitemOpen
  \bibfield  {author} {\bibinfo {author} {\bibfnamefont {M.}~\bibnamefont
  {Freedman}}, \bibinfo {author} {\bibfnamefont {C.}~\bibnamefont {Nayak}},
  \bibinfo {author} {\bibfnamefont {K.}~\bibnamefont {Shtengel}}, \bibinfo
  {author} {\bibfnamefont {K.}~\bibnamefont {Walker}}, \ and\ \bibinfo {author}
  {\bibfnamefont {Z.}~\bibnamefont {Wang}},\ }\href@noop {} {\bibfield
  {journal} {\bibinfo  {journal} {Ann. Phys. (NY)}\ }\textbf {\bibinfo {volume}
  {310}},\ \bibinfo {pages} {428} (\bibinfo {year} {2004})},\ \Eprint
  {http://arxiv.org/abs/cond-mat/0307511} {cond-mat/0307511} \BibitemShut
  {NoStop}%
\bibitem [{\citenamefont {Levin}\ and\ \citenamefont {Wen}(2005)}]{LWstrnet}%
  \BibitemOpen
  \bibfield  {author} {\bibinfo {author} {\bibfnamefont {M.}~\bibnamefont
  {Levin}}\ and\ \bibinfo {author} {\bibfnamefont {X.-G.}\ \bibnamefont
  {Wen}},\ }\href@noop {} {\bibfield  {journal} {\bibinfo  {journal} {Phys.
  Rev. B}\ }\textbf {\bibinfo {volume} {71}},\ \bibinfo {pages} {045110}
  (\bibinfo {year} {2005})},\ \Eprint {http://arxiv.org/abs/cond-mat/0404617}
  {cond-mat/0404617} \BibitemShut {NoStop}%
\bibitem [{\citenamefont {Gu}\ \emph {et~al.}(2010)\citenamefont {Gu},
  \citenamefont {Wang},\ and\ \citenamefont {Wen}}]{GWW1017}%
  \BibitemOpen
  \bibfield  {author} {\bibinfo {author} {\bibfnamefont {Z.-C.}\ \bibnamefont
  {Gu}}, \bibinfo {author} {\bibfnamefont {Z.}~\bibnamefont {Wang}}, \ and\
  \bibinfo {author} {\bibfnamefont {X.-G.}\ \bibnamefont {Wen}},\ }\href@noop
  {} {\  (\bibinfo {year} {2010})},\ \Eprint
  {http://arxiv.org/abs/arXiv:1010.1517} {arXiv:1010.1517} \BibitemShut
  {NoStop}%
\bibitem [{\citenamefont {Kitaev}\ and\ \citenamefont {Kong}(2012)}]{KK1251}%
  \BibitemOpen
  \bibfield  {author} {\bibinfo {author} {\bibfnamefont {A.}~\bibnamefont
  {Kitaev}}\ and\ \bibinfo {author} {\bibfnamefont {L.}~\bibnamefont {Kong}},\
  }\href {\doibase 10.1007/s00220-012-1500-5} {\bibfield  {journal} {\bibinfo
  {journal} {Commun. Math. Phys.}\ }\textbf {\bibinfo {volume} {313}},\
  \bibinfo {pages} {351 } (\bibinfo {year} {2012})},\ \Eprint
  {http://arxiv.org/abs/arXiv:1104.5047} {arXiv:1104.5047} \BibitemShut
  {NoStop}%
\bibitem [{\citenamefont {Gu}\ \emph {et~al.}(2013)\citenamefont {Gu},
  \citenamefont {Wang},\ and\ \citenamefont {Wen}}]{GWW1332}%
  \BibitemOpen
  \bibfield  {author} {\bibinfo {author} {\bibfnamefont {Z.-C.}\ \bibnamefont
  {Gu}}, \bibinfo {author} {\bibfnamefont {Z.}~\bibnamefont {Wang}}, \ and\
  \bibinfo {author} {\bibfnamefont {X.-G.}\ \bibnamefont {Wen}},\ }\href@noop
  {} {\  (\bibinfo {year} {2013})},\ \Eprint
  {http://arxiv.org/abs/arXiv:1309.7032} {arXiv:1309.7032} \BibitemShut
  {NoStop}%
\bibitem [{\citenamefont {Verstraete}\ and\ \citenamefont
  {Cirac}(2004)}]{VC0466}%
  \BibitemOpen
  \bibfield  {author} {\bibinfo {author} {\bibfnamefont {F.}~\bibnamefont
  {Verstraete}}\ and\ \bibinfo {author} {\bibfnamefont {J.~I.}\ \bibnamefont
  {Cirac}},\ }\href@noop {} {\  (\bibinfo {year} {2004})},\ \Eprint
  {http://arxiv.org/abs/arXiv:cond-mat/0407066} {arXiv:cond-mat/0407066}
  \BibitemShut {NoStop}%
\bibitem [{\citenamefont {Gu}\ \emph {et~al.}(2009)\citenamefont {Gu},
  \citenamefont {Levin}, \citenamefont {Swingle},\ and\ \citenamefont
  {Wen}}]{GLS0918}%
  \BibitemOpen
  \bibfield  {author} {\bibinfo {author} {\bibfnamefont {Z.-C.}\ \bibnamefont
  {Gu}}, \bibinfo {author} {\bibfnamefont {M.}~\bibnamefont {Levin}}, \bibinfo
  {author} {\bibfnamefont {B.}~\bibnamefont {Swingle}}, \ and\ \bibinfo
  {author} {\bibfnamefont {X.-G.}\ \bibnamefont {Wen}},\ }\href@noop {}
  {\bibfield  {journal} {\bibinfo  {journal} {Phys. Rev. B}\ }\textbf {\bibinfo
  {volume} {79}},\ \bibinfo {pages} {085118} (\bibinfo {year} {2009})},\
  \Eprint {http://arxiv.org/abs/arXiv:0809.2821} {arXiv:0809.2821} \BibitemShut
  {NoStop}%
\bibitem [{\citenamefont {Buerschaper}\ \emph {et~al.}(2009)\citenamefont
  {Buerschaper}, \citenamefont {Aguado},\ and\ \citenamefont
  {Vidal}}]{BAV0919}%
  \BibitemOpen
  \bibfield  {author} {\bibinfo {author} {\bibfnamefont {O.}~\bibnamefont
  {Buerschaper}}, \bibinfo {author} {\bibfnamefont {M.}~\bibnamefont {Aguado}},
  \ and\ \bibinfo {author} {\bibfnamefont {G.}~\bibnamefont {Vidal}},\
  }\href@noop {} {\bibfield  {journal} {\bibinfo  {journal} {Phys. Rev. B}\
  }\textbf {\bibinfo {volume} {79}},\ \bibinfo {pages} {085119} (\bibinfo
  {year} {2009})},\ \Eprint {http://arxiv.org/abs/arXiv:0809.2393}
  {arXiv:0809.2393} \BibitemShut {NoStop}%
\bibitem [{\citenamefont {Moore}\ and\ \citenamefont {Read}(1991)}]{MR9162}%
  \BibitemOpen
  \bibfield  {author} {\bibinfo {author} {\bibfnamefont {G.}~\bibnamefont
  {Moore}}\ and\ \bibinfo {author} {\bibfnamefont {N.}~\bibnamefont {Read}},\
  }\href@noop {} {\bibfield  {journal} {\bibinfo  {journal} {Nucl. Phys. B}\
  }\textbf {\bibinfo {volume} {360}},\ \bibinfo {pages} {362} (\bibinfo {year}
  {1991})}\BibitemShut {NoStop}%
\bibitem [{\citenamefont {Costantino}(2005)}]{C0527}%
  \BibitemOpen
  \bibfield  {author} {\bibinfo {author} {\bibfnamefont {F.}~\bibnamefont
  {Costantino}},\ }\href@noop {} {\bibfield  {journal} {\bibinfo  {journal}
  {Math. Z.}\ }\textbf {\bibinfo {volume} {251}},\ \bibinfo {pages} {427}
  (\bibinfo {year} {2005})},\ \Eprint {http://arxiv.org/abs/arXiv:math/0403014}
  {arXiv:math/0403014} \BibitemShut {NoStop}%
\bibitem [{\citenamefont {Hu}\ \emph {et~al.}(2013)\citenamefont {Hu},
  \citenamefont {Wan},\ and\ \citenamefont {Wu}}]{HWW1295}%
  \BibitemOpen
  \bibfield  {author} {\bibinfo {author} {\bibfnamefont {Y.}~\bibnamefont
  {Hu}}, \bibinfo {author} {\bibfnamefont {Y.}~\bibnamefont {Wan}}, \ and\
  \bibinfo {author} {\bibfnamefont {Y.-S.}\ \bibnamefont {Wu}},\ }\href@noop {}
  {\bibfield  {journal} {\bibinfo  {journal} {Phys. Rev. B}\ }\textbf {\bibinfo
  {volume} {87}},\ \bibinfo {pages} {125114} (\bibinfo {year} {2013})},\
  \Eprint {http://arxiv.org/abs/arXiv:1211.3695} {arXiv:1211.3695} \BibitemShut
  {NoStop}%
\bibitem [{\citenamefont {Dijkgraaf}\ and\ \citenamefont
  {Witten}(1990)}]{DW9093}%
  \BibitemOpen
  \bibfield  {author} {\bibinfo {author} {\bibfnamefont {R.}~\bibnamefont
  {Dijkgraaf}}\ and\ \bibinfo {author} {\bibfnamefont {E.}~\bibnamefont
  {Witten}},\ }\href@noop {} {\bibfield  {journal} {\bibinfo  {journal} {Comm.
  Math. Phys.}\ }\textbf {\bibinfo {volume} {129}},\ \bibinfo {pages} {393}
  (\bibinfo {year} {1990})}\BibitemShut {NoStop}%
\bibitem [{OCo(2013)}]{OCob}%
  \BibitemOpen
  \href@noop {} {\bibfield  {journal} {\bibinfo  {journal}
  {http://www.map.mpim-bonn.mpg.de/Oriented\_bordism}\ } (\bibinfo {year}
  {2013})}\BibitemShut {NoStop}%
\bibitem [{\citenamefont {Randal-Williams}(2011)}]{R11}%
  \BibitemOpen
  \bibfield  {author} {\bibinfo {author} {\bibfnamefont {O.}~\bibnamefont
  {Randal-Williams}},\ }\href@noop {} {\bibfield  {journal} {\bibinfo
  {journal} {http://mathoverflow.net/questions/75374}\ } (\bibinfo {year}
  {2011})}\BibitemShut {NoStop}%
\bibitem [{\citenamefont {Brown}(1982)}]{B8283}%
  \BibitemOpen
  \bibfield  {author} {\bibinfo {author} {\bibfnamefont {E.~H.}\ \bibnamefont
  {Brown}},\ }\href@noop {} {\bibfield  {journal} {\bibinfo  {journal}
  {Proceedings of the American mathematical society}\ }\textbf {\bibinfo
  {volume} {85}},\ \bibinfo {pages} {283} (\bibinfo {year} {1982})}\BibitemShut
  {NoStop}%
\bibitem [{UOC(2013)}]{UOCob}%
  \BibitemOpen
  \href@noop {} {\bibfield  {journal} {\bibinfo  {journal}
  {http://www.map.mpim-bonn.mpg.de/Unoriented\_bordism}\ } (\bibinfo {year}
  {2013})}\BibitemShut {NoStop}%
\bibitem [{Note2()}]{Note2}%
  \BibitemOpen
  \bibinfo {note} {For a simple introduction on classifying space, see the Wiki
  article ``Classifying space''. For a continuous group $G$, the classifying
  space $BG$ in this paper is defined with real manifold topology on
  $G$.}\BibitemShut {Stop}%
\bibitem [{\citenamefont {Thomas}(1960)}]{T6067}%
  \BibitemOpen
  \bibfield  {author} {\bibinfo {author} {\bibfnamefont {E.}~\bibnamefont
  {Thomas}},\ }\href@noop {} {\bibfield  {journal} {\bibinfo  {journal}
  {Transactions of the American Mathematical Society}\ }\textbf {\bibinfo
  {volume} {96}},\ \bibinfo {pages} {67} (\bibinfo {year} {1960})}\BibitemShut
  {NoStop}%
\bibitem [{PS(2011)}]{PS}%
  \BibitemOpen
  \href@noop {} {\bibfield  {journal} {\bibinfo  {journal}
  {http://www.encyclopediaofmath.org/index.php/Pontryagin\_square}\ } (\bibinfo
  {year} {2011})}\BibitemShut {NoStop}%
\bibitem [{\citenamefont {Spanier}(1966)}]{Spa66}%
  \BibitemOpen
  \bibfield  {author} {\bibinfo {author} {\bibfnamefont {E.~H.}\ \bibnamefont
  {Spanier}},\ }\href@noop {} {\emph {\bibinfo {title} {Algebraic Topology}}}\
  (\bibinfo  {publisher} {McGraw-Hill},\ \bibinfo {address} {New York},\
  \bibinfo {year} {1966})\BibitemShut {NoStop}%
\bibitem [{\citenamefont {Lyndon}(1948)}]{L4871}%
  \BibitemOpen
  \bibfield  {author} {\bibinfo {author} {\bibfnamefont {R.~C.}\ \bibnamefont
  {Lyndon}},\ }\href {\doibase 10.1215/S0012-7094-48-01528-2} {\bibfield
  {journal} {\bibinfo  {journal} {Duke Mathematical Journal}\ }\textbf
  {\bibinfo {volume} {15}},\ \bibinfo {pages} {271 } (\bibinfo {year}
  {1948})}\BibitemShut {NoStop}%
\bibitem [{\citenamefont {Hochschild}\ and\ \citenamefont
  {Serre}(1953)}]{HS5310}%
  \BibitemOpen
  \bibfield  {author} {\bibinfo {author} {\bibfnamefont {G.}~\bibnamefont
  {Hochschild}}\ and\ \bibinfo {author} {\bibfnamefont {J.-P.}\ \bibnamefont
  {Serre}},\ }\href {\doibase 10.2307/1990851} {\bibfield  {journal} {\bibinfo
  {journal} {Transactions of the American Mathematical Society (American
  Mathematical Society)}\ }\textbf {\bibinfo {volume} {74}},\ \bibinfo {pages}
  {110 } (\bibinfo {year} {1953})}\BibitemShut {NoStop}%
\end{thebibliography}%

\end{document}